\documentclass[12pt]{article}
\usepackage[sort&compress,square,comma,numbers]{natbib}
\usepackage{amsmath,amssymb,graphicx,slashed,bm}
\usepackage{array}
\usepackage{feynmp-auto}
\usepackage{subcaption}
\usepackage{placeins}
\usepackage{tikz}
\usetikzlibrary{arrows.meta,calc}
\usepackage{simpler-wick}
\usepackage[normalem]{ulem}

\usepackage[
	colorlinks=true,
	citecolor=black,
	linkcolor=black,
	urlcolor=blue,
	hypertexnames=false]{hyperref}

\makeatletter
\def\endfmffile{%
	\fmfcmd{\p@rcent\space the end.^^J%
		end.^^J%
		endinput;}%
	\if@fmfio
	\immediate\closeout\@outfmf
	\fi
	\ifnum\pdfshellescape>\z@
	\immediate\write18{mpost \thefmffile}%
	\fi}
\makeatother

\newcommand{\arXiv}[2]{\href{http://arxiv.org/pdf/#1}{{\tt #2/#1}}}
\newcommand{\arXivold}[1]{\href{http://arxiv.org/pdf/#1}{{\tt #1}}}

\def\tilde#1{\widetilde{#1}}

\newcommand{\beq}{\begin{eqnarray}}
\newcommand{\eeq}{\end{eqnarray}}

\newcommand{\dual}{{}^{\star}\!}
\newcommand{\Cmat}{\mathbf{C}}
\newcommand{\Gmat}{\mathbf{G}}
\newcommand{\Qvec}[1]{\bm{q}_{#1}}
\newcommand{\epsmat}{\boldsymbol{\epsilon}}
\newcommand{\betamat}{\boldsymbol{\beta}}
\newcommand{\Gammaem}{\Gamma_{\rm em}}
\newcommand{\Gammadyn}{\Gamma_{\rm dyn}}
\newcommand{\dlat}{N}

\newcommand{\Proj}{\mathcal P}

\begin{document}
\begin{center} 
{\huge \bf  Running Couplings in \\
	 \vskip 6pt
	 CP Violating  Abelian Theories
} 
\\
\end{center}
\vskip 8pt 
\begin{center} 

{\bf Joshua Newey}$^1$, {\bf John Terning}$^1$, and {\bf Christopher B. Verhaaren}$^2$ \\
\end{center}
\vskip 8pt
\begin{center} 
$^1${\it Center for Quantum Mathematics and Physics (QMAP)\\Department of Physics, University of California, Davis, CA 95616}\\
$^2${\it Department of Physics and Astronomy, Brigham Young University, \\ Provo, UT 84602, USA}
\end{center}

\vspace*{0.1cm}
\begin{center} 
{\tt 
 \href{mailto:jjnewey@ucdavis.edu}{jnewey@ucdavis.edu}\,
 \href{mailto:jterning@gmail.com}{jterning@gmail.com}\,
 \href{mailto:verhaaren@physics.byu.edu}{verhaaren@physics.byu.edu}}

\end{center}
\centerline{\large\bf Abstract}
\begin{quote}
We calculate the one- and two-loop running couplings for a massless photon coupled to dyons, including CP violating couplings. The electric and magnetic source vertices in the most general theory form a real coupling matrix $\Cmat$, while a current-current interaction depends on $\Cmat^T\Cmat$ and the quantized determinant $\det\Cmat$. These determine, respectively, the continuously running Coulombic interactions and the fixed antisymmetric Dirac--Schwinger--Zwanziger charge quantization condition which also determines pairwise helicity. The running preserves $\det\Cmat$, showing that the electric and magnetic couplings run inversely while every pairwise helicity remains constant. We comment on how the running of the CP-odd coupling $\theta$ relates to physical CP violation, which also depends on the spectrum, mass phases, and threshold corrections. We also show that an invertible field redefinition moves the CP-odd coupling between the current source terms and the photon kinetic terms;  these are two descriptions of the same local interaction. Finally, we show that the result agrees with the Argyres--Douglas $\beta$ function ansatz.
\end{quote}

\section{Introduction\label{s.Intro}}

In standard quantum electrodynamics (QED) there is no CP violation. The Maxwell equations are
\beq
\partial_\mu F^{\mu\nu}=eJ^\nu~, \ \ \ \ \partial_\mu\dual F^{\mu\nu}=0~,\label{e.firstMax}
\eeq
where $e$ and $J^\mu$ are the electric coupling and current, respectively. We take the Minkowski metric to have the form $\eta_{\mu\nu}=\operatorname{diag}(+,-,-,-)$ and the Hodge dual to be
\beq
\dual F^{\mu\nu}=\frac12\varepsilon^{\mu\nu\alpha\beta}F_{\alpha\beta}~,
\eeq 
with the Levi-Civita tensor specified by $\varepsilon^{0123}=+1$. Because it satisfies the Bianchi identity, $\partial_\mu\dual F^{\mu\nu}=0$, the field strength can be expressed in terms of a vector potential $A_\mu$ as
\beq
F_{\mu\nu}=\partial_\mu A_\nu-\partial_\nu A_\mu\equiv F^A_{\mu\nu}~.
\eeq
When the Lagrangian
\beq
\mathcal{L}=-\frac14 F_{\mu\nu}F^{\mu\nu}-eA_\mu J^\mu~,
\eeq
is varied with respect to the potential, the correct Maxwell equations~\eqref{e.firstMax} are produced. Adding a CP violating $\theta$ term to the Lagrangian is innocuous, because it is a total derivative
\beq
\dual F^{\mu\nu}F_{\mu\nu}=4\frac12\varepsilon^{\alpha\beta\gamma\delta}\partial_\alpha\left(A_\beta\right)\partial_\gamma A_\delta=2\partial_\alpha\left(\varepsilon^{\alpha\beta\gamma\delta}A_\beta \partial_\gamma A_\delta\right)~.
\eeq

Theories with magnetic currents $K^\mu$ \cite{Dirac:1948um,Zwanziger:1970hk,Blagojevic:1985sh} are more interesting: $\theta$ becomes observable through the Witten effect~\cite{Witten:1979ey}. In this case
\beq 
\partial_\mu\dual F^{\mu\nu}=bK^\nu~,
\eeq
where $b$ is the magnetic coupling. This implies that $F_{\mu\nu}\neq F^A_{\mu\nu}$, but following Dirac~\cite{Dirac:1948um}, one can still define a vector potential because 
\beq
\partial_\mu\left(\dual F^{\mu\nu}-b\Sigma_K^{\mu\nu}\right)=0~,
\eeq
where the Dirac surface $\Sigma_K^{\mu\nu}$ \cite{Dirac:1948um,Hull:2024uwz,Newey:2025ygq} satisfies
\beq
\partial_\mu\Sigma_K^{\mu\nu}=K^\nu~.
\eeq
One possibility is
\beq
\Sigma_K^{\mu\nu}=\frac{1}{n\cdot\partial}\left(n^\mu K^\nu-n^\nu K^\mu \right)~,
\label{eq:SigmanDef}
\eeq
for a constant vector $n^\mu$ and $1/(n\cdot\partial)$ a suitably defined integral operator (see Appendix~\ref{app:WittenSurfaceRepresentatives}). With this choice it is straightforward to show within the action how the Witten effect arises:
\beq
\frac{1}{4}\frac{\theta}{2\pi}\frac{e}{b}\dual F^{\mu\nu}F_{\mu\nu}\rightarrow-eA_\mu\frac{\theta}{2\pi}K^\mu~,\label{e.WittenOp}
\eeq
where the minus sign come from integration by parts. This analysis is somewhat schematic, but a more careful\textemdash and general\textemdash derivation is provided in Appendix~\ref{app:WittenSurfaceRepresentatives}. The operator on the left side of~\eqref{e.WittenOp} integrates to a winding number. The above analysis shows that this winding number has physical significance. This is directly related to the nontrivial Abelian instantons that arise in the presence of magnetic monopoles~\cite{Csaki:2024ajo,GarciaGarcia:2025uub}.

In short, a dyon with integral electric and magnetic charges $(q,g)$ carries the Witten-shifted electric charge $q+g\theta/2\pi$, and vacuum polarization can renormalize both the ordinary gauge coupling and $\theta$. In addition, dyonic theories of this type satisfy the Dirac--Schwinger--Zwanziger (DSZ) quantization condition~\cite{Dirac:1931kp,Dirac:1948um,Schwinger:1966mag,Zwanziger:1968rs}
\beq
eb\left(q_ig_j-q_jg_i\right)=2\pi\mathfrak{n}~,
\eeq
where $\mathfrak{n}$ is an integer. The central question of this paper is: how do the couplings run while preserving the DSZ condition?

This question is also the source of a long-standing disagreement. Schwinger and others calculated that electric $e$ and magnetic $b$ couplings renormalize proportionally, while Coleman argued that the two couplings must run inversely so that their quantized product is preserved~\cite{Schwinger:1966zza,Schwinger:1966zzb,Brandt:1977fa,Deans:1981qs,Panagiotakopoulos:1982fp,Coleman:1982cx}. The difficulty in reconciling these two calculations can be seen from the two dyon scattering portion of the 1PI effective action $\Gamma_{d_id_j}$. For dyons $d_i$ and $d_j$, the effective action contains two qualitatively different structures
\beq
\Gamma_{d_id_j}=\left[e^2\left(q_i+g_i\frac{\theta}{2\pi} \right) \left(q_j+g_j\frac{\theta}{2\pi} \right)+b^2g_ig_j \right]\Gamma_\text{C}+eb\left(q_ig_j-q_jg_i \right)\Gamma_\text{DSZ}~,
\eeq
where $\Gamma_\text{C}$ accounts for Coulomb-like interactions and $\Gamma_\text{DSZ}$ accounts for the cross product part of the Lorentz force. The ordinary Coulombic exchange term is a symmetric bilinear in the charges and can renormalize continuously. The second term is the antisymmetric DSZ interaction. Its coefficient is quantized and cannot become an independent running coupling.  

When considering mutually nonlocal interactions one must carefully define the charges and the theory. The DSZ charge quantization implies that all possible charges in a U(1) theory are isolated points. It can be useful to consider the charges as populating a lattice with orthogonal basis vectors of unit length. In an oriented, integral basis for this electromagnetic charge lattice the charge of particle species $i$ can be written as a vector pointing to a lattice position
\begin{align}
	\Qvec{i}=\begin{pmatrix}q_i\\ g_i\end{pmatrix}~.
\end{align}
The combination of charges that appears in the DSZ condition is
\begin{align}
	\Qvec{i}^{T}\epsmat\,
	\Qvec{j}=q_ig_j-g_iq_j~,
	\qquad
	\epsmat=\begin{pmatrix}0&1\\-1&0\end{pmatrix}~.
	\label{eq:introChargePairing}
\end{align}
This is exactly the oriented area obtained from taking the wedge product of the two charge vectors.

The rationalized electric and magnetic couplings obey $eb=2\pi\dlat$, where $\dlat\in\mathbb Z_{>0}$ is the normalization of the charge lattice. The cases $\dlat=1$ (often referred to as the primitive lattice) and $\dlat=2$ therefore describe different physical charge lattices. The different normalizations are employed to map a theory with \emph{rational} charges to an equivalent description with \emph{integer} charges.

While a given charge lattice contains infinitely many possible particle charge combinations, only charges realized as dynamical states enter loop sums. In this work we focus our attention on these loop effects. However, the full lattice and the spin assignment of each charged field determine the allowed line operators, exact dualities, and the physical periodicity of $\theta$~\cite{Witten:1995gf,Aharony:2013hda,Ang:2019line}.

In a CP preserving theory the coupling of the photon to electric or magnetic currents is fixed~\cite{Weinberg:1965rz}. If we allow violations of CP then a continuous family of possible couplings is allowed, as described in Appendix~\ref{app:helicityParityCP}. The most general couplings of a single photon to electromagnetic sources are efficiently organized by a real $2\times2$ coupling matrix $\Cmat$. This can be displayed without referring to a specific Lagrangian formulation because the couplings are determined by the equations of motion
\beq
\partial_\mu\boldsymbol{F}^{\mu\nu}
=\Cmat\boldsymbol{J}^{\nu}~,\label{eq:MatMax}
\eeq
where
\begin{align}
	\boldsymbol{F}^{\mu\nu}
	&\equiv
	\begin{pmatrix}F^{\mu\nu}\\\dual F^{\mu\nu}\end{pmatrix}~,
	\quad
	\boldsymbol{J}^{\nu}
	\equiv
	\begin{pmatrix}J^\nu\\K^\nu\end{pmatrix}~.\label{eq:introMatrixMaxwell}
\end{align}
Note, however, that the field strength and current vectors belong to different spaces. The first describes the photon field while the second deals with the charged sources. We also observe that $\boldsymbol{F}^{\mu\nu}$ satisfies the constraint
\beq
{}^\star\! \boldsymbol{F}^{\mu\nu}= \begin{pmatrix}\dual F^{\mu\nu}\\- F^{\mu\nu}\end{pmatrix}=
\boldsymbol{\epsilon} \boldsymbol{F}^{\mu\nu}~.
\eeq

As discussed in Sec.~\ref{sec:couplingMatrix}, CP violating theories allow a continuum of photon helicity phase choices, each of which can be used as a basis for calculations. Rephasing the photon helicities amounts to a rotation in the space of CP eigenstates $F_{\mu\nu}$ and ${}^\star\! F_{\mu\nu}$. Consequently, $\Cmat$ is best understood as a zweibein\footnote{A two-dimensional vielbein.} $C^a_{\phantom{a}z}$ where Latin indices from the beginning of the alphabet correspond to the space of CP eigenstates, while Latin indices from the end of the alphabet relate to the source space, which contains the charge lattice. One can then rewrite the Maxwell equations as
\beq
\partial_\mu F^{a\mu\nu}=C^a_{\phantom{a}z} J^{z\nu}~.
\eeq
While choosing a different helicity basis \cite{Colwell:2015wna} has no effect upon the charged particle currents, it does cause $\Cmat$ to transform. The Witten effect discussion above employed a helicity basis in which
\beq
\Cmat_0&=
\begin{pmatrix}
	e&\displaystyle \frac{e\theta}{2\pi}\\[3mm]
	0&b
\end{pmatrix}~,
\eeq
which we refer to as the Witten basis. Physical observables (as well as currents and charges) are independent of basis choice, so it is useful to recognize combinations of $\Cmat$ that are basis invariant.

As shown below, the basis-independent, one-photon exchange amplitude is determined by
\begin{align}
	\Gmat\equiv\Cmat^T\Cmat~,
	\qquad
	\Delta\equiv\det\Cmat=2\pi\dlat~,
	\label{eq:introGram}
\end{align}
where the value of $\Delta$ follows from DSZ charge quantization on a charge lattice with normalization $\dlat$. The matrix $\Gmat$
\beq
G^{y}_{\phantom{y}z}=\left(C^T\right)^{y}_{\phantom{y}a}C^a_{\phantom{a}z}
\eeq
acts only on the charge space.\footnote{$\Gmat$ can also be understood as an induced metric on the charge space.} It controls the symmetric, Coulombic photon-exchange interaction, while $\Delta\epsmat$ describes the antisymmetric DSZ interaction. 

In the Witten basis and the electric duality frame\footnote{The frame in which all electric couplings are local in a single gauge potential formalism.} the holomorphic coupling is
\begin{align}
	\tau&=\frac{\theta}{2\pi}+i\frac{b}{e}=\frac{\theta}{2\pi}+i\frac{2\pi\dlat}{e^2}~,
	\label{eq:introElectricFrameNormalization}
\end{align}
which relates to $\Gmat$ according to
\begin{align}
	\Gmat
	=\frac{\Delta}{\operatorname{Im}\tau}
	\begin{pmatrix}
		1&\operatorname{Re}\tau\\[1mm]
		\operatorname{Re}\tau&|\tau|^2
	\end{pmatrix}~.
	\label{eq:introGTauMap}
\end{align}
Since $\Gmat$ and $\Delta$ are independent of helicity basis we see that $\Cmat$ is constrained. There is not a third independent coupling (in addition to $e$ and $\theta$) in the CP violating Maxwell equations. Using $\Cmat$ is simply a bookkeeping device that keeps the renormalization group (RG) running of the Coulombic interaction separate from the quantized DSZ term. This is consistent with the determinant of $\Cmat$ being fixed by charge quantization and so should not run.

\begin{center}
	\begin{tikzpicture}[x=1cm,y=1cm,>=Latex]
		\node (C) at (0,0) {$(\Qvec{},\Cmat)$};
		\node (GD) at (3.1,0) {$(\Qvec{},\Gmat,\Delta)$};
		\node (Gamma) at (6.9,0) {$\Gamma_{\rm tree}+\Gamma_{\rm VP}$};
		\node (betaG) at (10.2,0) {$\bm{\beta}_{\Gmat}$};
		\node (betaTau) at (12.3,0) {$\beta_\tau$};
		\draw[-{Latex[length=2.2mm]},thick] (C) -- (GD);
		\draw[-{Latex[length=2.2mm]},thick] (GD) -- (Gamma);
		\draw[-{Latex[length=2.2mm]},thick] (Gamma) -- (betaG);
		\draw[-{Latex[length=2.2mm]},thick] (betaG) -- (betaTau);
	\end{tikzpicture}
	
	\smallskip
	\parbox{0.88\textwidth}{\small\emph{Logical flow of calculations. The charges and coupling matrix determine the invariant one-photon data; the tree and vacuum-polarization (VP) two-current contributions to the effective action then determine how $\Gmat$ runs, which can be expressed in terms of $\tau$.}}
\end{center}

In this work we explicitly calculate the ultraviolet-divergent, local part of the 1PI, two-current part of the effective action at both one- and two-loop order. Unless stated otherwise, the theory we consider contains electric and magnetic charges coupled to one standard massless Maxwell photon represented by two Zwanziger potentials. We refer to this restriction on the two gauge potentials as the ``one-photon condition." 

To calculate the running of $\Gmat$ we, in  Sec.~\ref{sec:sourceQuadraticMain}, generalize Zwanziger's formalism to account for CP violation. We show that, within the Lagrangian, the CP-odd coupling can be confined to the photon kinetic terms, only appear in the photon source term, or be shared between the two. Invertible field redefinitions relate these Lagrangians. Therefore, within the one-photon class of theories studied here, each constitutes a local description of the single CP-violating interaction rather than separate contributions to the $\beta$-function. 

In Sec.~\ref{sec:Feynman} we use this formalism to calculate the tree-level scattering of two dyons. To obtain the $\beta$-function, we must calculate the local, mass-independent 1PI contribution to the vacuum-polarization effective action rather than the complete on-shell one-loop amplitude. Finite threshold matching can be treated separately. Section~\ref{sec:localCounterterm} describes how to isolate the local ultraviolet part of the effective action that governs the running. This includes a discussion of how to recognize, isolate, and remove topological terms that can confuse the perturbative calculation.

In Sec.~\ref{sec:runningMain} we calculate our first loop-level result. Namely, in theories of vectorlike Dirac fermions in a fixed chiral convention with a perturbative conserved-current correlator, the one-loop coupling matrix runs as 
\begin{align}
	\betamat_{\Gmat}
	=\frac{1}{6\pi^2}\sum_\ell
	\left[
	\Gmat\Qvec{\ell}\Qvec{\ell}^{T}\Gmat
	+\Delta^2\epsmat\Qvec{\ell}\Qvec{\ell}^{T}\epsmat
	\right]~,
	\label{eq:introBetaG}
\end{align}
which is equivalent to
\begin{align}
	\beta_\tau
	=-\frac{i\,\dlat}{3\pi}\sum_\ell
	\left(q_\ell+g_\ell\tau\right)^2~.
	\label{eq:introBetaTau}
\end{align}
We also show that $\det\Gmat=\Delta^2$ is preserved by one-loop running. This ensures that the electric and magnetic couplings run inversely, while $eb$, every DSZ pairing, and every pairwise helicity remain fixed. In these calculations we assume a perturbative expansion, which we justify in Sec.~\ref{sec:perturbativeRegimes}. We also check this running against several known results. Our calculation is clearest when using $\Gmat$ and $\Delta$ while recognizing their qualities and invariances.

Of course, as $\tau$ depends on duality frame, so does its running. In particular, the $\theta$ running in Eq.~\eqref{eq:introBetaTau} is a statement about couplings in a fixed  duality frame. The advantage of Eq. \eqref{eq:introBetaG} is that it transforms covariantly under duality making it straightforward to determine the running in any frame. 

Theories with $\theta\neq0$ may or may not violate CP.  The duality-invariant question, which we address in Sec.~\ref{sec:cpSpectrum}, is whether the renormalization group flow remains tangent to the CP invariant curve in theory space. That is, CP invariance depends on more than the value of $\theta$. One must also consider the spectrum of states, their masses and mass phases and multiplicities, their interactions, threshold corrections, and the vacuum. We show explicitly that if a theory exhibits CP symmetry at tree-level then the one-loop running preserves that symmetry.

In addition to several limiting case checks on our results, we also compare our calculation to the protected Wilsonian contribution of an $\mathcal N=2$ hypermultiplet in Sec.~\ref{sec:ADComparison}. Most interestingly, we find exact agreement with Argyres--Douglas (AD). As we explain in that section, the degree-two AD isogeny\footnote{A degree-two isogeny is a two-to-one map between elliptic curves that preserves their group structure.} explains the normalization of each single-state term in the $\beta$-function, but does not prove additivity across mutually nonlocal sectors, as assumed by AD. Consequently, the agreement of our results with the AD ansatz is quite nontrivial and worthy of further investigation.

Section~\ref{sec:twoLoopDualityPrediction} extends our calculation to two-loop order. The two-loop running of $\Gmat$, and the fact that $\Delta$ does not run at two loops, is also expressed in terms of the holomorphic coupling $\tau$ and the more familiar $e$, $b$, and $\theta$. We also show that the two-loop RG flow preserves the CP symmetry if it is already present. Finally, we discuss the scheme independence of these results and comment on extensions beyond two-loop order, before concluding in Sec.~\ref{s.con}.

To increase the readability of our work we have confined many detailed calculations to the appendices. We hope the main body of the paper provides a quick, but complete, path through our most significant results. The appendices contain transformation formulas, propagator algebra, worldline derivations, and additional detail to aid interested readers in reproducing or extending our work.

\section{Coupling Matrix and Physical Invariants}\label{sec:couplingMatrix}

As described in the introduction, the Witten source term
\begin{align}
	\mathcal{L}_{\text{int}}
	= -eA_\mu\left(J^\mu+\frac{\theta}{2\pi}K^\mu\right)~,
	\label{eq:WittenSource}
\end{align}
describes the couplings in the electric duality frame and a particular helicity basis. To make calculations that are independent of helicity phase, we write
\begin{align}
\Cmat=
\begin{pmatrix}x_J&x_K\\y_J&y_K\end{pmatrix}
=\bigl(\mathbf c_J\ \mathbf c_K\bigr)~.
\label{eq:GenMax}
\end{align}
The column vectors $\mathbf c_J$ and $\mathbf c_K$ are made from the components of the one-photon helicity vertices; that is, we can write the Maxwell equations as
\begin{align}
\partial_\mu F^{\mu\nu}_+=\partial_\mu\left(F^{\mu\nu}+i\,\dual F^{\mu\nu} \right)=\left(x_J+i\,y_J\right)J^\nu+\left(x_K+i\,y_K \right)K^\nu~.\label{eq:HeliEq}
\end{align}
Charge quantization fixes the oriented area $\mathbf c_J\wedge \mathbf c_K$, or in other words
\begin{align}
2\pi\dlat=\Delta=\det\Cmat
=\mathbf c_J^T\epsmat\,\mathbf c_K~.
\label{eq:DeltaDef}
\end{align}
The different values of $\dlat$ correspond to different electromagnetic charge lattices, not different notations for the same lattice.

A helicity rephasing $F_+^{\prime\mu\nu}=e^{-i\phi}F_+^{\mu\nu}$ changes the real coordinates used to display the same one-photon amplitude. 
However, the currents, charge labels, and theory parameters are fixed. 
This distinguishes these transformations from the well-known SL$(2,\mathbb{R})$ duality transformations of classical electromagnetism.
This rephasing produces the transformed quantities 
\begin{align}
F^{\prime a\mu\nu}=R^a_{\phantom{a}b}(\phi)F^{b\mu\nu}~,
\qquad
C^{\prime a}_{\phantom{\prime a}z}=R^a_{\phantom{a}b}(\phi)C^{b}_{\phantom{b}z}~,
\qquad
\mathbf{R}(\phi)=
\begin{pmatrix}\cos\phi&-\sin\phi\\\sin\phi&\cos\phi\end{pmatrix}~,
\label{e.paramRot}
\end{align}
so that
\begin{align}
\partial_\mu\boldsymbol{F}^{\prime\mu\nu}
=\Cmat'\boldsymbol{J}^{\nu}
\end{align}
Since $\mathbf{R}(\phi)$ is orthogonal, amplitudes can only depend on the $\phi$-independent combinations
\begin{align}
\Gmat=\Cmat^T\Cmat
=\begin{pmatrix}
\mathbf c_J^2&\mathbf c_J\cdot\mathbf c_K\\
\mathbf c_J\cdot\mathbf c_K&\mathbf c_K^2
\end{pmatrix}~,
\qquad
\Cmat^T\epsmat\,\Cmat=\Delta\epsmat~.
\label{eq:GramAndArea}
\end{align}
As shown in Sec.~\ref{sec:treeDyonScattering}, $\Gmat$ determines tree-level Coulombic photon exchange, while $\Delta\epsmat$ controls the DSZ interaction. 
By construction 
\begin{align}
\det\Gmat=\Delta^2~,
\label{eq:GramConstraint}
\end{align}
is fixed. Therefore, only two entries of $\Gmat$ are independent. These can be discovered by associating each column of $\Gmat$ with the complex couplings that appear in Eq.~\eqref{eq:HeliEq} and looking at the helicity invariant ratio 
\begin{align}
	\tau = \frac{
	x_K + i y_K
	}{
	x_J + iy_J
	} = \frac{
	\mathbf{c}_J\cdot \mathbf{c}_K + i\Delta }
	{
	\mathbf{c}_J^2
	}~.
\end{align}
This gives the holomorphic coupling of the theory (see for example \cite{Terning:2018lsv}).

The charge lattice is not modified by changing the helicity basis, but it is affected by SL$(2,\mathbb{Z})$ duality transformations. These are given by 
\begin{align}
	\bm{q} = 
	\mathbf{D} \bm{q}'~, \ \ \ 
	\boldsymbol{J}^\mu = 
	\mathbf{D} \bm{J}^{\prime\mu}~, \ \ \ 
	\mathbf{D} = 
	\begin{pmatrix}
		\delta & \beta \\
		\gamma & \alpha
	\end{pmatrix}
	\in \text{SL}(2,\mathbb{Z})~.
\end{align}
Under these transformations we obtain the standard transformation \cite{GaillardZumino:1981} of the holomorphic coupling from
\begin{align}
	\Cmat' = \Cmat \mathbf{D}~,
	\ \ \
	\Gmat' = \mathbf{D}^T \Gmat \mathbf{D}~,
	\ \ \ 
	\tau' = 
	\frac{\alpha\tau + \beta}{\gamma \tau + \beta}~.
\end{align}
We see that $\Gmat$ transforms as a tensor under duality, while the Maxwell equations~\eqref{eq:MatMax} imply that $\boldsymbol{F}^{\mu\nu}$ is invariant.
This is a choice and other conventions differ from this by a helicity rephasing or a change of normalization (see for example~\cite{Csaki:2010rv,Terning:2018lsv}).

Dyon charge vectors $\Qvec{i}$ and $\Qvec{j}$ align with the photon vertices for certain helicity phases. 
Their antisymmetric product gives the pairwise helicity~\cite{Zwanziger:1972sx,Csaki:2020inw,Csaki:2022pairwise},
\begin{align}
h_{ij}^{\rm pair}
&\equiv
\frac{\Delta}{4\pi}\,\Qvec{i}^{T}\epsmat\,\Qvec{j}
 =
 \frac{\dlat}{2}\left(q_i g_j-q_j g_i\right)
 ~.
\label{eq:pairwiseHelicityDSZ}
\end{align}
The minimal electric and magnetic generators ($q_i=g_j=1$ and $q_j=g_i=0$) carry pairwise helicity $\dlat/2$. The same invariant controls the additional pairwise angular-momentum quantum number of a mutually nonlocal two-particle state. In Sec.~\ref{sec:Feynman}
we find that the quantized antisymmetric invariant $\Delta\Qvec{i}^T\epsmat\,\Qvec{j}$ and the symmetric bilinear $\Qvec{i}^T\Gmat\,\Qvec{j}$ are precisely the two structures that appear in the physical amplitudes. 
This is not surprising, as these are the two duality invariants which can be constructed from $\Gmat$, $\bm{q}_i$, and $\bm{q}_j$.

\subsection{Witten-Basis Variables}
We can choose helicity coordinates in which $\mathbf c_J$ lies along the electric axis. Defining $e=|\mathbf c_J|$ and $b=\Delta/e$, we find the usual Witten effect form
\begin{align}
\Cmat_0=
\begin{pmatrix}
 e&\displaystyle \frac{e\theta}{2\pi}\\[3mm]
 0&b
\end{pmatrix}~,
\qquad
eb=\Delta=2\pi\dlat~,
\label{eq:Cstandard}
\end{align}
and the basis invariant
\begin{align}
\Gmat=
\begin{pmatrix}
 e^2&e\displaystyle \frac{e\theta}{2\pi} \\[3mm]
 e\displaystyle \frac{e\theta}{2\pi}&\displaystyle  b^2+\frac{e^2\theta^2}{4\pi^2}
\end{pmatrix}~.
\label{eq:Gstandard}
\end{align}
Equivalently, the same information can be expressed in the holomorphic coupling
\begin{align}
\tau
=\frac{\theta}{2\pi}+i\frac{b}{e}~.
\label{eq:tauFromMatrix}
\end{align}
For fixed $\Delta$, the invariants $\Gmat$ and $\tau$ contain the same two continuous couplings, $e$ and $\theta$. The matrix notation, however, keeps the symmetric interaction (which runs) separate from the RG invariant DSZ pairing (which does not). The condition $\mathbf c_J\cdot\mathbf c_K=0$ indicates $\theta=0$ in the Witten basis, but physical CP invariance also requires the spectral conditions discussed in Sec.~\ref{sec:cpSpectrum}.

\section{A Generalized Zwanziger Lagrangian}\label{sec:sourceQuadraticMain}

Zwanziger showed how theories with electric and magnetic charges can be described by a Lagrangian with a local coupling to potentials at the cost of introducing dependence on a constant four vector $n^\mu$ \cite{Zwanziger:1970hk}. As shown in \cite{Brandt:1977be}, this dependence is an artifact of the Lagrangian description and does not contribute to the quantum theory. We summarize and generalize this argument in Appendix~\ref{app:worldlineTopology}.

We now generalize Zwanziger's formalism to include CP violating couplings. As sketched in the introduction (and shown in Appendix~\ref{app:WittenSurfaceRepresentatives}) the $\theta$ source term in the one-potential formalism is equivalent to a $\theta$ dependent kinetic term. We show below that the same is true of two-potential Lagrangians.

To compare source and kinetic CP violation, we must first generalize Zwanziger to include a CP-odd kinetic term. That is, we need a two-potential theory of a single photon that produces the CP violating Maxwell equations. In Appendix~\ref{app:Zwan} we determine a class of two-potential theories\textemdash which includes CP preserving Zwanziger\textemdash that are single photon theories. These have Lagrangians of the form
\begin{align}
	\mathcal{L}_{Z}&= 
	-\frac{1}{4} \bm{F}_{\bm{A}}^T{}_{\alpha\beta}\, {}^{\alpha\beta}\mathbf{K}^{\gamma\delta}\, \bm{F}_{\bm{A}}{}_{\gamma\delta}
	- \bm{A}^T_\mu \mathbf{S}	\bm{J}^\mu~,	\label{eq:GenZwanLag}
\end{align}
which produces the equations of motion
\begin{align}
\partial_{\mu}\,  {}^{\mu\nu}\mathbf{K}^{\alpha\beta}\, \bm{F}_{\bm{A}}{}_{\alpha\beta}=\mathbf{S}\bm{J}^\nu~.
\label{eq:GenEOMZwan}
\end{align}
To produce the correct Maxwell equations~\eqref{eq:MatMax} we see that the physical field strength is related to the potentials by
\begin{align}
\bm{F}^{\mu\nu}=\Cmat\mathbf{S}^{-1}\,{}^{\mu\nu}\mathbf{K}^{\alpha\beta}\, \bm{F}_{\bm{A}}{}_{\alpha\beta}~.
\end{align}

In the above we have used the notation
\begin{align}
	\bm{A}_{\mu} = 
	\begin{pmatrix}
		A_\mu \\
		B_\mu
	\end{pmatrix}, 
	 \ \ \ \ 
	\bm{F}^{\mu\nu}_{\bm{A}}
	= \partial^\mu \bm{A}^{\nu}
	- \partial^\nu \bm{A}^{\mu}~,
\end{align}
and
\begin{align}
	{}^{\alpha\beta}\mathbf{K}^{\gamma\delta}= 
	{}_{(n)}^{\alpha\beta}\mathcal{P}_{(n)}^{\gamma\delta}
	\,
	\mathbf{H}
	- \kappa
	{}^\star{}_{(n)}^{\alpha\beta}\mathcal{P}_{(n)}^{\gamma\delta}
	\,
	\epsmat~.
\end{align}
Here $\mathbf{H}$ is an arbitrary real, positive definite, $2 \times 2$, symmetric matrix, $\kappa$ is a real number, and $\mathbf S$ describes what combination of the currents is locally coupled to each potential. The objects ${}^{\mu\nu}_{(v)}\mathcal{P}_{(u)}^{\alpha\beta}$ are projectors on the space of two-forms and their properties are described in Appendix~\ref{app:Zwan} (see also~\cite{Newey:2024gug}).
For instance, given a two-form $X_{\rho\sigma}$, the action of the operator is
\begin{align}
 {}^{\mu\nu}_{(v)}\!\mathcal{P}_{(u)}^{\rho\sigma}X_{\rho\sigma}=\frac{1}{u\cdot v}\left(v^\mu u_\alpha X^{\alpha\nu}-v^\nu u_\alpha X^{\alpha \mu} \right)~.
\end{align}
 These projectors are built out of vectors $v^\mu$ and $u^\mu$, the former associated with the left indices and the latter with the right indices. In the Lagrangian and amplitude calculations these objects link various two-forms, sources or field strengths. The notational choice of having indices on the left of the and right of the symbol is to help illustrate this linking structure. 

Because each kinetic term includes exactly one projector we refer to this class of theories as projector-linear. Appendix~\ref{app:Zwan} shows that these theories describe a single photon, with transverse polarizations. In addition, when the condition
\begin{align}
\det\mathbf H=\kappa^2
\label{eq:mainZwanzigerLightCone}
\end{align}
is satisfied the photon obeys the correct on-shell dispersion relation, $\omega=|\vec{p}\,|$.

The usual (canonically normalized) Zwanziger theory, without CP violation in the kinetic terms, has $\mathbf H$ equal to the identity matrix, $\kappa=1$ and $\mathbf{S}=\Cmat$. The source terms may or may not include CP violation, but if they do not the theory has no $\theta$ term. In what follows we show that off-diagonal couplings in the source term can be removed, at the cost of a nondiagonal $\mathbf H$, by a change of potential variables.

\subsection{Field-Redefinitions}

We now consider how the Lagrangian is changed by global, linear field redefinitions. We define an invertible change of potential variables by 
\begin{align}
\bm{A}_\mu=\mathbf M\,\bm{A}'_\mu~.
\end{align} 
The form of the Lagrangian is preserved when we make the definitions
\begin{align}
\mathbf H'&=\mathbf M^T\mathbf H\,\mathbf M~,
&
\kappa'&=(\det\mathbf M)\kappa~,
&
\mathbf S'&=\mathbf M^T\mathbf S~,
\label{eq:mainZwanzigerRedefinition}
\end{align}
so that
\begin{align}
{}^{\alpha\beta}\mathbf{K}^{\prime\gamma\delta}=\mathbf M^T\,{}^{\alpha\beta}\mathbf{K}^{\gamma\delta}\,\mathbf M~.
\end{align}
Note that these transformation rules preserve the on-shell photon condition $\det \mathbf{H} = \kappa^2$ and that $\bm{F}^{\mu\nu}$ is unchanged. Two other invariants are
\begin{align}
\mathbf{S}^T\mathbf{H}^{-1}\mathbf{S} \ \ \text{and} \ \ \frac{\det\mathbf{S}}{\kappa}~.\label{eq:invariants}
\end{align}

We are particularly interested in two choices of potential variables. We can transform a general theory to Zwanziger with CP-violation only in the source terms by taking $\mathbf M=\mathbf H^{-1/2}$, which produces
\begin{align}
\mathbf H_{\rm src}=\mathbf 1~,
\qquad
\kappa_{\rm src}=1~.
\label{eq:mainSourceRepresentative}
\end{align}
The equations of motion imply that 
\begin{align}
	\mathbf S_{\rm src}=\Cmat~.
\end{align}
We refer to this as the source-coupling Lagrangian, in which the field strength is simply
\begin{align}
	\bm{F}^{\mu\nu}&=
	\left({}_{(n)}^{\mu\nu}
	\mathcal{P}_{(n)}^{\alpha\beta}
	\,\mathbf{1}- 
	\,\,
	{}_{(n)}^{\star\mu\nu}
	\mathcal{P}_{(n)}^{\alpha\beta}
	\,\epsmat\right)
	\bm{F}_{\bm{A}} {}_{\alpha\beta}~,
	\quad \text{(source-coupling)}
\end{align}
which agrees with the definition of the field strength given by Zwanziger~\cite{Zwanziger:1970hk,Newey:2024gug}. This result also agrees with the CP violating Lagrangian in \cite{Terning:2018lsv} after the removal of total derivative terms.

By evaluating the invariants~\eqref{eq:invariants} in this basis we find the general results
\begin{align}
\mathbf{S}^T\mathbf{H}^{-1}\mathbf{S} =\Cmat^T\Cmat=\Gmat~, \ \ \ \  \frac{\det\mathbf{S}}{\kappa}=\det\Cmat=\Delta~.\label{eq:H2Grel}
\end{align}
We can also invert the relationships in~\eqref{eq:H2Grel} to find
\begin{align}
\mathbf{H}=\mathbf{S}\Gmat^{-1}\mathbf{S}^T=\mathbf{S}\Cmat^{-1}\left(\mathbf{S}\Cmat^{-1} \right)^T~, \ \ \ \ \kappa=\frac{\det\mathbf{S}}{\Delta}=\det\left( \mathbf{S}\mathbf{C}^{-1}\right)~.
\end{align}
Because for any two-dimensional matrix $\mathbf{O}$
\begin{align}
\mathbf{O}^T\epsmat\mathbf{O}=\mathbf{O}\epsmat\mathbf{O}^T=\det\left(\mathbf{O} \right)\,\epsmat~,\label{eq:EpsDetIdent}
\end{align}
we find
\begin{align}
\bm{F}^{\mu\nu}&=\Cmat\mathbf{S}^{-1}\left({}_{(n)}^{\alpha\beta}\mathcal{P}_{(n)}^{\gamma\delta}
	\,\mathbf{H}- \kappa{}^\star{}_{(n)}^{\alpha\beta}\mathcal{P}_{(n)}^{\gamma\delta}\,\epsmat \right)\bm{F}_{\bm{A}}{}_{\alpha\beta}\nonumber\\
	&=\Cmat\mathbf{S}^{-1}\left({}_{(n)}^{\alpha\beta}\mathcal{P}_{(n)}^{\gamma\delta}
	\,\mathbf{S}\Cmat^{-1}\left(\mathbf{S}\Cmat^{-1} \right)^T- {}^\star{}_{(n)}^{\alpha\beta}\mathcal{P}_{(n)}^{\gamma\delta}\,\mathbf{S}\Cmat^{-1}\epsmat\left(\mathbf{S}\Cmat^{-1} \right)^T \right)\bm{F}_{\bm{A}}{}_{\alpha\beta}\nonumber\\
	&=\left({}_{(n)}^{\alpha\beta}\mathcal{P}_{(n)}^{\gamma\delta}\mathbf{1}- {}^\star{}_{(n)}^{\alpha\beta}\mathcal{P}_{(n)}^{\gamma\delta}\epsmat \right)\left(\mathbf{S}\Cmat^{-1} \right)^T\,\bm{F}_{\bm{A}}{}_{\alpha\beta}~.
\end{align}

With these results in mind, we consider the family of transformations
\begin{align}
\mathbf M=\mathbf S^{-T}\mathbf{V}~,
\end{align}
where $\mathbf{V}$ is a diagonal matrix. These produce
\begin{align}
\mathbf{H}'&=\mathbf{V}\mathbf{S}^{-1}\mathbf{H}\mathbf{S}^{-T}\mathbf{V}=\mathbf{V}\Gmat^{-1}\mathbf{V}~,\\
\kappa'&=\frac{\det\mathbf{V}}{\det\mathbf{S}}\kappa=\frac{\det\mathbf{V}}{\Delta}~,\\
\mathbf{S}'&=\mathbf{V}~.
\end{align}
Because $\mathbf{S}'$ is diagonal, each gauge field is only sourced by one type of current, either electric or magnetic. Therefore, all CP violating couplings must be in $\mathbf{H}$. We refer to these as kinetic-coupling Lagrangians. One simple realization takes $\mathbf{V}$ to be the identity matrix. In this case
\begin{align}
	\mathbf{H}_{\text{kin}}= \mathbf{G}^{-1}~,
	\quad
	\kappa_{\text{kin}}= \frac{1}{\Delta}~,
	\quad
	\mathbf{S}_{\text{kin}}=   \mathbf{1}~, \ \ \ \ \text{(holomorphic)}
\end{align}
and corresponds to the holomorphic normalization of the gauge fields. As shown below, in this normalization the Feynman rules take a particularly simple form.

The canonical normalization of the gauge fields is obtained by choosing
\begin{align}
\mathbf{V}=\left(\begin{array}{cc}
e & 0\\
0 & b
\end{array} \right)~,
\end{align}
which means that $\det\mathbf{V}=\Delta$. The Lagrangian is described by
\begin{align}
\mathbf{H}_{\text{kin}}= \left( \begin{array}{cc}
1+\displaystyle\frac{e^2\theta^2}{4\pi^2 b^2} & \displaystyle-\frac{e\theta}{2\pi b}\\[0.3cm]
\displaystyle-\frac{e\theta}{2\pi b} & 1
\end{array}\right)~, \ \ 
\kappa_{\text{kin}}=1~, \ \ 
\mathbf{S}_{\text{kin}}=  \left(\begin{array}{cc}
e & 0\\
0 & b
\end{array} \right)~,  \ \ \text{(canonical)}~,
\end{align}
which can also be written 
\begin{align}
	\mathcal L={}&-\frac{1+e^2\theta^2/(4\pi^2b^2)}4F^{\mathcal A}_{\mu\nu} 
	\,
	{}_{(n)}^{\mu\nu} \Proj_{(n)}^{\alpha\beta}
	\,
	F^{\mathcal A}_{\alpha\beta}
	-\frac14F^{\mathcal B}_{\mu\nu}
	\,
	{}_{(n)}^{\mu\nu} \Proj_{(n)}^{\alpha\beta}
	\,
	F^{\mathcal B}_{\alpha\beta}
	+\frac12F^{\mathcal A}_{\mu\nu}
	\,
	{}_{(n)}^{\star \mu\nu} \Proj_{(n)}^{\alpha\beta}
	\,
	F^{\mathcal B}_{\alpha\beta}
	\notag\\
	&+\frac{e\theta}{4\pi b}F^{\mathcal A}_{\mu\nu}
	\,
	{}_{(n)}^{\mu\nu} \Proj_{(n)}^{\alpha\beta}
	\,
	F^{\mathcal B}_{\alpha\beta}
	-e\mathcal A_\mu J^\mu-b\mathcal B_\mu K^\mu.
	\label{eq:GenZwanNorm}
\end{align}
Note that the term linear in $\theta$ is accompanied by a term with quadratic dependence on $\theta$. Keeping the linear term alone does not produce the correct CP violating Maxwell equations. This result agrees with the constant-axion, zero-dilaton specialization of the duality-symmetric action of Lee and Min~\cite{Lee:2013ewa}.

This Lagrangian, with all CP violation in the kinetic term is simply a field redefinition of the Lagrangian with all CP violation in the source terms
\begin{align}
\mathbf{M}=\Cmat^{-T}\mathbf{V}~.
\end{align}
In the Witten basis this is the field redefinition 
\begin{align}
\mathcal A_\mu=A_\mu~,
\quad
\mathcal B_\mu=B_\mu- \frac{e\theta}{2\pi b} 
A_\mu~,
\label{eq:mainThetaToSource}
\end{align}
so that the field strength is
\begin{align}
	F^{\mu\nu}
	={}_{(n)}^{\mu\nu}\Proj_{(n)}^{\alpha\beta}
	F^{\alpha\beta}_{\mathcal{A}}
	-{}_{(n)}^{\star\mu\nu}\Proj_{(n)}^{\alpha\beta}
	\left(F^{\mathcal B\alpha\beta}-\frac{e\theta}{2\pi b}F^{\mathcal A\alpha\beta}\right)~.
	\label{eq:FmatrixReconstruct}
\end{align}
By choosing a basis it becomes clear that the $\theta$ dependent part of the Lagrangian is not proportional to $F_{\mu\nu}{}^\star F^{\mu\nu}$; the coefficients of the quadratic and linear terms differ from this by a factor of $\frac{1}{2}$. However, differentiating the complete action, including the required $\theta^2$ term, and writing the result in terms of the physical field strength gives
\begin{align}
\frac{\partial\mathcal L}{\partial\theta}
&=\frac{e}{8\pi b}F_{\mu\nu}\dual F^{\mu\nu}
=\frac{e^2}{16\pi^2\dlat}F_{\mu\nu}\dual F^{\mu\nu},
\label{e.dLdtheta}
\end{align}
where $eb=2\pi\dlat$. Thus, the \(F\dual F\) counterterm contains the same running information as the \(\theta\)-dependent \(A_\mu K^\mu\) vertex in the source-coupling form. For compact gauge fields, equivalence of the complete theories additionally requires the global charge and line-operator data to be transformed.

\section{Feynman Rules}\label{sec:Feynman}
Having determined the Lagrangian for a CP violating theory with electric and magnetic currents, we turn to the Feynman rules. At this level the equivalence of the kinetic-coupling and source-coupling forms of the Lagrangian is obvious; it corresponds to whether we include $\Cmat$ in the photon propagator or in the vertices. 
The Feynman rules are simplest in kinetic-coupling with holomorphic normalization which convention we use below. 
Following \cite{Newey:2024gug} we state the Feynman rules in terms of the Zwanziger potential field strengths ($F_A$ and $F_B$), which couple to the nonlocal, $n$-dependent $\Sigma_J$ and $\Sigma_K$. 
This field-strength focused method enables easy identification of the topological numbers which exponentiate out of physical amplitudes (see Appendix \ref{app:worldlineTopology}).

The photon propagators can be expressed as a $2\times 2$ matrix of correlators.
As derived in Appendix \ref{app:singlePhotonProp}, the correlators for the source-coupling Lagrangian are given by 
\begin{align}
	\langle \bm{F}^{\alpha\beta}_{\bm{A}}
	\bm{F}^{\gamma\delta}_{\bm{A}} 
	\rangle_k^{\text{src}} 
	&=
	\begin{bmatrix}
		\langle F_A^{\alpha\beta} 
		F_A^{\gamma\delta} \rangle_k
		& \langle F_A^{\alpha\beta} 
		F_B^{\gamma\delta} \rangle_k
		\\
		\langle F_B^{\alpha\beta} 
		F_A^{\gamma\delta} \rangle_k
		& \langle F_B^{\alpha\beta} 
		F_B^{\gamma\delta} \rangle_k
	\end{bmatrix}
	\notag\\
	&= 
	-2i {}_{(k)}^{\alpha\beta}
	\mathcal{P}_{(k)}^{\gamma\delta}
	\,
	\mathbf{H}^{-1}
	-2i
	\frac{1}{\kappa}
	{}_{(k)}^{\alpha\beta}\mathcal{P}_{(k)}^{\mu\nu}\,{}^{(n)}_{\star\mu\nu}\mathcal{P}_{(k)}^{\gamma\delta }
	\,
	\epsmat~.
\end{align}
For the kinetic-coupling basis in holomorphic normalization this takes the simple form
\begin{align}
	\langle \bm{F}^{\alpha\beta}_{\bm{A}}\bm{F}^{\gamma\delta}_{\bm{A}} \rangle_k 
	= 
	-2i {}_{(k)}^{\alpha\beta}\mathcal{P}_{(k)}^{\gamma\delta}\,\Gmat
	-2i \Delta\,{}_{(k)}^{\alpha\beta}\mathcal{P}_{(k)}^{\mu\nu}\,{}^{(n)}_{\star\mu\nu}\mathcal{P}_{(k)}^{\gamma\delta }\,\epsmat~.
\end{align}

\subsection{Fermion Vertices and Surface Invariance}
\label{sec:SurfaceInvariance}

The kinetic-coupling Lagrangian with holomorphic normalization features currents coupled locally to the potentials 
\begin{align}
	\mathcal{L}_{\text{int}}
	= 
	-\bm{A}_\mu^T
	\bm{J}^\mu~.  
\end{align}
Current conservation implies the existence of a two-form satisfying 
\begin{align}
	\partial_\mu \bm{\Sigma}^{\mu\nu} =\bm{J}^\nu~,
	\label{eq:SigmaDef}
\end{align} 
which can be integrated by parts to find 
\begin{align}
	\mathcal{L}_{\text{int}}
	= \frac{1}{2}
	\bm{F}_{\bm{A}\mu\nu}^T
	\boldsymbol{\Sigma}^{\mu\nu}~.
\end{align}
In the worldline formalism the $\boldsymbol{\Sigma}^{\mu\nu}$ correspond to surfaces bounded by particle worldlines, so we refer to them as surface operators.  Because the couplings are to derivatives of the potentials the theory is independent of the particular choice of $\boldsymbol{\Sigma}^{\mu\nu}$ so long as it satisfies \eqref{eq:SigmaDef}. 

We note that the restriction \eqref{eq:SigmaDef} is unaffected by transformations of the form
\begin{align}
	\bm{\Sigma}^{\mu\nu}
	\mapsto 
	\bm{\Sigma}^{\mu\nu}
	+ \epsilon^{\mu\nu\alpha\beta}
	\partial_\alpha \Lambda^\beta~, 
\end{align}
for an arbitrary vector field $\Lambda^\mu$. 
By Poincare's Lemma on flat spacetime this parametrizes the surface operator's entire redundancy. 
From this argument we see that this redundancy is a one-form gauge symmetry as was originally stated in \cite{Hull:2024uwz}.

While $\boldsymbol{\Sigma}^{\mu\nu}$ is not unique, a convenient choice makes the momentum space Feynman rules simple
\begin{align}
	\boldsymbol{\Sigma}^{\mu\nu}
	= \frac{1}{n\cdot \partial}
	\left(
	n^\mu \bm{J}^\nu 
	- n^\nu \bm{J}^\mu 
	\right)~.
\end{align}
This can be seen directly from the photon propagator. Looking at the spacetime part we see that each pair of external indices is associated with the photon momentum $k^\mu$. Consequently, the Fourier transformed version of the identity \eqref{eq:SigmaDef} guarantees that calculations made with these Feynman rules only depend on $\bm{J}$. The propagators, however, are $n$ dependent, and the removal of $n$-dependence involves a careful understanding of the quantum theory.

Let $J_i^\mu$ be the conserved unit-charge matter current of species $i$. That is, we have
\begin{align}
	\bm{J}_i^{\mu}
	= \bm{q}_i J_i^{\mu}~,
\end{align}
with no sum over $i$.
The vertex for a particle of species $i$ with charge vector $\bm{q}_i$ is given by 
\begin{align}
	\frac{1}{2}\bm{\Sigma}_i^{\mu\nu}
	=
	\bm{q}_{i} 
	\frac{1}{2n\cdot k}
	\left(
	n^\mu  J^\nu_i - n^\nu J^{\mu}_i
	\right)~.
\end{align}
Notice that the electric-magnetic current doublet and photon vertex factorize, and that the $i$ from the definition of $\Sigma^{\mu\nu}$ cancels with the normal factor of $-i$ with the coupling in the feynman rules. 
Thus $\Qvec{i}$ contains the electric and magnetic charges, while $ J_i^\mu$ contains the kinematics and spin of the state.
For a vector-like fermion this gives the following vertex
\begin{align}
	\frac{1}{2n\cdot k}
	\left(
	n^\mu \gamma^\nu  - n^\nu \gamma^\mu 
	\right)
	\, \bm{q}_i~.
\end{align}

\begin{figure}
	\centering
	\begin{tabular}{cc}
		\begin{tabular}{c}
			\begin{fmffile}{treeFSFeyn1}
				\begin{fmfgraph*}(120,75)
					\fmfpen{1.0}
					\fmfstraight
					\fmfset{arrow_len}{3mm}
					\fmfleft{p4,p1,p2,i1,p5} \fmfright{o1,p3,o2}
					\fmf{phantom,tension=0.2}{p2,v1}
					\fmf{phantom,tension=0.922}{v1,p3}
					\fmf{plain,tension=0.9,fore=(1,,1,,1),rubout=0.9}{o1,o2}
					\fmffreeze
					\fmf{photon,tension=0.5,rubout=0.9,label=$k\rightarrow$,label.side=left}{i1,v2}
					\fmf{phantom,tension=0.85}{o2,v2}
					\fmf{phantom,tension=0.8}{o1,v2}
					\fmffreeze
					\fmf{phantom,tension=0.53}{p2,v3}
					\fmf{phantom,tension=0.47}{v3,p3}
					\fmf{dashes,tension=0.0,rubout=0.9}{v3,v2}
					\fmf{phantom_arrow,tension=0.0,rubout=0.9}{o1,v3,o2}
					\fmfv{decor.shape=triangle,decor.size=85,decoration.angle=-30,background=(0.875,,0.875,,0.875),fore=(0.0,,0.0,,0.0),decor.fill=0}{v1} 
					\fmfv{decor.shape=circle,decor.filled=full,decor.size=1.5thick,fore=(0.0,,0.0,,0.0),l=$\hspace{0.05cm}\bm{q}$,l.a=0,l.d=3}{v2} 
					\fmfv{decor.shape=circle,decor.filled=full,decor.size=thick,fore=(0.0,,0.0,,0.0)}{v3} 
					\fmfv{l=$F_{\mu\nu}$}{i1} 
					\fmfv{l=$J_i^\alpha$,l.a=-45,l.d=1}{o1} 
					\fmfv{l=$J_i^\alpha$,l.a=45,l.d=1}{o2} 
				\end{fmfgraph*}
			\end{fmffile}
		\end{tabular}
		\begin{tabular}{c}
			$\phantom{A}\displaystyle \,\bm{q}\,\frac{\delta^\mu_\alpha n^\nu-\delta^\nu_\alpha n^\mu}{2 n\cdot k}\gamma^\alpha$
		\end{tabular}
	\end{tabular}
	\caption{\label{f.fsFeynRule} Field-strength vertex for a Dirac fermion of charge $\bm{q}$. The shaded region denotes the Stokes surface bounded by the worldline. The field strength couples to the surface, while the dashed line represents propagation along the surface through the factor $1/(n\cdot k)$. The overall sign depends on whether the photon momentum flows into or out of the vertex.}
\end{figure}

\subsection{ Tree-level Dyon Scattering}
\label{sec:treeDyonScattering}
We are, at last, ready to calculate the interactions between dyons. We begin with the tree-level two-current amplitude for dyon $i$ interacting with dyon $j$
\begin{align}
	\mathcal M_{ij}
	=\,& \frac{1}{2}\left(\Qvec{i}^T\Gmat\Qvec{j}\right)
	\Sigma_{i\alpha\beta}(k)
	\,
	{}_{(k)}^{\alpha\beta}
	\Proj {}_{(k)}^{\gamma\delta}
	\,
	\Sigma_{j\gamma\delta}^*(k)
	\notag\\
	&-
	\frac{1}{2}
	\Delta\left(\Qvec{i}^T\epsmat\Qvec{j}\right)
	\Sigma_{i\alpha\beta}(k)\,
	{}_{(k)}^{\alpha\beta}\mathcal{P}_{(k)}^{\mu\nu}\,{}^{(n)}_{\star\mu\nu}\mathcal{P}_{(k)}^{\gamma\delta }\,
	\Sigma_{j\gamma\delta}^*(k)~.\label{eq:DyonScatSigma}
\end{align}
This can also be expressed in terms of $J_i^\mu$ 
\begin{align}
	\mathcal M_{ij}
	=&\left(\Qvec{i}^T\Gmat\Qvec{j}\right)
	\frac{J_i^\mu(k) J_{j\mu}^*(k)
	}{k^2}
	-\frac{\Delta}{2}\left(\Qvec{i}^T\epsmat\Qvec{j}\right)
	\frac{
	\varepsilon^{\alpha\beta\gamma\delta}
	J_{i\alpha}(k)
	J_{j\beta}^*(k) 
	k_\gamma n_\delta}
	{k^2(n\cdot k)}~.\label{eq:DyonScatJ}
\end{align}
In the Witten basis this is 
\begin{align}
&\mathcal M_{ij}=\nonumber\\
	&\left[e^2\left(q_i+g_i\frac{\theta}{2\pi}\right)
	\left(q_j+g_j\frac{\theta}{2\pi}\right)+b^2g_i g_j\right]
	\frac{J_i^\mu J_{j\mu}^*}{k^2}
	-\frac{eb}{2}(q_i g_j-q_j g_i)
	\frac{\varepsilon^{\alpha\beta\gamma\delta}
	J_{i\alpha} 
	J_{j\beta}^*
	k_\gamma n_\delta}
	{k^2(n\cdot k)}~.
	\label{e.ijDyonScat}
\end{align}
Notice that this has two distinct physical structures. The symmetric coefficient $\Qvec{i}^T\Gmat\Qvec{j}$ is the Coulombic interaction that can run.  The antisymmetric coefficient is
\begin{align}
	\Delta\Qvec{i}^T\epsmat\Qvec{j}
	=4\pi h_{ij}^{\rm pair},
	\label{eq:treeTopologicalPairwiseHelicity}
\end{align}
so the second term is fixed by the quantized pairwise helicity.

In the classical limit the antisymmetric DSZ term is responsible for the cross product part of Lorentz force. It is transverse to the velocity, does no work, and changes the mechanical orbital angular momentum by the amount carried in the electromagnetic field.  It therefore does not define an additional static potential.  Its $1/(n\cdot k)$ pole is not physical, but rather comes from the topological intersection numbers which exponentiate to a trivial phase ( $e^{2\pi iN} = 1$) in the full nonperturbative amplitude.  

This is most easily seen by looking at the amplitude in \eqref{eq:DyonScatSigma}. Because the $\Sigma^{\mu\nu}$s can be chosen in a way which is $n$ independent (see Appendix \ref{app:SurfCorr} for a specific example), all of the $n$-dependence in the amplitude comes from 
\begin{align}
	\frac{1}{2}
	\Delta\left(\Qvec{i}^T\epsmat\Qvec{j}\right)
	\Sigma_{i\alpha\beta}(k)
	\,
	{}_{(k)}^{\alpha\beta}
	\Proj_{(n)}^{\gamma\delta\star}
	\,
	\Sigma_{j\gamma\delta}^*(k)~, \label{eq:nTopologicalTerm}
\end{align}
where we have used the first identity in
\begin{align}
	{}_{(k)}^{\alpha\beta}\mathcal{P}_{(k)}^{\mu\nu}\,
	{}^{(n)}_{\star\mu\nu}\mathcal{P}_{(k)}^{\gamma\delta }&=
	{}_{(k)}^{\alpha\beta}\mathcal{P}_{(k)}^{\star\gamma\delta }- 
	{}_{(k)}^{\alpha\beta}\mathcal{P}_{(n)}^{\star\gamma\delta }\nonumber\\
	&=-{}_{(k)}^{\star\alpha\beta}\mathcal{P}_{(k)}^{\gamma\delta }+
	{}_{(n)}^{\star\alpha\beta}\mathcal{P}_{(k)}^{\gamma\delta }~.\label{eq:hodgeStarChoice}
\end{align}

In Appendix \ref{app:worldlineTopology} we show how gauge invariant correlation functions can be computed in the worldline formalism (generalizing the arguments of \cite{Brandt:1977be} to the theories discussed in this paper). In this approach, we find that on particular particle paths in the worldline measure \eqref{eq:nTopologicalTerm} contributes a phase of 
\begin{align}
	\exp\left[i\Delta \bm{q}_i \epsmat \bm{q}_j N_{ij}^{\text{(top)}}\right]
	= 1~,
\end{align}
where $N_{ij}^{\text{(top)}}$ is a topological intersection number.\footnote{
Reference \cite{Newey:2024gug} refers to certain topological numbers as linking numbers and others as intersection numbers. However, every linking number can also be realized as an intersection number, so we refer to all of these topological numbers as intersection numbers.
}

The other part of the DSZ amplitude
\begin{align}
	-\frac{1}{2}\Delta\left(\Qvec{i}^T\epsmat\Qvec{j}\right)
	\Sigma_{i\alpha\beta}(k)
	\,
	{}_{(k)}^{\alpha\beta}\mathcal{P}_{(k)}^{\mu\nu}\,{}^{(n)}_{\star\mu\nu}\mathcal{P}_{(k)}^{\gamma\delta }
	\,
	\Sigma_{j\gamma\delta}^*(k)~,
\end{align}
can be made $n$ independent so long as the $\Sigma$s are chosen appropriately. As the identities in~\eqref{eq:hodgeStarChoice} suggest, however, it is also ambiguous. Projector identities imply that
\begin{align}
	&\frac{\Delta}{2}\left(\Qvec{i}^T\epsmat\Qvec{j}\right)
	\Sigma_{i\alpha\beta}(k)\,{}_{(k)}^{\alpha\beta}
	\Proj_{(k)}^{\gamma\delta\star}\,
	\Sigma_{j\gamma\delta}^*(k)
	+\frac{\Delta}{2}
	\left(\Qvec{i}^T\epsmat\Qvec{j}\right) 
	\Sigma_{i\alpha\beta}(k)\,{}_{(k)}^{\star\alpha\beta}\Proj_{(k)}^{\gamma\delta}\,\Sigma_{j\gamma\delta}^*(k)
	\notag\\
	&= 
	\frac{\Delta}{2}\left(\Qvec{i}^T\epsmat\Qvec{j}\right)
	\Sigma_{i\alpha\beta}(k)\,{}^\star\Sigma_{j}^{*\alpha\beta}(k)~.
\end{align}
Therefore, we can interchange the position of the  Hodge dual at the cost of a minus sign and the extra term
\begin{align}
	\frac{\Delta}{2}\left(\Qvec{i}^T\epsmat\Qvec{j}\right)
	\Sigma_{i\alpha\beta}(k)
	\,
	{}^\star\Sigma_{j}^{*\alpha\beta}(k)~,
\end{align}
which only contributes a topological number that does not enter the physical amplitude (see Appendix \ref{app:worldlineTopology}). 

The argument about the exponentiation of topological numbers was first given in \cite{Brandt:1977be}. 
The $n$-independence of the theory can also be verified by resumming soft corrections to hard scattering amplitudes; this was explicitly computed in \cite{Terning:2018udc}, where a similar exponentiation of $n$- dependent topological numbers was found.
Therefore, formally, the physical part of the mixed photon propagator can be written with the Hodge dual on either side (that is, the full nonperturbative amplitude is the same). However, as shown below, when calculating finite loop corrections it does matter which convention is used.


\section{Isolating the Local Counterterm}\label{sec:localCounterterm}

In this section we determine the ultraviolet part of the one-loop surface-current correlator that can renormalize $\Gmat$ and $\Delta$.  
Our focus on renormalization means that we are primarily interested in the local counterterm associated with the one-loop amplitude. This local counterterm must be gauge invariant, polynomial in momentum, and independent of the arbitrary Dirac-string direction $n^\mu$. We use and build upon the methods outlined in~\cite{Newey:2024gug} to determine the counterterm. 
The formal justification for these methods comes from the worldline argument given in Appendix \ref{app:worldlineTopology}. 
We do not use worldlines to make the actual calculations, but it is the foundation for our momentum space prescription and Feynman rules.

\subsection{Removal of topological terms}

Following the discussion of the previous section, when computing the physical loop corrections to dyon scattering we should not include the terms proportional to ${}_{(k)}^{\alpha\beta}\Proj_{(n)}^{\gamma\delta}$. 
These exponentiate out of any physical amplitude because they are due to an overall phase. This phase is associated with the intersection number of the surfaces bounded by dyon worldlines. 

However, the process of removing these terms can be subtle, because the DSZ part of the propagator can be written in two equivalent ways~\eqref{eq:hodgeStarChoice}. The $n^\mu$ dependent portions need to be removed, but we must determine which indices should carry the Hodge star, which also determines the sign of that term. The answer comes from isolating the contributions of the physical field strength in the one-loop amplitude.

In the worldline approach of Appendix \ref{app:worldlineTopology} the one-loop corrections are computed by integrating over worldline loops in the potential field created by the external particles. The one-loop corrections can be found by computing the one-loop connected generating functional 
\begin{align}
	W_{\text{1-loop}}[J_i,J_j] = 
	\text{tr}
	\ln \left[
	\frac{
	i\gamma^\mu \partial_\mu 
	-
	\bm{q}_\ell^T
	\bm{A}_\mu^{\text{ext}}[J_i,J_j]
	- m_\ell 	
	}{
	i\gamma^\mu \partial_\mu 
	- m_\ell 
	}
	\right]
	~
	\label{eq:effectiveAction}
\end{align}
where $\bm{A}^{\text{ext}}[J_i,J_j]$ is the potential created by the external currents of the scattered particles; explicitly we can write
\begin{align}
	\bm{A}^{\text{ext}}_\mu[J_i,J_j ](x)
	= -i\int d^4y
	\langle \bm A_\mu(x) \bm A_\nu(y) \rangle
	\bm q_i
	J^\nu_i(y)
	-i
	\int d^4y
	\langle \bm A_\mu(x) \bm A_\nu(y) \rangle \bm q_j J^\nu_j(y)~. 
\end{align}
Expanding $W$ at second order in $\bm{J}_{i,j}$ produces the scattering process computed in this paper. 
This functional determinant could in principle be computed by solving the Dirac equation in the background of the external particles\footnote{A calculation of this sort was done in \cite{Goebel:1983we} for the specific example of scalar charged fields and static charges.}, or by the worldline calculations described in Appendix \ref{app:worldlineTopology}.
In either approach, we see that the one-loop correction comes from understanding the behavior of the loop dyon in the field generated by the external particles.

The potential, $\bm{A}_\text{ext}$, and its derivatives, $\bm{F}_{\bm{A}}^{\mu\nu}$, sourced by the external currents include a physical piece (which enters the definition of $\bm{F}^{\mu\nu}$), and a topological piece which does not affect the physical amplitudes. But $W^{\text{1-loop}}$ only depends on the physical field strength, as is made clear from the worldline expression given in \eqref{eq:funcDetWL} of Appendix \ref{app:worldlineTopology}.\footnote{
This also should be clear from any hypothetical method based on solving the Dirac equation in this background. The topological terms enter as an Aharanov Bohm phase which is trivial due to the quantization condition. 
}
Therefore, we should choose our prescription to preserve the physical field strength of the external currents and to remove topological artifacts created by truncating the expression for $W^{\text{1-loop}}$ to finite order. 

We isolate the physical part of the propagator by finding how the physical field strength is sourced by external currents. For an electric particle in the Witten basis Maxwell's equations imply
\begin{align}
	F^{\mu\nu} = \frac{1}{\partial^2}
	\left(
	\partial^\mu J^\nu - \partial^\nu J^\mu 
	\right)~,
\end{align}
and ${}^\star F^{\mu\nu}$ given by the Hodge dual.

We generalize this result by writing the electric particle in the Witten basis as
\begin{align}
	\bm{q}_i =
	\bm{q}^{\text{ele}}
	= 
	\begin{pmatrix}
		1\\
		0
	\end{pmatrix},
	\quad
	\Cmat = \Cmat_0
	=\begin{pmatrix}
		e & \frac{e\theta}{2\pi} \\
		0 & b
	\end{pmatrix}~,
\end{align}
and find
\begin{align}
	\bm{F}^{\mu\nu} &= 
	\Cmat \bm{q}_i
	\frac{1}{\partial^2}
	\left(
	\partial^\mu J^\nu - \partial^\nu J^\mu 
	\right)
	- 
	\epsmat \Cmat 
	\bm{q}_i
	\varepsilon^{\mu\nu\alpha\beta}
	\frac{1}{\partial^2} \partial_\alpha J_\beta~,
	\notag\\
	&= 
	\Cmat\bm{q}_i
	\,
	{}_{(\partial)}^{\mu\nu}
	\mathcal{P}
	{}_{(\partial)}^{\alpha\beta} \Sigma_{\alpha\beta}
	-
	\epsmat
	\Cmat\bm{q}_i
	\,
	{}_{(\partial)}^{\star\mu\nu}
	\mathcal{P}
	{}_{(\partial)}^{\alpha\beta} \Sigma_{\alpha\beta}~.	
\end{align}
This covariant way of expressing the result indicates that the last equality describes the field strength due to a general dyon $\bm{q}_i$ in any basis.

The component of the physical field strength sourced by an external dyon of charge $q_i$ which is felt by the loop dyon $\bm{q}_\ell$ is 
\begin{align}
	\bm{q}_\ell^T \mathbf{C}^T \bm{F}^{\mu\nu}
	&=
	\bm{q}_\ell^T
	\Gmat\bm{q}_i
	\,
	{}_{(k)}^{\mu\nu}
	\mathcal{P}
	{}_{(k)}^{\alpha\beta} \Sigma_{i\,\alpha\beta}
	-
	\Delta\,\bm{q}_\ell^T
	\epsmat
	\bm{q}_i
	\,
	{}_{(k)}^{\star\mu\nu}
	\mathcal{P}
	{}_{(k)}^{\alpha\beta} \Sigma_{i\,\alpha\beta}~.
\end{align}
The second term gives the physical DSZ interaction. It is proportional to ${}_{(k)}^{\star \mu\nu}\Proj_{(k)}^{\sigma\rho}$ applied to $\bm{\Sigma}_i^{\sigma\rho}$ \emph{not} ${}_{(k)}^{ \mu\nu}\Proj_{(k)}^{\sigma\rho\star}$ applied to $\bm{\Sigma}_i^{\sigma\rho}$.
Thus, the physical field strength that couples to the dyon loop has the Hodge star applied on the opposite side to the external current.
The remaining term, which may appear in the gauge dependent $\bm{F}_{\bm{A}}^{\mu\nu}$, is a topological artifact. Therefore, the convention that produces the physical answer in perturbation theory is contracting the Hodge star with the fluctuating loop.\footnote{Another way to arrive at this prescription \cite{Newey:2024gug} is to retain the $n$-dependent part of the propagator. If one follows the above prescription, all $n$ dependence appears in the topological part of the mixed propagator, which does not affect the amplitude. The opposite prescription leaves $n$ dependence from $\bm{\Sigma}^{\mu\nu}$ in observable results. }

This prescription was first presented in \cite{Calucci:1982wy}, which argued that otherwise contributions from small loops include topological number pieces which do not vanish as the loop shrinks to zero. We emphasize that this prescription is consistent with SL$(2,\mathbb{Z})$ duality while naively calculating without removing topological terms is not, as shown in \cite{Newey:2024gug}.

\subsection{The shrinking-surface correlator}

In general, the structure of a one-loop amplitude has external currents with Dirac surfaces $\bm{\Sigma}_i^{\mu\nu}$ connected by  propagators to a loop-term ${}^{\alpha\beta}\bm{\Pi}^{\gamma\delta}(k)$ as
\begin{equation}
	\mathcal{M}\propto \bm{\Sigma}_i^{\mu\nu}(k)
	\langle \bm{F}_{\boldsymbol{A}\mu\nu}\bm{F}_{\boldsymbol{A}\alpha\beta} \rangle_k \big|_{\text{phys}}
	{}^{\alpha\beta}\bm{\Pi}^{\gamma\delta}(k) 
	\langle \bm{F}_{\boldsymbol{A}\gamma\delta}\bm{F}_{\boldsymbol{A}\sigma\rho} \rangle_k \big|_{\text{phys}}
	\bm{\Sigma}_j^{*\sigma\rho}(k)~,
\end{equation}
where
\begin{align}
	{}^{\alpha\beta}\bm{\Pi}^{\gamma\delta}(k)=
	\int \frac{d^4 k}{(2\pi)^4}
	\left\langle
	\boldsymbol{\Sigma}_{\ell}^{\alpha\beta}
	\boldsymbol{\Sigma}_{\ell}^{\gamma\delta}\right\rangle~,
\end{align}
see Appendix \ref{app:SurfCorr} for an explicit definition.

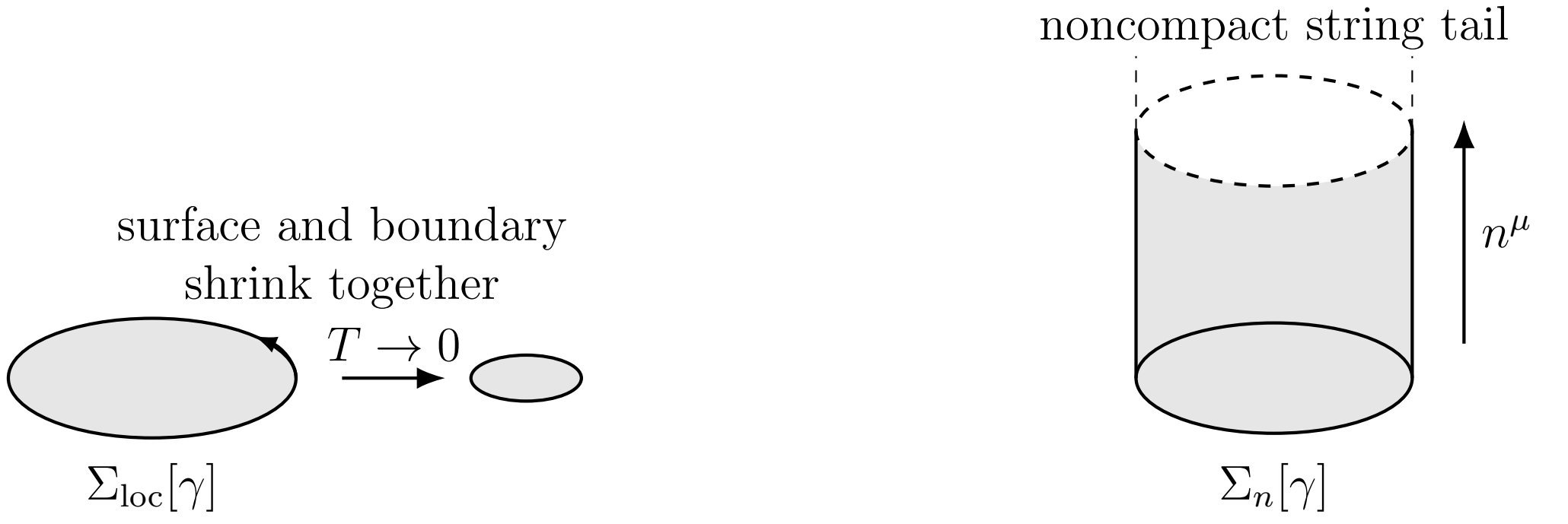
\begin{figure}[t]
\centering
\begin{subfigure}[t]{0.47\textwidth}
\centering
\begin{tikzpicture}[x=1cm,y=1cm,>=Latex]
  \fill[black!10] (0,0) ellipse (1.25 and 0.52);
  \draw[thick] (0,0) ellipse (1.25 and 0.52);
  \draw[-{Latex[length=2mm]},thick] (1.25,0) arc[start angle=0,end angle=45,x radius=1.25,y radius=0.52];
  \node at (0,-0.95) {$\Sigma_{\rm loc}[\gamma]$};
  \draw[-{Latex[length=2.5mm]},thick] (1.65,0) -- (2.55,0) node[midway,above] {$T\to0$};
  \fill[black!10] (3.25,0) ellipse (0.48 and 0.20);
  \draw[thick] (3.25,0) ellipse (0.48 and 0.20);
  \node[align=center] at (1.65,1.05) {surface and boundary\\shrink together};
\end{tikzpicture}
\end{subfigure}
\hfill
\begin{subfigure}[t]{0.47\textwidth}
\centering
\begin{tikzpicture}[x=1cm,y=1cm,>=Latex]
  \fill[black!10] (-1.2,0) -- (-1.2,2.15) arc[start angle=180,end angle=360,x radius=1.2,y radius=0.48] -- (1.2,0) arc[start angle=0,end angle=-180,x radius=1.2,y radius=0.48] -- cycle;
  \draw[thick] (0,0) ellipse (1.2 and 0.48);
  \draw[thick] (-1.2,0) -- (-1.2,2.15);
  \draw[thick] (1.2,0) -- (1.2,2.15);
  \draw[dashed,thick] (0,2.15) ellipse (1.2 and 0.48);
  \draw[dashed] (-1.2,2.15) -- (-1.2,2.8);
  \draw[dashed] (1.2,2.15) -- (1.2,2.8);
  \draw[-{Latex[length=2.5mm]},thick] (1.65,0.3) -- (1.65,2.25) node[midway,right] {$n^\mu$};
  \node at (0,-0.95) {$\Sigma_n[\gamma]$};
  \node[align=center] at (0,3.05) {noncompact string tail};
\end{tikzpicture}
\end{subfigure}
\caption{Two Dirac surfaces with the same worldline boundary.  On the left, a surface that collapses with the proper-time loop and isolates the $n$-independent local counterterm. On the right, a fixed-string that extends along $n^\mu$ and retains dependence on the auxiliary string choice. }
\label{fig:surfaceRepresentatives}
\end{figure}

The divergent part of the loop function determines how couplings run.  In Appendix~\ref{app:SurfCorr} we evaluate the two-current correlator of vectorlike Dirac fermions using a smooth spanning surface that shrinks with the worldline in the proper-time ultraviolet limit, like the one shown on the left of Fig.~\ref{fig:surfaceRepresentatives}. Current conservation, ultraviolet power counting, and Euclidean invariance then fix its complete logarithmically divergent local part to be
\begin{align}
	\left.{}^{\alpha\beta}\boldsymbol{\Pi}_{\mathrm{loc}}^{\gamma\delta}(k)\right|_{\rm UV\,div}
	=
	\left.2\Pi(k^2)  \right|_{\rm UV\,div}\,\,\bm{q}_\ell \bm{q}_\ell^T\,\,{}^{\alpha\beta}\mathbb{I}^{\gamma\delta}~.
	\label{eq:loopPrescription}
\end{align}
The scalar function $\Pi(k^2)$ is the ordinary unit-current vacuum polarization, with the standard Dirac coefficient used below~\cite{Schwartz:2014sze}. Equation~\eqref{eq:loopPrescription} is therefore the local tensor prescription needed below. Terms with explicit external momentum or dependence on the detailed local spanning surface are ultraviolet finite.

Surfaces that vanish with the loop size are useful for making the UV identification above, but quite difficult to use in momentum space calculations. The corollary of this is that surfaces that are simple in momentum space are troublesome in position space. To illustrate this point we consider the two surfaces in Fig.~\ref{fig:surfaceRepresentatives}. These have the same worldline boundary but very different ultraviolet behavior. The local surface (on the left) collapses with the loop and isolates Eq.~\eqref{eq:loopPrescription} capturing the physical local physics. The fixed cylinder (on the right) has a simple momentum space description but retains a noncompact tail in physical space, reminiscent of the Dirac string. Consequently, even in the UV its correlator contains structures tied to the auxiliary direction $n^\mu$, as shown in~\cite{Newey:2024gug}. 

While these surfaces are very different, the two descriptions only differ by topological terms which exponentiate out of the full quantum theory. Using the fixed string correlator directly leads to $n$ dependent loop corrections. When contracted with external currents, however, the fixed string correlator factors into an $n$ independent structure~\eqref{eq:loopPrescription} and clearly identifiable topological terms~\cite{Newey:2024gug}. This procedure is detailed in Appendix~\ref{app:worldlineTopology}.

In short, the UV divergent part of the one-loop calculation is surface independent and given by \eqref{eq:loopPrescription}. This is the quantity that renormalizes $\Gmat$ and $\Delta$. These results also suggest that the remainder of the loop corrections may have the same form as \eqref{eq:loopPrescription}, but this has not been proven. 

The separation of topological terms also resolves the historical sign ambiguity in magnetic-charge renormalization.  Retaining the $n^\mu$-dependent fixed-string terms assigns ordinary running to the intersection sector and gives equal-sign electric and magnetic renormalization~\cite{Schwinger:1966zza,Schwinger:1966zzb,Brandt:1977fa,Deans:1981qs,Panagiotakopoulos:1982fp,Panagiotakopoulos:1982ne}.  
Isolating the local $n^\mu$-independent quantities gives the $\Delta$-preserving flow anticipated by Coleman and found in related analyses~\cite{Coleman:1982cx,Calucci:1982wy,Jengo:1982wx,Goebel:1983we,Laperashvili:1999pu,Aggarwal:2025fiq}.

\section{One-loop running of $\Gmat$, $\tau$, and $\theta$}\label{sec:runningMain}
Having determined the correct Feynman rules and established a method for extracting the UV divergent part of the amplitude we are ready to calculate. After obtaining the one-loop running we discuss physical scenarios in which the perturbative result is justified. The section ends with a series of duality checks on the running.

\subsection{Main result}\label{sec:mainOneLoopResult}

While calculating the one-loop running we treat $\Gmat$ as a general symmetric matrix. That is, we do not enforce the quantization condition relating $e$ and $b$. This allows us to obtain as an output that the one-loop running of $\Gmat$ preserves the quantization condition. 

Using Feynman rules and the prescription above for removing intersection number terms outlined above, we find a somewhat simple, compact result for the scattering of dyons $i$ and $j$. The particle in the loop is taken to be a dyon of type $\ell$ with unit-current vacuum polarization $\Pi_\ell$. The one-loop correction due to this field is given by 
\begin{align}
\Gamma_{ij,{\rm VP}}^{(\ell)}
&=
2
\Sigma_{i\alpha\beta}
\bm{q}_i^T 
\left({}_{(k)}^{\alpha\beta}\mathcal{P}_{(k)}^{\gamma\delta} \Gmat
+ 
\Delta{}_{(k)}^{\alpha\beta}\mathcal{P}_{(k)}^{\gamma\delta \star}\epsmat\right) 
\,
\Pi_\ell
\bm{q}_\ell \bm{q}_\ell^T
\,
{}_{\gamma\delta}\mathbb{I}_{\mu\nu}
\,\left(
{}_{(k)}^{\mu\nu}\mathcal{P}_{(k)}^{\sigma\rho} 
\Gmat
+ \Delta{}_{(k)}^{\star\mu\nu}\mathcal{P}_{(k)}^{\sigma\rho}
\epsmat
\right) 
\bm{q}_j
\Sigma_{j\sigma\rho}^{*}
\notag\\
&=\Pi_\ell\frac{ J_{i\mu} J_j^{*\mu}}{k^2}
\Qvec{i}^T\left[
\left(\Gmat\Qvec{\ell}\right)
\left(\Qvec{\ell}^T\Gmat\right)
+\Delta^2
\left(\epsmat\Qvec{\ell}\right)
\left(\Qvec{\ell}^T\epsmat\right)
\right]\Qvec{j}~,
\label{eq:matrixOneLoop}
\end{align}
where we have used the identity 
\begin{align}
	{}_{(k)}^{\alpha\beta}\mathcal{P}_{(k)}^{\gamma\delta}
	\,\,
	{}^{ (k)}_{\star \gamma\delta}\mathcal{P}_{(k)}^{\mu\nu}
	=0~.
\end{align}

We can compare this result to the tree-level amplitude~\eqref{eq:DyonScatJ} with $\Gmat\to\Gmat+\delta\Gmat$ and $\Delta\to\Delta+\delta\Delta$. The first and most conspicuous result is that the one-loop calculation has no part that contributes to the $\delta\Delta$ part of the tree-level form. This indicates that $\Delta$ does not run at one loop. In other words, at one loop
\begin{align}
\beta_\Delta=0~,
\end{align}
where the $\beta$-function of a renormalized coupling coordinate $X$ is given by The $\beta$ function 
\begin{align}
	\beta_X\equiv
	\mu\frac{dX(\mu)}{d\mu}~.
	\label{eq:betaDefinition}
\end{align}

The same method is applied to the $\delta\Gmat$ part of the tree-level amplitude. In this case the entire one-loop result matches the $\delta\Gmat$ form. Specifically, we find 
\begin{align}
	\delta\Gmat
	=\sum_\ell\Pi_\ell
	\left[
	\Gmat\Qvec{\ell}\Qvec{\ell}^T\Gmat
	+\Delta^2\epsmat\Qvec{\ell}\Qvec{\ell}^T\epsmat
	\right]~.
	\label{eq:deltaGmatrix}
\end{align}
The first term is the ordinary product of the two Coulombic vertices while the second is the mutually nonlocal, DSZ term.  At one loop, the scale dependence comes from the renormalized 1PI vacuum polarization; the proper-time cutoff in Appendix~\ref{app:SurfCorr} extracts its ultraviolet logarithm. For a Dirac fermion,
\begin{align}
	\left.
	\mu\frac{\partial}{\partial\mu}
	\Pi_\ell^{\rm ren}(k^2,m_\ell;\mu)
	\right|_{k,m_\ell}
	=\frac{1}{6\pi^2}~.
	\label{eq:diracVacPolDerivativeMain}
\end{align}
For $m_\ell\neq0$, we can read this coefficient at $k^2=0$.  For a massless state, $k^2=0$ is infrared singular, so we keep nonzero Euclidean momentum or extract the coefficient directly from the ultraviolet pole. Both procedures give the same local logarithm.  

It follows that for a perturbative theory of vectorlike Dirac fermions the corresponding mass-independent RG flow of $\Gmat$ is
\begin{align}
	\bm{\beta}_{\Gmat}
	=\frac{1}{6\pi^2}\sum_\ell
	\left[
	\Gmat\Qvec{\ell}\Qvec{\ell}^T\Gmat
	+\Delta^2\epsmat\Qvec{\ell}\Qvec{\ell}^T\epsmat	\right]~.
	\label{eq:betaGmatrix}
\end{align}
Because $\tau=\theta/(2\pi)+ib/e$ and $\Delta=2\pi\dlat$ this result is equivalent to
\begin{align}
	\beta_\tau
	=-\frac{i\Delta}{6\pi^2}\sum_\ell
	\left(q_\ell+g_\ell\tau\right)^2
	=-\frac{i\,\dlat}{3\pi}\sum_\ell
	\left(q_\ell+g_\ell\tau\right)^2~.
	\label{eq:betaTauMatrixGeneralD}
\end{align}
This result can be further separated into
\begin{align}
	\beta_e&=\frac{e}{12\pi^2}\sum_\ell
	\left[e^2\left(q_\ell+g_\ell\frac{\theta}{2\pi}\right)^2-b^2g_\ell^2\right]~,
	\label{e.eRun}\\
	\beta_b&=-\frac{b}{12\pi^2}\sum_\ell
	\left[e^2\left(q_\ell+g_\ell\frac{\theta}{2\pi}\right)^2-b^2g_\ell^2\right]~,\\
	\beta_\theta&=\frac{2b^2}{3\pi}\sum_\ell
	\left(q_\ell+g_\ell\frac{\theta}{2\pi}\right)g_\ell~.
\end{align}

\subsection{Checks on the RG Flow}\label{sec:matrixFlowChecks}
It is worthwhile to make a few checks of the above results. For $g_\ell=0$ and $\theta=0$, Eq.~\eqref{eq:betaGmatrix} reduces to
\begin{align}
\beta_e=\frac{e^3}{12\pi^2}\sum_\ell q_\ell^2~,
\qquad
\beta_\theta=0~,
\label{eq:checkElectricQED}
\end{align}
which is the ordinary Dirac-fermion QED result. This comparison tests the determinant-preserving electric--magnetic part of the flow, but does not test the off-diagonal $\theta$ term. More generally, if all charges lie on one mutually local ray, a single SL$(2,\mathbb Z)$ transformation makes them electric, and the complete flow reduces to QED.  In short, the covariant result above introduces no additional dynamics in a mutually local sector.

A second check is that the running of $\Gmat$ preserves the quantization condition. In other words, is $\det \Gmat $ preserved by the matrix flow described above? Using $\det\Gmat=\Delta^2$, one finds that the second term of Eq.~\eqref{eq:betaGmatrix} can be written as
\begin{align}
\Delta^2\bm{\epsilon}\bm{q}_\ell\bm{q}_\ell^T\bm{\epsilon}=\Gmat\bm{\epsilon}\Gmat\bm{q}_\ell\bm{q}_\ell^T\bm{\epsilon}~.
\end{align}
The above result, along with the cyclic nature of the trace and the fact that $\bm{\epsilon}\bm{\epsilon}=-\bm{1}$, leads to
\begin{align}
	\operatorname{Tr}\!\left(\Gmat^{-1}\bm{\beta}_{\Gmat}\right)=0~.
\end{align}
Applying  Jacobi's formula 
\begin{align}
	\frac{\partial}{\partial\lambda}
	\det \mathbf{M} 
	= \det \mathbf{M}
	\,\,
	\text{Tr} \left(
	\mathbf{M}^{-1}
	\frac{\partial}{\partial \lambda}
	\mathbf{M}
	\right)~,
\end{align}
we arrive at
\begin{align}
	\frac{\partial}{\partial \mu } \det\Gmat=0~.
\label{eq:checkDetGPreserved}
\end{align}
This confirms that while the symmetric photon-exchange matrix runs, $\Delta$, every DSZ pairing, and every pairwise helicity remain fixed. The RG flow changes the Coulombic interaction without changing the pairwise Poincar\'e representation of the asymptotic dyon states.

The $\dlat=2$ specialization of Eq.~\eqref{eq:betaTauMatrixGeneralD} reproduces the conventional $eb=4\pi$ charge polynomial obtained from duality and anomaly arguments~\cite{Csaki:2010rv}, while $\dlat=1$ gives the primitive lattice.  The matrix form makes the common content transparent: the Coulombic interaction runs, the DSZ pairing does not, and the result can be expressed in any integral charge basis.

\subsection{Assumptions and domain of validity}\label{sec:perturbativeRegimes}

The coupling matrix $\Gmat$ follows from the one-photon coupling structure and does not require a free or perturbative matter sector. In contrast, the $1/(6\pi^2)$ coefficient in Eq.~\eqref{eq:betaGmatrix} assumes that each species in the sum is a vectorlike Dirac fermion with a real mass in the chosen chiral convention and that its conserved-current correlator is perturbative. For an interacting matter sector, $\Gmat$ must be contracted with the full current correlator. The formulas above give the mass-independent local 1PI flow; masses enter through threshold weights or finite matching.

In this section we discuss a few specific perturbative scenarios. One straightforward application, would be to look at dyon loop corrections for the correlation functions of static external probes (e.g Wilson and t'Hooft lines).
For example, our methods could be applied to rederive and generalize the results of \cite{Goebel:1983we} which used a semiclassical method to compute the one-loop corrections to the potential between two static monopoles. 
Another straightforward application would be to supersymmetric theories where the perturbative parts of $\beta$-functions are one loop exact. 
A full understanding of supersymmetric theories with mutually nonlocal charges would require a nonrenormalization theorem for a supersymmetric version of the Zwanziger theory, which is beyond the scope of this work.

The next section outlines two other regimes in which the nonsupersymmetric current correlator entering the one-loop counterterm can be evaluated perturbatively.  In both cases the charge dependence factorizes from the current correlator. If the matter sector is strongly coupled, the same charge factors remain, but the free Dirac function must be replaced by the full conserved-current correlator; kinetic mixing in strongly coupled sectors is discussed in Ref.~\cite{DelZotto:2016fju}. Genuinely nonperturbative magnetic-worldline sectors are beyond the scope of this paper~\cite{Csaki:2024ajo,GarciaGarcia:2025uub} .

The Dirac condition constrains electric and magnetic generators of the same unmixed compact U(1): $eb=2\pi\dlat$, so $e$ and $b$ cannot both be small. This does not prevent weak projected couplings in multi-U(1) theories, nor does it prevent a mutually local monopole sector from being weakly coupled in an appropriate region of $(b,\theta)$.

\subsubsection{Pure-monopole sector}

One simple system with only perturbative couplings is one in which the light dynamical particles have $q_\ell=0$ and mutually local magnetic charges $g_\ell$.  We use the electric frame because it displays separately the magnetic vertex and the Witten-induced electric vertex. Since $e=2\pi\dlat/b$, the parameter $e$ is large when $b\ll1$, but no light particle has a vertex proportional to $e$ alone.  The two monopole vertices are
\begin{align}
 b g_\ell~,
 \quad \quad
 \frac{e\theta}{2\pi}g_\ell
 =\frac{\dlat\theta}{b}g_\ell~,
\label{eq:pureMonopoleTwoVertices}
\end{align}
 or equivalently
\begin{align}
\Cmat \bm{q}_\ell
=
g_\ell
\begin{pmatrix}
 \dlat\theta/b\\[1mm]
 b
\end{pmatrix}~,
\text{ which leads to }
\bm{q}_\ell^T
\Gmat \bm{q}_\ell
=
g_\ell^2
\left[
b^2+\left(\frac{\dlat\theta}{b}\right)^2
\right]~.
\label{eq:pureMonopoleCouplingVector}
\end{align}
The perturbative expansion is under control as long as
\begin{align}
 b|g_\ell|\ll1~,
 \qquad
 \frac{\dlat|\theta g_\ell|}{b}\ll1~.
\label{eq:smallThetaConditions}
\end{align}
For charges of order one, a simple parametric regime is $\dlat|\theta|\ll b\ll1$.

Before contraction with external states, the loop vertices give
\begin{align}
\Cmat 
\bm{q}_\ell 
\bm{q}_\ell^T \Cmat^T
=
g_\ell^2
\begin{pmatrix}
 \dlat^2\theta^2/b^2 & \dlat\theta\\[1mm]
 \dlat\theta & b^2
\end{pmatrix}.
\label{eq:pureMonopoleVertexOuterProduct}
\end{align}

Thus the $\theta^0$, $\theta^1$, and $\theta^2$ terms are separately visible in one perturbative electric-frame calculation.
They determine the magnetic renormalization, the running of the $\theta$ coordinate, and the renormalization of the Witten-induced electric interaction.
We see that because we can Taylor expand in $\theta$ we recover all the information required to understand the mutually nonlocal case.

Electromagnetic duality provides a check on this analysis. In the system described above
\begin{align}
\tau
=
\frac{\theta}{2\pi}
+i\frac{b^2}{2\pi\dlat}~.
\end{align}
An $S$ transformation maps every light magnetic monopole into an ordinary electric monopole. The transformed coupling is given by $\tau'=-1/\tau$, so the dual electric coupling is
\begin{align}
 e^{\prime2}
 =
 \frac{\Delta}{\operatorname{Im}\tau'}
 =
 b^2+\frac{\dlat^2\theta^2}{b^2}~.
\label{eq:dualMonopoleCoupling}
\end{align}
A magnetic monopole of charge $g_\ell$ therefore has electric coupling $e'\,g_\ell$ in the dual frame. This confirms that internal photon lines do not reintroduce an uncompensated power of the large electric-frame parameter $e$. The dual description is a check rather than a replacement for the electric-frame calculation, since the latter keeps the terms linear and quadratic in $\theta$ separate.

\subsubsection{Kinetic mixing}

A perturbative theory that is not dual to a purely electric one is provided by kinetic mixing. Consider a visible U(1$)_V$ with tiny kinetic mixing $\epsilon\ll1$ to a dark U(1$)_D$~\cite{Holdom:1985ag}. The labels $I,J=V,D$ denote the visible and dark photon eigenstates. We define their symmetric kinetic matrix by\footnote{The sketch below considers mixing using the physical field strength directly. A careful treatment of mixing using the Zwanziger formalism is given in~\cite{Terning:2018lsv}}
\begin{align}
\mathcal L_{\rm kin}
&=-\frac14 F_I^{\mu\nu}\mathcal K_{IJ}F_{J\mu\nu}~,
&
\boldsymbol{\mathcal K}
&=
\begin{pmatrix}
\mathcal K_{VV}&\mathcal K_{VD}\\
\mathcal K_{VD}&\mathcal K_{DD}
\end{pmatrix}~.
\label{eq:twoUKineticMatrixDefinition}
\end{align}

After diagonalizing the two photon fields, a dark monopole couples to the visible and dark photon eigenstates with vertices
\begin{align}
 b_V\equiv b_{\rm vis}^{\rm eff}=O(\epsilon b_D)~,
 \qquad
 b_D~,
\label{eq:kineticMixingMonopoleVertices}
\end{align}
where precise coefficient in the first relation depends on the convention for the kinetic-mixing parameter. In the conventions of~\cite{Terning:2018lsv},
$b_V=2\pi\dlat\,\epsilon e$, which is perturbative for sufficiently small $\epsilon$. Such projected milli-magnetic couplings underlie phenomenological studies of monopole and string dark matter, magnetar bounds, Aharonov--Bohm detection, and galactic magnetic-field constraints~\cite{GomezSanchez:2011orv,Hook:2017vyc,Terning:2019bhg,Graesser:2021vkr,Hiramatsu:2021kvu,Chitose:2023bnd,Dai:2024zim,Graesser:2026zzs}.

The two external vertices give
\begin{align}
 \Pi_{IJ}^{\mu\nu}(p)
 &=
 b_Ib_J
 \left(p^2\eta^{\mu\nu}-p^\mu p^\nu\right)
 \Pi_M(p^2)~,
 &
 (b_V,b_D)&=(b_{\rm vis}^{\rm eff},b_D)~,
\label{eq:twoUOneLoopPolarization}
\end{align}
where $\Pi_M(p^2)$ denotes the unit-coupling monopole current correlator. Thus, the logarithmically divergent correction has the form
\begin{align}
 \left.\delta\boldsymbol{\mathcal K}\right|_{\log}
 =
 \left.\Pi_M\right|_{\log}
 \begin{pmatrix}
  b_V^2 & b_Vb_D\\
  b_Vb_D & b_D^2
 \end{pmatrix}~.
\label{eq:twoUOneLoopKineticMatrix}
\end{align}
The three entries renormalize different operators composed of the visible or dark photon. In particular, the two visible photon operator is renormalized by
\begin{align}
 \left.\delta\mathcal K_{VV}\right|_{\log}
 =
 \left.\Pi_M\right|_{\log}
 \left(b_{\rm vis}^{\rm eff}\right)^2~.
\label{eq:visibleProjectedCounterterm}
\end{align}
The unsuppressed dark vertex renormalizes $\mathcal K_{DD}$, while $b_Vb_D$ renormalizes the mixed entry. Neither produces an additional (nonperturbative) contribution to $\mathcal K_{VV}$. Consequently, when the dark current correlator is perturbative, $\Pi_M$ is the free
Dirac function.

Charge quantization is imposed before kinetic and mass diagonalization. In the original compact
U(1$)_V\times$ U(1$)_D$ basis, the electric and magnetic charge vectors can be taken to lie on an integral lattice. Under a real-valued redefinition of the photon fields
 \begin{align}
 \left( \begin{array}{c}
 A_{V\mu}'\\
 A_{D\mu}'
 \end{array}\right)=\mathbf{R}_A \left( \begin{array}{c}
 A_{V\mu}\\
 A_{D\mu}
 \end{array}\right)
 \end{align}
the charge vectors become
\begin{align}
 \left( \begin{array}{c}
 q_{V}'\\
 q_{D}'
 \end{array}\right)=\mathbf{R}^{-T}_A \left( \begin{array}{c}
 q_{V}\\
 q_{D}
 \end{array}\right)~,
\qquad
 \left( \begin{array}{c}
 g_{V}'\\
 g_{D}'
 \end{array}\right)=\mathbf{R}_A \left( \begin{array}{c}
 g_{V}\\
 g_{D}
 \end{array}\right)~.
\end{align}
Therefore, the individual mass-eigenstate couplings \emph{need not be integral}. This allows both couplings in a single sector to be perturbative. At the same time
\begin{align}
q_{V\,i}'g_{V\,j}'+q_{D\,i}'g_{D\,j}'-g_{V\,i}'q_{V\,j}'-g_{D\,i}'q_{D\,j}'=
q_{V\,i}g_{V\,j}+q_{D\,i}g_{D\,j}-g_{V\,i}q_{V\,j}-g_{D\,i}q_{D\,j}~,
\end{align}
so the full DSZ pairing is unchanged. 
In other words, the visible electron and the induced visible coupling of the dark monopole are  projections of the complete two-U(1) charge data, not primitive generators of an autonomous one-photon charge lattice~\cite{Brummer:2009cs,Bruemmer:2009ky}. Therefore, the effective theory describing the visible photon's interactions with visible electric charges and dark monopoles is perturbative in both $e_V$ and the $b_V$. The addition of a nonzero $\theta$ in the visible sector realizes perturbative running of all three couplings.

\section{Running and CP}\label{sec:cpSpectrum}

The $\beta$-functions above describe the running of coupling coordinates in a fixed basis for the charge lattice and the fermion mass-phase convention specified in Sec.~\ref{s.Intro}. They do not by themselves diagnose physical CP invariance  \cite{Choi:2025duality}, which also depends on the charge spectrum, masses, multiplicities, interactions, and vacuum. 

\subsection{CP Invariant Curve and RG Tangency}\label{sec:CPThresholdImplications}

Assuming that all fermion masses can be made real using an anomalous chiral symmetry rotation, exact CP transformations act on the gauge coupling as an anti-holomorphic fractional-linear map. We outline this below.

We define CP in a specific charge basis, and generalize using duality. In general, there is a different choice of CP transformation for each starting charge basis. In the electric charge basis CP acts as $\theta \mapsto - \theta$. In terms of the holomorphic coupling we have 
\begin{align}
	\tau \mapsto - \tau^*~,
\end{align}
and in terms of $\Gmat$ this implies 
\begin{align}
	\Gmat \mapsto \Gmat'=
	\mathbf{P} \Gmat \mathbf{P},
	\quad \mathbf{P}
	=\begin{pmatrix}
		1 & 0 \\
		0 & -1
	\end{pmatrix}~.
\end{align}
We can find the effect of a CP transformation \cite{Ramsey:1958gvj} in a general charge basis by performing a duality transformation $\mathbf{D}_e$ to the electric charge basis, acting with CP, and then performing the inverse duality transformation:
\begin{align}
	 \Gmat 
	\mapsto \Gmat'=
	\mathbf{D}^{-T}_e
	\mathbf{P}
	\mathbf{D}^T_e
	\Gmat \mathbf{D}_e
	\mathbf{P} \mathbf{D}^{-1}_e~.    \label{eq:Gcp}
\end{align}
CP invariance requires that the physical theory\textemdash not the Lagrangian\textemdash is invariant under CP. Indeed, the most general possibility is 
that $\Gmat'$ is related to $\Gmat$ by a duality transformation.

The condition for CP invariance is then 
\begin{align}
	\Gmat \mapsto \Gmat'=
	\mathbf{D}_\text{CP}^T\Gmat \mathbf{D}_\text{CP} 
	=
	\mathbf{D}^{-T}_e
	\mathbf{P}
	\mathbf{D}^T_e
	\Gmat \mathbf{D}_e
	\mathbf{P} \mathbf{D}^{-1}_e,
\end{align}
for some duality transformation $\mathbf{D}_\text{CP}$. This shows how we can alter $\Gmat$'s transformation rule to include the duality transformation. This ensures the Lagrangian is invariant under CP. This modified transformation rule is
\begin{align}
	\Gmat \mapsto 
	\mathbf{D}_\text{CP}^T
	\mathbf{D}^{-T}_e
	\mathbf{P}
	\mathbf{D}^T_e
	\Gmat \mathbf{D}_e
	\mathbf{P} \mathbf{D}^{-1}_e 
	\mathbf{D}_\text{CP}.
\end{align}
We can write this as 
\begin{align}
	\Gmat \mapsto \mathbf{U}_\text{CP}^T\Gmat 
	\mathbf{U}_\text{CP}~,
	\quad \mathbf{U}_\text{CP} \in \text{GL}(2,\mathbb{Z})~, \ \ 
	\det \mathbf{U}_\text{CP} = -1~,
	\label{eq:Gcptrans}
\end{align}
for a specific choice of $\mathbf{U}_\text{CP}$. 
Each $\mathbf{U}_\text{CP}$ gives a different CP transformation rule for $\Gmat$.\footnote{
Note that different $\mathbf{U}_\text{CP}$ may come from the same original CP transformation rule.
For example, in the electric charge basis both $\theta=0$ and $\theta=\pi$ preserve CP.
However, when $\theta = \pi$ the CP transformation takes us to a theory that is related to the original by a $T$ symmetry, whereas in the $\theta =0$ case the Lagrangian is invariant.} 

There is no choice of $\Gmat$ that is invariant under every possible CP transformation rule. However, given a choice of $\mathbf{U}_\text{CP}$ we can understand the invariance properties of $\Gmat$ under that CP transformation rule. In terms of $\tau$, the transformation rule takes the form
\begin{align}
	\tau\mapsto
	\Theta_{U_\text{CP}}(\tau)   
	&=
	\frac{A\tau^\ast+B}{C\tau^\ast+D}~,
	\quad
	\mathbf{U}_\text{CP}=
	\begin{pmatrix}
	A&B\\
	C&D
	\end{pmatrix}~,
	\label{eq:generalCPAntiHolomorphic}
\end{align}
and the condition for CP invariance is
\begin{align}
\tau=\Theta_{U_\text{CP}}(\tau)~.
\label{eq:CPFixedLocusGeneral}
\end{align} 
Explicitly, the fixed-point equation is
\begin{align}
	C|\tau|^2+D\tau-A\tau^\ast-B=0~.
	\label{eq:CPFixedEquationExpanded}
\end{align}
Writing $\tau=x+iy$ with $y>0$, its imaginary part gives
\begin{align}
	(A+D)y=0~.
\end{align}
Consequently, to have a fixed point at nonzero coupling, $A+D=0$. 
Together with $AD-BC=-1$, this implies $U_\text{CP}^2=\mathbf 1$. The real part of Eq.~\eqref{eq:CPFixedEquationExpanded} is then
\begin{align}
	C(x^2+y^2)-2Ax-B=0~.
	\label{eq:CPFixedGeometry}
\end{align}
For $C\neq0$, the couplings $x$ and $y$ must lie on the semicircle solving
\begin{align}
	\left(x-\frac{A}{C}\right)^2+y^2=\frac{1}{C^2}~,
	\label{eq:CPFixedSemicircle}
\end{align}
while $C=0$ gives the vertical line $x=-B/(2A)$.

The CP invariance of the theory also depends on the charge spectrum of our theory. In the electric charge basis CP transforms a charge doublet as
\begin{align}
	(q,\,g) \mapsto (-q,\,g)~.
\end{align}
In terms of Weyl spinors, this means that for every left-handed spinor with charge $(q,\,g)$ there must be another with charge $(q,\,-g)$ \cite{Csaki:2010rv}
\footnote{
Following with the rest of the paper we assume all fermions are Dirac spinors, however the arguments described here also hold for chiral theories so long as the charges are chosen such that there is no gauge anomaly; the sole difference being that the $\beta$-functions differ by a factor of two if written in terms of the number of Weyl spinors.
}.

This condition implies that $\mathbf{Q} \equiv \sum_\ell \bm{q}_\ell \bm{q}_\ell^T$ being diagonal (because every term odd in $g_\ell$ cancels in the sum) or
\begin{align}
	\mathbf{P} \mathbf{Q} \mathbf{P} = \mathbf{Q}~.
\end{align}
An equivalent condition must hold for each mass. 
Following the above reasoning regarding $\Gmat$ and using the duality transformation rule $\bm{q}_{\ell} \mapsto \mathbf{D}^{-1} \bm{q}_\ell$ the charge spectrum must be invariant under $\mathbf{U}_\text{CP}$ i.e.
\begin{align}
	\{
	\bm{q}_\ell
	\} = 
	\{
	\mathbf{U}_\text{CP}^{-1} \bm{q}_\ell 
	\}~.
\end{align}
This implies the condition 
\begin{align}
	\mathbf{U}_\text{CP}^{-1}\mathbf{Q} \mathbf{U}_\text{CP}^{-T}= \mathbf{Q}~,
\end{align}
or more generally 
\begin{align}
	\sum_\ell f(\bm{q}_\ell) 
	= \sum_\ell f(\mathbf{U}_\text{CP}^{-1} \bm{q}_\ell)~,
	\label{eq:GeneralCPCondition} 
\end{align}
for any function $f$.

The RG flow preserves CP symmetry when its vector field is tangent to the CP invariant curve in coupling space. This is most easily checked by writing the transformation in terms of $\Gmat$, see \eqref{eq:Gcptrans}, because (unlike for $\tau$) the transformation rule for $\Gmat$ is linear. The condition for the RG flow to preserve CP comes from differentiating \eqref{eq:Gcptrans} with respect to $\ln\mu$:
\begin{align}
	\boldsymbol{\beta}_{\Gmat}
	= \mathbf{U}_\text{CP}^T \boldsymbol{\beta}_{\Gmat}
	\mathbf{U}_\text{CP}
	\bigg|_{
	\mathbf{U}_\text{CP}^T \Gmat
	\mathbf{U}_\text{CP} = \Gmat,
	\,\,\,
	\mathbf{U}_\text{CP}^{-1}\mathbf{Q} \mathbf{U}_\text{CP}^{-T}=\mathbf{Q}}~.  \label{eq:CPtangencyCond}
\end{align}

We can check this explicitly for the $\beta$-function 
\begin{align}
	\boldsymbol{\beta}_{\Gmat}
	= \frac{1}{6 \pi^2}
	\left[
	\Gmat \mathbf{Q} \Gmat
	+ \Delta^2
	\epsmat \mathbf{Q} \epsmat
	\right],
\end{align}
obtained in Sec.~\ref{sec:mainOneLoopResult}. Because $\mathbf{U}_\text{CP}^T\epsmat \mathbf{U}_\text{CP} = -\epsmat$ we find 
\begin{align}
	\mathbf{U}_\text{CP}^T\boldsymbol{\beta}_{\Gmat}
	\mathbf{U}_\text{CP}
	&= 
	\frac{1}{6 \pi^2}
	\left[
	\mathbf{U}_\text{CP}^T\Gmat 
	\mathbf{Q} 
	\Gmat \mathbf{U}_\text{CP}
	+ \Delta^2
	\mathbf{U}_\text{CP}^T
	\epsmat \mathbf{Q} \epsmat
	\mathbf{U}_\text{CP}
	\right]
	\notag\\
	&=
	\frac{1}{6 \pi^2}
	\left[
	\mathbf{U}_\text{CP}^T\Gmat 
	\mathbf{U}_\text{CP}
	\mathbf{U}_\text{CP}^{-1}
	\mathbf{Q} 
	\mathbf{U}_\text{CP}^{-T}
	\mathbf{U}_\text{CP}^T
	\Gmat \mathbf{U}_\text{CP}
	+ \Delta^2
	\mathbf{U}_\text{CP}^T
	\epsmat
	\mathbf{U}_\text{CP}
	\mathbf{U}_\text{CP}^{-1}
	\mathbf{Q} 
	\mathbf{U}_\text{CP}^{-T}
	\mathbf{U}_\text{CP}^T
	\epsmat
	\mathbf{U}_\text{CP}
	\right]\notag\\
	&= \boldsymbol{\beta}_{\Gmat}~.
\end{align}
This confirms that the one-loop running preserves CP symmetry if it is already present.

This can also be expressed in terms of $\tau$. By Differentiating Eq.~\eqref{eq:generalCPAntiHolomorphic} we find condition for CP tangency of the RG flow
\begin{align}
	\beta_\tau = - \frac{1}{(C\tau^* + D)^2}
	\beta_\tau^*~.
	\label{eq:TauCPFlowTangency}
\end{align}

\section{Application to the Argyres--Douglas Ansatz}\label{sec:ADComparison}

In this section we compare to the results \cite{Argyres:1995jj} in a $\mathcal{N}=2$ supersymmetric theory. 
Readers not familiar with technical details may wish to skip ahead to the following section.
The first ingredient is the protected Wilsonian contribution of one light $\mathcal N=2$ hypermultiplet to the running of $\tau$. For a state with primitive AD charge $(q_{\rm e}^{\rm AD},q_{\rm m}^{\rm AD})$, an integral change of basis makes the state purely electric in a local duality frame.  Since the Wilsonian prepotential receives no perturbative corrections beyond one loop in that frame, transforming back gives~\cite{SeibergWitten,SeibergWittenQCD,Argyres:1995jj,Argyres}
\begin{align}
\left.\beta_{\tau_{\rm AD}}^{\rm W}\right|_{(q_{\rm e}^{\rm AD},q_{\rm m}^{\rm AD})}
=-\frac{i}{2\pi}
\left(q_{\rm e}^{\rm AD}+q_{\rm m}^{\rm AD}\tau_{\rm AD}\right)^2~.
\label{eq:ArgyresDouglasBeta}
\end{align}
This equation fixes one protected single-state contribution.  It does not require a local Lagrangian in which several mutually nonlocal hypermultiplets are simultaneously elementary.

For several mutually nonlocal light states, no single duality frame makes every state electric. Argyres and Douglas therefore proposed the additional Wilsonian additivity ansatz~\cite{Argyres:1995jj}
\begin{align}
\beta_{\tau_{\rm AD}}^{\rm W,\,AD\,ansatz}
=-\frac{i}{2\pi}
\sum_{\ell\in\mathrm{light}}
\left(q_{{\rm e},\ell}^{\rm AD}
+q_{{\rm m},\ell}^{\rm AD}\tau_{\rm AD}\right)^2~.
\label{eq:ADAdditiveAnsatz}
\end{align}
Each term is protected in its natural electric frame, but summing the terms is an extra dynamical assumption.  Thus Eq.~\eqref{eq:ADAdditiveAnsatz} is not implied by the one-loop exactness of Eq.~\eqref{eq:ArgyresDouglasBeta}.

We next compare the primitive AD normalization with the conventional $eb=4\pi$, or $\dlat=2$, normalization. 
The degree-two period map (described in Appendix \ref{app:ADNormalization}) keeps the electric period fixed and doubles the magnetic period,
\begin{align}
a&=a^{\rm AD}~,
& a_D&=2a_D^{\rm AD}~,
& \tau&=2\tau_{\rm AD}~.
\label{eq:ADMainPeriodMap}
\end{align}
Preserving the central charge of the extended supersymmetry algebra then requires
\begin{align}
Q_{\rm e}=q_{\rm e}^{\rm AD}~,
\qquad
Q_{\rm m}=\frac12q_{\rm m}^{\rm AD}~.
\label{eq:ADMainChargeMap}
\end{align}
This non-unimodular map is the degree-two isogeny discussed in Appendix~\ref{app:ADNormalization}.  Since $\beta_\tau^{\rm W}=2\beta_{\tau_{\rm AD}}^{\rm W}$ and
\begin{align}
Q_{\rm e}+Q_{\rm m}\tau
=q_{\rm e}^{\rm AD}+q_{\rm m}^{\rm AD}\tau_{\rm AD}~,
\end{align}
Eq.~\eqref{eq:ArgyresDouglasBeta} becomes
\begin{align}
\left.\beta_{\tau}^{\rm W}\right|_{(Q_{\rm e},Q_{\rm m})}^{\rm hyper}
=-\frac{i}{\pi}
\left(Q_{\rm e}+Q_{\rm m}\tau\right)^2~.
\label{eq:CanonicalHyperBeta}
\end{align}
Thus the factor of two in the conventional hypermultiplet coefficient comes from the degree-two period and charge map, not from an additional loop contribution.  Applying the same map term by term converts Eq.~\eqref{eq:ADAdditiveAnsatz} to the corresponding conventional-normalization ansatz.

The isogeny is a local comparison of periods, charges, and one protected Wilsonian contribution.  It does not embed the complete mutually nonlocal Argyres--Douglas spectrum into one integral $\dlat=2$ electromagnetic charge lattice, prove the additive ansatz, or determine the full strongly coupled 1PI $\beta$ function.  Appendix~\ref{app:ADNormalization} gives the detailed derivation of the DSZ pairings and the normalization of the Seiberg–Witten periods.
It would be interesting to pursue a fully supersymmetric calculation to confirm the expectation that higher loop corrections vanish and to explore nonperturbative corrections.


\section{Two-loop Running}
\label{sec:twoLoopDualityPrediction}

The one-loop running~\eqref{eq:betaGmatrix} comes from a single matter loop attached to two external currents. At two loops the same matter loop is dressed by one internal photon. Because both ends of that internal photon attach to the same species, its mixed electric--magnetic charge factor vanishes identically. The remaining loop integral is the ordinary two-loop QED vacuum polarization, while the external charge structure is unchanged from one loop. This determines the two-loop running and shows directly that the DSZ normalization remains fixed.


To begin, for dyon species $\ell$, we define the one-loop matrix structure
\begin{align}
\mathbf b_\ell(\Gmat)
&\equiv
\Gmat\Qvec{\ell}\Qvec{\ell}^T\Gmat+\Delta^2\epsmat\Qvec{\ell}\Qvec{\ell}^T\epsmat\nonumber\\
&=\Gmat\Qvec{\ell}\Qvec{\ell}^T\Gmat+\frac{\Delta^2}{\det\Gmat}\Gmat\epsmat\Gmat\Qvec{\ell}\Qvec{\ell}^T\epsmat~,
\label{eq:TwoLoopAllowedMatrixStructure}
\end{align}
where we have used~\eqref{eq:EpsDetIdent} and the symmetry of $\Gmat$. It follows that
\begin{align}
\operatorname{tr}\!\left(\Gmat^{-1}\mathbf b_\ell(\Gmat)\right)
&=\operatorname{tr}\left( \Qvec{\ell}\Qvec{\ell}^T\Gmat\right)+\frac{\Delta^2}{\det\Gmat}\operatorname{tr}\left(\Gmat\Qvec{\ell}\Qvec{\ell}^T\epsmat\epsmat \right)\nonumber\\
&=\operatorname{tr}\left( \Gmat\Qvec{\ell}\Qvec{\ell}^T\right)-\frac{\Delta^2}{\det\Gmat}\operatorname{tr}\left(\Gmat\Qvec{\ell}\Qvec{\ell}^T \right)=\Qvec{\ell}^T\Gmat\Qvec{\ell}\left(1-\frac{\Delta^2}{\det\Gmat} \right)~.\label{eq:TwoLoopbTraceless}
\end{align}
As discussed in Sec.~\ref{sec:matrixFlowChecks}, this shows that when $\det\Gmat=\Delta^2$ that $\operatorname{tr}(\Gmat^{-1}\mathbf b_\ell)=0$, 
which is the condition for the corresponding RG flow to preserve $\det\Gmat$.

\subsection{The two-loop vacuum polarization}
\label{sec:TwoLoopVP}

The photon has no self-interactions, and the two-potential formulation does not change the topology of the 1PI graphs. At two loops the photon self-energy contains one matter loop of species $\ell$ dressed by one internal photon, together with the usual self-energy and vertex counterterm insertions. Graphs containing two distinct matter loops are one-particle reducible at this order. Therefore the two-loop 1PI current correlator is a sum over all possible particles running in the fermion loop, just like the one-loop calculation.

In the holomorphic kinetic-coupling basis of Sec.~\ref{sec:Feynman}, the field-strength propagator is the sum of a Coulomb term proportional to $\Gmat$ and a mixed electric--magnetic term proportional to $\Delta\epsmat$:
\begin{align}
\Qvec{\ell}^T\!\left[
{}_{(k)}^{\alpha\beta}\mathcal{P}_{(k)}^{\gamma\delta}\,\Gmat
+\Delta\,{}_{(k)}^{\alpha\beta}\mathcal{P}_{(k)}^{\mu\nu}\,{}^{(n)}_{\star\mu\nu}\mathcal{P}_{(k)}^{\gamma\delta }\,\epsmat
\right]\!\Qvec{\ell}
&=
\left(\Qvec{\ell}^T\Gmat\Qvec{\ell}\right)
{}_{(k)}^{\alpha\beta}\mathcal{P}_{(k)}^{\gamma\delta}\,
+\Delta\left(\Qvec{\ell}^T\epsmat\Qvec{\ell}\right)\,{}_{(k)}^{\alpha\beta}\mathcal{P}_{(k)}^{\mu\nu}\,{}^{(n)}_{\star\mu\nu}\mathcal{P}_{(k)}^{\gamma\delta }
\nonumber\\
&=
\left(\Qvec{\ell}^T\Gmat\Qvec{\ell}\right)
{}_{(k)}^{\alpha\beta}\mathcal{P}_{(k)}^{\gamma\delta}\,~.
\label{eq:TwoLoopInternalLine}
\end{align}
The $n^\mu$-dependent mixed propagator drops out because $\Qvec{\ell}^T\epsmat\Qvec{\ell}=0$. Consequently, the internal line reduces to QED photon exchange within a current loop, with squared coupling
\begin{align}
\Qvec{\ell}^T\Gmat\Qvec{\ell}
=
\frac{\Delta\,|q_\ell+g_\ell\tau|^2}{\operatorname{Im}\tau}~.
\label{eq:TwoLoopEffectiveCoupling}
\end{align}
This combination is duality invariant. The physical coupling controls the QED subdiagram and therefore the perturbative expansion of species $\ell$.

The same conclusion holds in the worldline description of App.~\ref{app:worldlineTopology}. A topological self-interaction of a single dyon worldline is proportional to $\Delta\,\Qvec{\ell}^T\epsmat\Qvec{\ell}=0$. 
No subtraction of a topological self-interaction is needed for the internal photon line.
The external field strengths attach to the species-$\ell$ loop exactly as at one loop. So, the prescription of Sec.~\ref{sec:localCounterterm}, with the Hodge star on the fluctuating-loop side, applies unchanged. 
The external contraction is the same as in Eq.~\eqref{eq:matrixOneLoop}; except the unit-current correlator is replaced by its two-loop value.

The self-energy and vertex subdivergences involve only the coupling in Eq.~\eqref{eq:TwoLoopEffectiveCoupling}, so the Ward identity $Z_1=Z_2$ removes them exactly as in QED. After these subdivergences are subtracted, the remaining divergence of the conserved-current correlator is local and has the same parity-even tensor structure as at one loop. In particular, 
nothing inside the loop supplies an $\varepsilon^{\mu\nu\rho\sigma}$ tensor.
The momentum integrals are therefore ordinary QED integrals. We can evaluate the unit-current correlator in dimensional regularization and $\overline{\rm MS}$ without continuing the four-dimensional $\varepsilon$ tensor. The latter appears only in the external Zwanziger contractions.

Let $\Pi^{(2)}_\ell(k^2)$ denote the two-loop QED correction to the unit-current correlator with one factor of $\Qvec{\ell}^T\Gmat\Qvec{\ell}$ removed. Then
\begin{align}
\Gamma^{(\ell)}_{ij,\mathrm{VP}}
=
\left[
\Pi_\ell(k^2)
+\left(\Qvec{\ell}^T\Gmat\Qvec{\ell}\right)\Pi^{(2)}_\ell(k^2)
\right]
\frac{J_{i\mu}J_j^{*\mu}}{k^2}\,
\Qvec{i}^T\mathbf b_\ell(\Gmat)\,\Qvec{j}
+\mathcal O(\text{three loops})~.
\label{eq:TwoLoopGamma}
\end{align}
The scale dependence of the renormalized two-loop QED function is~\cite{JostLuttinger:1950,MachacekVaughn:1983,KataevLarin:2012}
\begin{align}
\mu\frac{\partial}{\partial\mu}\Pi^{(2),\rm ren}_\ell
=
\frac{1}{32\pi^4}~.
\label{eq:TwoLoopPiDerivative}
\end{align}
In the normalization of Eq.~\eqref{eq:diracVacPolDerivativeMain}, this reproduces
\begin{align}
\beta_{e^2}
=
\frac{e^4}{6\pi^2}
+\frac{e^6}{32\pi^4}
+\cdots
\end{align}
for one unit-charge Dirac fermion.

Comparing Eq.~\eqref{eq:TwoLoopGamma} with the tree-level amplitude gives
\begin{align}
\boxed{
\betamat_{\Gmat}
=
\frac{1}{6\pi^2}\sum_\ell\mathbf b_\ell(\Gmat)
+
\frac{1}{32\pi^4}\sum_\ell
\left(\Qvec{\ell}^T\Gmat\Qvec{\ell}\right)
\mathbf b_\ell(\Gmat)
+\mathcal O(\text{three loops})~.
}
\label{eq:TwoLoopDyonPredictionSum}
\end{align}
Thus, each species receives the familiar QED two-loop correction, with its electric charge squared replaced by the duality-invariant coupling $\Qvec{\ell}^T\Gmat\Qvec{\ell}$. 

\subsection{Consequences}
\label{sec:TwoLoopConsequences}

Using Eq.~\eqref{eq:TwoLoopbTraceless} in Eq.~\eqref{eq:TwoLoopDyonPredictionSum} gives
\begin{align}
\frac{d}{d\log\mu}\det\Gmat
&=
\left(\det\Gmat-\Delta^2\right)
\sum_\ell
\left[
\frac{1}{6\pi^2}
+\frac{\Qvec{\ell}^T\Gmat\Qvec{\ell}}{32\pi^4}
\right]
\Qvec{\ell}^T\Gmat\Qvec{\ell}
+\mathcal O(\text{three loops})~.
\end{align}
Hence $\det\Gmat=\Delta^2$ is preserved by the RG flow through two loops. This is an output of the calculation: the mixed propagator is the only part of the internal photon that could change the antisymmetric electric--magnetic coupling, and its charge factor vanishes. The Coulomb couplings run, while $eb$, every DSZ pairing, and every pairwise helicity remain fixed.

For a single species, or for a spectrum whose charges are all mutually local, Eq.~\eqref{eq:TwoLoopDyonPredictionSum} reduces by an SL$(2,\mathbb Z)$ change of charge basis to ordinary multi-flavor two-loop QED. The new result is that the same species-by-species sum remains valid when the spectrum contains mutually nonlocal charges.

At fixed $\Delta$, Eq.~\eqref{eq:betaTauMatrixGeneralD} maps $\mathbf b_\ell(\Gmat)$ to $-i\Delta(q_\ell+g_\ell\tau)^2$. Using Eq.~\eqref{eq:TwoLoopEffectiveCoupling}, we obtain
\begin{align}
\beta_\tau^{(2)}
&=
-\frac{i\Delta}{32\pi^4}\sum_\ell
\left(\Qvec{\ell}^T\Gmat\Qvec{\ell}\right)
(q_\ell+g_\ell\tau)^2
\nonumber\\
&=
-\frac{i\dlat^2}{8\pi^2}\sum_\ell
\frac{|q_\ell+g_\ell\tau|^2}{\operatorname{Im}\tau}
(q_\ell+g_\ell\tau)^2~.
\label{eq:TwoLoopBetaTau}
\end{align}
Unlike the one-loop result, $\beta_\tau^{(2)}$ depends on both $\tau$ and $\tau^*$, as generally expected in a nonsupersymmetric theory. The factor $|q+g\tau|^2/\operatorname{Im}\tau$ is duality invariant, so $\beta_\tau^{(2)}$ has the same modular weight as the one-loop result.

The component $\beta$-functions are
\begin{align}
\beta_e^{(2)}
&=
\frac{e}{64\pi^4}\sum_\ell
\left[e^4\left(q_\ell+g_\ell\frac{\theta}{2\pi}\right)^4-b^4g_\ell^4\right]~,\nonumber\\
\beta_b^{(2)}
&=
-\frac{b}{64\pi^4}\sum_\ell
\left[e^4\left(q_\ell+g_\ell\frac{\theta}{2\pi}\right)^4-b^4g_\ell^4\right]~,\nonumber\\
\beta_\theta^{(2)}
&=
\frac{b^2}{8\pi^3}\sum_\ell
\left[e^2\left(q_\ell+g_\ell\frac{\theta}{2\pi}\right)^2+b^2g_\ell^2\right]\left(q_\ell+g_\ell\frac{\theta}{2\pi}\right) g_\ell~.
\label{eq:TwoLoopBetaTheta}
\end{align}
The combination $\beta_e/e+\beta_b/b$ vanishes, as required by fixed $eb$. For $g_\ell=0$, these expressions reduce to the standard two-loop QED result.

Under a general CP transformation $\mathbf U_{\rm CP}$, the CP-invariant
couplings obey
$\mathbf U_{\rm CP}^T\Gmat\mathbf U_{\rm CP}=\Gmat$.
Since $\det\mathbf U_{\rm CP}=-1$, Eq.~\eqref{eq:Gcptrans} also implies
$\mathbf U_{\rm CP}^T\epsmat\mathbf U_{\rm CP}=-\epsmat$.
It then follows directly from
Eq.~\eqref{eq:TwoLoopAllowedMatrixStructure} that
\begin{align}
\mathbf U_{\rm CP}^T\,
\mathbf b_\ell(\Gmat)\,
\mathbf U_{\rm CP}
&=
\left.
\mathbf b_\ell(\Gmat)
\right|_{\Qvec{\ell}\to
\mathbf U_{\rm CP}^{-1}\Qvec{\ell}},
\qquad
\Qvec{\ell}^T\Gmat\Qvec{\ell}
=
\left(\mathbf U_{\rm CP}^{-1}\Qvec{\ell}\right)^T
\Gmat
\left(\mathbf U_{\rm CP}^{-1}\Qvec{\ell}\right).
\end{align}
Therefore the two-loop term in
Eq.~\eqref{eq:TwoLoopDyonPredictionSum} transforms as
\begin{align}
\mathbf U_{\rm CP}^T
\betamat_{\Gmat}^{(2)}
\mathbf U_{\rm CP}
&=
\frac{1}{32\pi^4}
\sum_\ell
\left[
\left(\mathbf U_{\rm CP}^{-1}\Qvec{\ell}\right)^T
\Gmat
\left(\mathbf U_{\rm CP}^{-1}\Qvec{\ell}\right)
\right]
\left.
\mathbf b_\ell(\Gmat)
\right|_{\Qvec{\ell}\to
\mathbf U_{\rm CP}^{-1}\Qvec{\ell}}
\nonumber\\
&=
\betamat_{\Gmat}^{(2)}~,
\end{align}
where the second equality follows from the spectral CP condition
\eqref{eq:GeneralCPCondition}.   
which is the matrix form of the CP tangency condition \eqref{eq:CPtangencyCond} derived in Sec.~\ref{sec:cpSpectrum} extended to two loops.
Thus, the two-loop flow preserves an exact CP
symmetry of the full spectrum and interactions.

\subsection{Scheme dependence}
\label{sec:TwoLoopScheme}

For one running coupling, the first two $\beta$-function coefficients are universal in mass independent schemes. Here both $e$ and $\theta$ can run, so a finite redefinition of the couplings can change the two-loop $\beta$-functions. We therefore specify the finite redefinitions under which Eq.~\eqref{eq:TwoLoopDyonPredictionSum} is unchanged.
Consider a coupling redefinition $\Gmat'=\Gmat+\mathbf h(\Gmat)$, with $\mathbf h$ of one-loop order. The two-loop $\beta$-function shifts by the commutator of the one-loop RG flow with $\mathbf h$. 
For the finite renormalizations relevant to the unit-current correlators, $\Pi_\ell\to\Pi_\ell+a_\ell$, one has $\mathbf h=\sum_\ell a_\ell\mathbf b_\ell(\Gmat)$.

Writing $z_\ell\equiv q_\ell+g_\ell\tau$, the corresponding one-loop terms are proportional to $z_\ell^2\partial_\tau$. Their commutator is
\begin{align}
\left[z_\ell^2\partial_\tau,\,z_m^2\partial_\tau\right]
=
2\left(\Qvec{\ell}^T\epsmat\Qvec{m}\right)
z_\ell z_m\,\partial_\tau~,
\label{eq:TwoLoopBracket}
\end{align}
so the induced change of the two-loop coefficient is
\begin{align}
\delta\beta^{(2)}_\tau
=
\frac{\Delta^2}{3\pi^2}\sum_{\ell,m}
a_\ell
\left(\Qvec{\ell}^T\epsmat\Qvec{m}\right)
z_\ell z_m~.
\label{eq:TwoLoopSchemeShift}
\end{align}
A universal finite subtraction, $a_\ell=a$, gives zero by antisymmetry. Equation~\eqref{eq:TwoLoopDyonPredictionSum} is therefore unchanged within the class of mass-independent prescriptions that differ by a universal finite subtraction of the unit-current correlators. It is also unchanged under these redefinitions for a mutually local spectrum. Species-dependent finite renormalizations can shift the two-loop coefficient through pairwise DSZ invariants. We therefore quote the two-loop result in $\overline{\rm MS}$, applied uniformly to the unit-current correlators.

\subsection{Beyond Two Loops}
\label{sec:TwoLoopBeyond}

The simplification in Eq.~\eqref{eq:TwoLoopInternalLine} is special to two loops. At three loops an internal photon on a species-$\ell$ loop can contain a one-loop vacuum-polarization insertion from species $m$. Furry's theorem removes the other candidate cross-species graphs with an odd number of vector-current insertions on a fermion loop.

The charge factor of the polarization-dressed internal line is
\begin{align}
\Qvec{\ell}^T\mathbf b_m(\Gmat)\Qvec{\ell}
=
\left(\Qvec{\ell}^T\Gmat\Qvec{m}\right)^2
-\Delta^2\left(\Qvec{\ell}^T\epsmat\Qvec{m}\right)^2~.
\label{eq:ThreeLoopChargeFactor}
\end{align}
For mutually local species the second term vanishes and the result reduces to the familiar QED charge factor. For mutually nonlocal species, the square of the DSZ pairing enters an internal photon line for the first time.

At this order the mixed electric--magnetic structure can no longer be eliminated by $\Qvec{\ell}^T\epsmat\Qvec{\ell}=0$. The separation of dynamical and topological terms must then be extended to photon lines involving two different fluctuating species. The argument of Sec.~\ref{sec:localCounterterm} does not directly cover that case.
Thus two loops is the last order for which the running follows directly from the one-loop external structure and the known QED vacuum polarization. 
Beginning at three loops, pairwise DSZ invariants of different species enter internal lines, so the running probes genuinely mutually nonlocal dynamics.

\section{Conclusion\label{s.con}}

The main result is the charge-covariant flow at one loop
\begin{align}
	\bm{\beta}_{\Gmat}^{(1)}
	&=\frac{1}{6\pi^2}\sum_\ell
	\left[
	\Gmat\Qvec{\ell}\Qvec{\ell}^T\Gmat
	+\Delta^2\epsmat\Qvec{\ell}\Qvec{\ell}^T\epsmat
	\right]=\frac{1}{6\pi^2}\sum_\ell \mathbf b_\ell(\Gmat)~,\nonumber\\
	\beta_\tau^{(1)}
	&=-\frac{i\,\dlat}{3\pi}\sum_\ell(q_\ell+g_\ell\tau)^2~,
\end{align}
and at two loops
\begin{align}
	\bm{\beta}_{\Gmat}^{(2)}&=\frac{1}{32\pi^4}\sum_\ell\left(\Qvec{\ell}^T\Gmat\Qvec{\ell} \right)\mathbf b_\ell(\Gmat)~,\nonumber\\
	\beta_\tau^{(2)}&=-\frac{i\dlat^2}{8\pi^2}\sum_\ell
	\frac{|q_\ell+g_\ell\tau|^2}{\operatorname{Im}\tau}
	(q_\ell+g_\ell\tau)^2~.
\end{align}
The matrix $\Gmat=\Cmat^T\Cmat$ is defined in terms of the coupling matrix, which converts, for each particle, integral charge labels  in the charge lattice into physical photon couplings. The RG flow preserves $\det\Cmat=\Delta=eb$, so the Coulombic interaction runs while $eb$, every DSZ pairing, and every pairwise helicity remain fixed. These results apply only to perturbative theories, in a strongly coupled matter sector one must consider the full conserved-current correlator. 

The running of $\theta$ changes the Witten-shifted electric charge carried by a monopole in a fixed charge basis.  Equivalently, the same parameter multiplies the local $F\dual F$ counterterm and the $\theta$-dependent electric coupling of the magnetic current.  This coordinate running is not by itself physical CP violation. Indeed, we have shown that the RG running through two loops does not introduce CP violation.  Moreover, exact CP also constrains the fermion mass phases, masses, multiplicities, interactions, thresholds, and the vacuum. Within the one-photon class of theories we consider in this work, moving the CP-odd term between the source vertices and quadratic kinetic terms is an invertible field redefinition, so these two placements do not define independent couplings.

For an $\mathcal N=2$ hypermultiplet, duality and perturbative nonrenormalization fix one protected Wilsonian contribution.   Our results do not establish additivity across mutually nonlocal light states, embed their complete spectrum into one integral charge lattice, or determine the full strongly coupled 1PI flow. So, while the present one-loop calculation agrees with the Argyres--Douglas $\beta$ function, their use of it remains an ansatz rather than a derived result. It would be very interesting to explore this connection further.

\section*{Acknowledgments}
We thank Markus Luty for enlightening discussions.  C.B.V. was supported in part by the National Science Foundation under Grant No.~PHY-2210067.  J.T. is supported by private donations from readers like you. Thank you. 

\appendix

\section{The Witten coupling and Dirac surfaces}
\label{app:WittenSurfaceRepresentatives}
In this appendix we provide a more general derivation of the Witten effect in abelian theories and show that it does not depend on the choice of Dirac surface $\Sigma_K$~\cite{Dirac:1955uv}. 
We show that changing the Dirac surface only redistributes intermediate surface terms, it does not change the physical field strength or the Witten coupling. 

In the absence of magnetic sources, the action
\begin{align}
\mathcal{L}=-\frac14 F_{\mu\nu}F^{\mu\nu}
+\frac14\frac{\theta}{2\pi}\frac{e}{b}
\dual F^{\mu\nu}F_{\mu\nu}-eA_\mu J^\mu
\label{e.MaxLagKinTheta}
\end{align}
contains a total-derivative $\dual F^{\mu\nu}F_{\mu\nu}$ term.  
For magnetic sources, with conserved current $K^\mu$, we can define a Dirac surface to satisfy 
\begin{align}
	\partial_\mu \Sigma_K^{\mu\nu}
	= K^\nu.
	\label{eq:sigmaKdef}
\end{align}
The general solution to  $\partial_\mu\left({}^\star\!F^{\mu\nu}-b\,\Sigma_K^{\mu\nu}\right)=0$ in flat spacetime can then be written as  $F^A_{\mu\nu}=\partial_\mu A_\nu-\partial_\nu A_\mu$, which is related to the physical field strength
\begin{align}
\dual F^{A\mu\nu}
&=\dual F^{\mu\nu}-b\Sigma_K^{\mu\nu}~,
\label{eq:FdefSinglePot}\\
F_{\mu\nu}&=F^A_{\mu\nu}-b\,\dual\hspace{0.1em}\Sigma_{K\mu\nu}~.
\label{eq:FphysicalSinglePot}
\end{align}

Substitution into the $\theta$ interaction gives
\begin{align}
\frac14\frac{\theta}{2\pi}\frac{e}{b}\dual F^{\mu\nu}F_{\mu\nu}
={}&\frac14\frac{\theta}{2\pi}\frac{e}{b}
\left(\dual F^{A\mu\nu} F^A_{\mu\nu}+2bF^A_{\mu\nu}\Sigma_K^{\mu\nu}
-b^2\Sigma_{K\mu\nu}\dual\hspace{0.1em}\Sigma_K^{\mu\nu}\right)~.
\label{eq:thetaSurfaceExpansion}
\end{align}
The first term on the right is a total derivative.
This may be relevant for global topological effects (e.g. instantons \cite{Csaki:2024ajo,GarciaGarcia:2025uub}), but does not affect the equation of motion. 
  
The magnetic self-surface contribution is
\begin{align}
I_K\equiv-\frac{\theta eb}{8\pi}
\int d^4x\,\Sigma_K^{\mu\nu}\dual\hspace{0.1em}\Sigma_{K\mu\nu}~.
\label{eq:WittenSelfSurfaceIK}
\end{align}
For the $n^\mu$-dependent $\Sigma^{\mu\nu}_K$ defined by Eq. \eqref{eq:SigmanDef} we have $I_K=0$ by symmetry (it has two $n^\mu$ contracted into a Levi-Civita).
Finally, to reach a larger class of objects, consider a general deformation of the surfaces considered above,
\begin{align}
	\Sigma_K^{\mu\nu}\longrightarrow
	\Sigma_K^{\mu\nu}+\delta\Sigma_K^{\mu\nu}~,
	\qquad
	\partial_\mu\delta\Sigma_K^{\mu\nu}=0~,
\end{align}
which can locally be written as $\dual\hspace{0.1em}\delta\Sigma_{K\mu\nu}=\partial_\mu\Lambda_\nu-\partial_\nu\Lambda_\mu$. 
This is a manifestation of the one-form symmetry of Maxwell's equations~\cite{Hull:2024uwz}. 

The simultaneous potential redefinition
\begin{align}
	A_\mu\longrightarrow A_\mu+b\,\Lambda_\mu
	\label{eq:WittenRepresentativePotentialShift}
\end{align}
leaves
\begin{align}
	F_{\mu\nu}=F^A_{\mu\nu}-b\,\dual\hspace{0.1em}\Sigma_{K\mu\nu}
\end{align}
unchanged. Because the field strength is unchanged, so too is the value of the whole ${}^\star\!F^{\mu\nu}F_{\mu\nu}$ term in the Lagrangian. 

Therefore, the value we found for the above case must also be its value for a general Dirac surface. 
In other words, a surface deformation can affect the individual values of $F^A_{\mu\nu}\Sigma_K^{\mu\nu}$ and $\Sigma_{K\mu\nu} {}^\star \Sigma_K^{\mu\nu}$ terms in Eq.~\eqref{eq:thetaSurfaceExpansion}, but it cannot change their sum.
 Evaluating $I_K=0$ in the longitudinal or fixed-string representative is therefore a convenient choice within a fixed formulation, not an additional restriction on the physical theory.

The remaining mixed term gives the Witten effect coupling. 
By integrating by parts and using Eq.~\eqref{eq:sigmaKdef},
\begin{align}
\int d^4x\,F^A_{\mu\nu}\Sigma_K^{\mu\nu}
=-2\int d^4x\,A_\mu K^\mu~.
\label{eq:potentialSurfaceEquivalence}
\end{align}
Combining Eqs.~\eqref{e.MaxLagKinTheta}--\eqref{eq:potentialSurfaceEquivalence} produces the usual Witten source interaction given in Eq.~\eqref{eq:WittenSource}.

One could instead solve for a dual potential $B_\mu$ that has local couplings to magnetic currents. 
In this approach, the dual field strength would take the form
\begin{align}
	\dual F_{\mu\nu}=F^B_{\mu\nu}+e\,\dual\hspace{0.1em}\Sigma_{J\mu\nu}~.
\end{align}
Clearly repeating the same argument given above gives a local coupling proportional to $\theta J^\mu K^\mu$.
This appears to be a dual version of the traditional Witten effect \cite{Heidenreich:2023yfc}. 
Following the arguments of Sec. \ref{sec:couplingMatrix}, however, by a  helicity rotation this coupling can be put in the standard Witten effect form.
Therefore, the dual construction of the $\theta$ term does not lead to a separate physical theory in that case.


\section{Helicity, Parity, and CP}\label{app:helicityParityCP}
Weinberg showed how Lorentz invariance and the two physical helicities of a massless spin-1 particle lead to Maxwell interactions~\cite{Weinberg:1965rz}. 
We use his definite-helicity field strengths as a basis, but we do not repeat the $S$-matrix derivation. 
Instead, we classify the most general Hermitian linear coupling of one photon to a polar electric current and an axial magnetic current. 
We first impose parity and then explain the additional conditions required for CP invariance of a theory containing dyons. 
This distinction is important: the tensor analysis constrains the gauge-current couplings, whereas CP is a symmetry only if it also maps the complete particle spectrum, masses, and interactions into themselves.

The definite-helicity field strengths used by Weinberg can be written as
\begin{align}
F_{\pm\mu\nu}=\int\frac{d^3p}{(2\pi)^3\sqrt{2\omega_p}}\left( a_{p\pm}\epsilon^\pm_{\mu\nu}e^{-ip\cdot x}+a^{\dag}_{p\pm}\epsilon^{\pm\ast}_{\mu\nu}e^{ip\cdot x}\right)~,
\end{align}
where $\omega_p=|\vec{p}\,|$ and the $\epsilon^\pm_{\mu\nu}$ are antisymmetric polarization tensors that satisfy
\begin{align}
{}^\star\epsilon^{\pm\alpha\beta}=\mp i\epsilon^{\pm\alpha\beta}~.
\end{align}
Note that the two helicity states are related by complex conjugation
\begin{align}
F_{\pm\mu\nu}^\ast=F_{\mp\mu\nu}~,
\end{align}
but are related to themselves by Hodge duality
\begin{align}
\dual F_{\pm\mu\nu}=\mp iF_{\pm\mu\nu}~.
\end{align}

It is also true that under parity transformations
\begin{align}
F_{\pm\mu\nu}\xrightarrow{\text{P}} (-1)^{(\mu)} (-1)^{(\nu)} F_{\mp\mu\nu}~,
\end{align}
where $(-1)^{(\mu)}=1$ for $\mu=0$ and $(-1)^{(\mu)}=-1$ for $\mu=1,2,3$. There is no implied summation. While the components of these fields are not eigenstates of parity, we can define
\begin{align}
F_{\mu\nu}=\frac12\left(F_{+\mu\nu}+F_{-\mu\nu} \right)~, \ \ \ \dual F_{\mu\nu}=\frac{-i}{2}\left( F_{+\mu\nu}-F_{-\mu\nu} \right)~,
\end{align}
whose components are either odd or even under parity:
\begin{align}
F_{\mu\nu}\xrightarrow{\text{P}} (-1)^{(\mu)} (-1)^{(\nu)} F_{\mu\nu}~, \ \ \ \ \dual F_{\mu\nu}\xrightarrow{\text{P}} -(-1)^{(\mu)} (-1)^{(\nu)} \dual F_{\mu\nu}~.
\end{align} 

For the standard parity assignment, $F_{\mu\nu}$ couples to $\Sigma_J^{\mu\nu}$, where the electric current $J^\mu$ transforms as a polar vector,
\begin{align}
J^\mu\xrightarrow{\text{P}} (-1)^{(\mu)} J^\mu~.
\end{align}
From the identity $\partial_\mu \Sigma^{\mu\nu}_J = J^\nu$ it clearly follows that
\begin{align}
\Sigma_{J\mu\nu}\xrightarrow{\text{P}} (-1)^{(\mu)} (-1)^{(\nu)} \Sigma_{J\mu\nu}~,
\end{align}
and therefore its contraction with $F_{\mu\nu}$ is parity even. 
Similarly, $\dual F_{\mu\nu}$ can couple to the magnetic current in a parity-even interaction. 
The magnetic current is an axial vector under parity:
\begin{align}
K^\mu\xrightarrow{\text{P}} -(-1)^{(\mu)} K^\mu~.
\end{align}
It follows that
\begin{align}
\Sigma_{K\mu\nu}\xrightarrow{\text{P}} -(-1)^{(\mu)} (-1)^{(\nu)} \Sigma_{K\mu\nu}~.
\end{align}

The following discussion extends these observations to the CP-violating case. Suppose the interaction Lagrangian contains the general couplings
\begin{align}
\mathcal L_{\rm int}=-\left(
\alpha F_{+\mu\nu}\Sigma^{\mu\nu}_J+\beta F_{+\mu\nu}\Sigma^{\mu\nu}_K+\gamma F_{-\mu\nu}\Sigma^{\mu\nu}_J+\delta F_{-\mu\nu}\Sigma^{\mu\nu}_K\right)~.
\end{align}
Because the Lagrangian must be Hermitian, we find that $\gamma$ and $\delta$ must be related to $\alpha$ and $\beta$ according to
\begin{align}
\mathcal L_{\rm int}=-\left(
\alpha F_{+\mu\nu}\Sigma^{\mu\nu}_J+\beta F_{+\mu\nu}\Sigma^{\mu\nu}_K+\alpha^\ast F_{-\mu\nu}\Sigma^{\mu\nu}_J+\beta^\ast F_{-\mu\nu}\Sigma^{\mu\nu}_K\right)~,\label{e.HelictyCouplings}
\end{align}
or equivalently
\begin{align}
\mathcal L_{\rm int}=-F_{\mu\nu}\left[(\alpha+\alpha^\ast)\Sigma_J^{\mu\nu}+(\beta+\beta^\ast)\Sigma_K^{\mu\nu}\right]
-i\dual F_{\mu\nu}\left[(\alpha-\alpha^\ast)\Sigma_J^{\mu\nu}+(\beta-\beta^\ast)\Sigma_K^{\mu\nu}\right]~.
\end{align}
A parity transformation of the interactions in Eq.~\eqref{e.HelictyCouplings} produces
\begin{align}
\alpha F_{-\mu\nu}\Sigma^{\mu\nu}_J-\beta F_{-\mu\nu}\Sigma^{\mu\nu}_K+\alpha^\ast  F_{+\mu\nu}\Sigma^{\mu\nu}_J-\beta^\ast  F_{+\mu\nu}\Sigma^{\mu\nu}_K~.
\end{align}

Therefore, a parity-preserving theory is one in which
\begin{align}
\alpha=\alpha^\ast ~, \ \ \ \ \beta=-\beta^\ast ~.
\end{align}
In other words, $\alpha=e/2$ is a real number and $\beta=-ib/2$ is purely imaginary. 
Substituting this into the preceding interaction gives
\begin{align}
\mathcal L_{\rm int}=-eF_{\mu\nu}\Sigma_J^{\mu\nu}-b\dual F_{\mu\nu}\Sigma_K^{\mu\nu}~.
\label{eq:parityEvenSurfaceSources}
\end{align}
Thus, we find that the standard field strength couples to the electric current while its dual couples to the magnetic current. 

This is the same coupling structure that appears in the one-potential formulation~\cite{Blagojevic:1979bm}, although there the magnetic source is obtained through the definition
\begin{align}
F_{\mu\nu}=\partial_\mu A_\nu-\partial_\nu A_\mu-b\,{}^\star\Sigma_{K\mu\nu}=F^A_{\mu\nu}-b\,{}^\star\Sigma_{K\mu\nu}~,
\end{align}
rather than added as an independent term. Substituting this relation into the parity-even source operators gives
\begin{align}
-eF_{\mu\nu}\Sigma_J^{\mu\nu}-b\dual F_{\mu\nu}\Sigma_K^{\mu\nu}
={}&-eF^A_{\mu\nu}\Sigma_J^{\mu\nu}
-b\dual F^A_{\mu\nu}\Sigma_K^{\mu\nu}
+2b^2\Sigma_K^{\mu\nu}\Sigma_{K\mu\nu}
\nonumber\\
&-2eb\,{}^\star\Sigma_K^{\mu\nu}\Sigma_{J\mu\nu}.
\label{e.CPSource}
\end{align}
Similar to the analysis in Appendix~\ref{app:WittenSurfaceRepresentatives}, this last term vanishes when the surface is defined by a fixed-string prescription.
\footnote{ This is a surface choice not a restriction on the physical CP structure. 
This term is not zero for arbitrary surface choices, but  as described in Appendix \ref{app:WittenSurfaceRepresentatives}
choosing a more general surface also requires a potential redefinition that leaves the physical field strength unchanged.}

We now drop the parity restriction on the source couplings. Consider again the general interaction terms in Eq.~\eqref{e.HelictyCouplings}. In theories without parity, $\alpha$ need not be real. However, by writing $\alpha$ in terms of its modulus and phase
\begin{align}
\alpha=\frac{e}{2}e^{i\phi}~,
\end{align}
we can make it real by rephasing the helicity fields according to
\begin{align}
F_{+\mu\nu}\to e^{-i\phi} F'_{+\mu\nu}~, \ \ \ \  F_{-\mu\nu}\to e^{i\phi} F'_{-\mu\nu}~,
\end{align}
which preserves their relationship as conjugate fields. This transformation leads to
\begin{align}
\mathcal L_{\rm int}=-\left[
\frac{e}{2} F'_{+\mu\nu}\Sigma^{\mu\nu}_J+\beta e^{-i\phi} F'_{+\mu\nu}\Sigma^{\mu\nu}_K+\frac{e}{2} F'_{-\mu\nu}\Sigma^{\mu\nu}_J+\beta^\ast e^{i\phi}F'_{-\mu\nu}\Sigma^{\mu\nu}_K\right]~,
\end{align}
which can also be written as
\begin{align}
\mathcal L_{\rm int}=-F'_{\mu\nu}\left[e\Sigma_J^{\mu\nu}+(\beta e^{-i\phi}+\beta^\ast e^{i\phi})\Sigma_K^{\mu\nu}\right]
-i\dual F'_{\mu\nu}(\beta e^{-i\phi}-\beta^\ast e^{i\phi})\Sigma_K^{\mu\nu}~.
\end{align}
With this choice of helicity phase, $F'_{\mu\nu}$ couples to the magnetic current, whereas $\dual F'_{\mu\nu}$ does not couple to the electric current. By writing 
\begin{align}
\beta e^{-i\phi}=\frac{e\theta}{4\pi}-i\frac{b}{2}=\frac{e}{2}\left(\frac{\theta}{2\pi}-i\frac{b}{e} \right)~,
\end{align}
we find the standard source terms
\begin{align}
\mathcal L_{\rm int}=-F'_{\mu\nu}\left[e\Sigma_J^{\mu\nu}+\frac{e\theta}{2\pi}\Sigma_K^{\mu\nu}\right]
-b\dual F'_{\mu\nu}\Sigma_K^{\mu\nu}~.
\end{align}

Clearly, this parameterization is simply convention. A different rephasing makes $\beta$ purely imaginary and leads to a coupling between $\dual F_{\mu\nu}$ and the electric current while removing the coupling of $F_{\mu\nu}$ to the magnetic current. In general, if we write
\begin{align}
\alpha=|\alpha|e^{i\phi_\alpha}~, \ \ \ \ \beta=|\beta|e^{i\phi_\beta}~,
\end{align}
and rephase the helicity fields by
\begin{align}
F_{\pm\mu\nu}\to e^{\mp i\phi_F}F'_{\pm\mu\nu}~,
\end{align}
we find the source terms
\begin{align}
\mathcal L_{\rm int}={}&-2F'_{\mu\nu}\left[|\alpha|\cos(\phi_\alpha-\phi_F)\Sigma_J^{\mu\nu}+|\beta|\cos(\phi_\beta-\phi_F)\Sigma_K^{\mu\nu}\right]
\nonumber\\
&-2\dual F'_{\mu\nu}\left[|\alpha|\sin(\phi_F-\phi_\alpha)\Sigma_J^{\mu\nu}+|\beta|\sin(\phi_F-\phi_\beta)\Sigma_K^{\mu\nu}\right]~.
\label{e.GenSourceTerms}
\end{align}
One can always rephase the helicity fields so that either $\phi_F=\phi_\alpha$ or $\phi_F=\phi_\alpha-\pi/2$, but not both.  The same argument applies to $\phi_\beta$.  Thus, in the general case, a helicity rephasing can eliminate at most one term in Eq.~\eqref{e.GenSourceTerms}.

\section{Worldline Formalism}
\label{app:worldlineTopology}

In this section we show that every term in the dyon-loop expansion of gauge-invariant correlation functions is independent of $n^\mu$, which establishes the Lorentz invariance of the theory. 
The argument given here is the straightforward generalization of the one given by Brandt, Neri, and Zwanziger \cite{Brandt:1977be,Brandt:1978wc}.
This is easiest to show in the world-line formalism, which we outline in the next section.
Our approach is similar to other Field-strength and closed-worldline formulations of electric and magnetic charge theories~\cite{Halpern:1978ik,Calucci:1981we,Calucci:1982fm}.  

\subsection{Derivation of Worldline Expressions}

In order to maintain gauge invariance, we study correlation functions of Wilson-'t Hooft loop operators of the form 
\begin{align}
	W_{\gamma_i}[\bm{A}]
	= 
	e^{-i\oint_{\gamma_i} \bm{q}^T_i\bm{A}}
	= 
	e^{
		-i \oint
		d\tau
		dz^\mu_i(\tau)
		\bm{q}_i^T
		\bm{A}_\mu}(z_i)
	=
	e^{
		\frac{i}{2} \int d^4x
		\Sigma_{iL}^{\mu\nu}
		\bm{q}_i^T
		\bm{F}_{\bm{A}\mu\nu}}~,
\end{align}
where $\gamma_i$ is a closed path in spacetime parametrized by $z^{\mu}_i(\tau)$.\footnote{
	Here we follow the holomorphic kinetic-coupling convention from Sec. \ref{sec:sourceQuadraticMain}. We use the notation $\int_\gamma \bm{q}^T \bm{A}$ and $\int d\tau \dot{z}^{\mu} \bm{q}^T \bm{A}_\mu$ interchangeably.}

For loops we define the loop Dirac surfaces as
\begin{align}
	\Sigma^{\mu\nu}_L
	= -\int_S ds^{\mu\nu}\,\delta^{(4)}\!\left(x-s\right),
	\quad
	ds^{\mu\nu}
	= 
	\left(
	\frac{\partial s^{\mu}}{\partial u}
	\frac{\partial s^{\nu}}{\partial v}
	-
	\frac{\partial s^{\nu}}{\partial u}
	\frac{\partial s^{\mu}}{\partial v}
	\right)
	\,du\,dv~,
	\label{eq:loopSurf}
\end{align}
where $s^\mu(u,v)$ parametrizes a choice of surface $S$ with boundary $\partial S_i  = \gamma_i$, and with $(u,v)$ ordered so that the induced orientation on $\partial S$ matches the orientation of $\gamma_i$.
Integrating $\Sigma^{\mu\nu}_L$ against $\partial_\mu \phi_\nu$ for a smooth test one-form $\phi$ and using Stokes' theorem shows that $\partial_\mu \Sigma^{\mu\nu}_L = J_L^{\nu}$, where
\begin{align}
	J^{\mu}_L(x) = \int_0^T d\tau \dot{z}^\mu 
	\delta^4(x-z(\tau)\,)~.
\end{align}
We can then write the Wilson-t'Hooft operators as 
\begin{align}
	W_{\gamma_i}
	[\bm{A}]
	= \exp \left\{
	\frac{i}{2}
	\int d^4x 
	\bm{q}_i^T\bm{F}_{\bm{A}\mu\nu}(x)
	\Sigma_{iL}^{\mu\nu}(x)
	\right\}~,
\end{align}
where $\Sigma^{\mu\nu}_{iL}(x)$ is supported on some surface with boundary $\gamma_i$.

Dynamical fermions enter through the worldline representation of the fermion determinant \cite{Strassler:1992zr,Schubert:1996jj} 
\begin{align}
	&\text{tr} \ln\left[
	i \slashed{D}_\ell - m_\ell 
	\right]
	= 
	\frac{i}{2}
	\mathcal{N}
	\int_0^{\infty}
	\frac{d T}{T} e^{-i(m^2_\ell - i\epsilon)T}
	\frac{1}{(4\pi T)^2}
	\int_{z(0) = z(T)} 
	\mathcal{D}z 
	\int_{\psi(0) = -\psi(T)}
	\mathcal{D} \psi
	\notag\\ 
	&\times \exp\left\{-i\int_0^T d\tau
	\left[ 
	\frac{1}{4}
	\dot{z}^2
	- \frac{i}{2} \psi_\mu \dot{\psi}^\mu
	+ \dot{z}^\mu \bm{q}^T_\ell \bm{A}_\mu 
	+ i \psi^\mu \psi^\nu \bm{F}_{\bm{A} \mu\nu}
	\right] 
	\right\}+\text{const}~,\label{eq:funcDetWL}
\end{align}
where $\psi^\mu(\tau)$ are additional Grassmann variables on the worldline which give the spin degrees of freedom.

This motivates the definition of the Dirac surface for fermions as $\Sigma^{\mu\nu} = \Sigma_L^{\mu\nu} + \Sigma^{\mu\nu}_S$ where the spin Dirac surface $\Sigma_S^{\mu\nu}$ is defined by
\begin{align}
	\Sigma^{\mu\nu}_S
	= -2i\int_0^T d\tau \psi^\mu(\tau) \psi^\nu(\tau)
	\delta^4(x-z(\tau)\,)~,
\end{align}
which satisfies $\partial_\mu \Sigma_S^{\mu\nu} = J_S^{\nu}$ for the spin current
\begin{align}
	J^{\nu}_S(z)
	&= 2i\int_0^T d\tau \psi^\nu(\tau) \psi^\mu(\tau)
	\partial_\mu
	\delta^4(x-z(\tau)\,)~.
\end{align}
It should be noted that $\Sigma_S^{\mu\nu}$ is localized on the fermion worldline, whereas $\Sigma_L^{\mu\nu}$ is not; their antiperiodic boundary conditions implement the trace over spinor indices.
The path integral is normalized according to $\int\mathcal{D}z e^{-\frac{i}{4}\dot{z}^2} = \int d^4z_0$, and $\int \mathcal{D}\psi e^{-\frac{1}{2}\int  \psi \dot{\psi}} = \text{tr} (I) = 4$.

It is useful to define a compact notation for the loop Feynman integrals. Writing the loop fermion worldline measure as 
\begin{align}
	\oint d\Gamma_\ell =
	\frac{i}{2}
	\int_0^{\infty}
	\frac{d T}{T} e^{-i(m^2_\ell - i\epsilon)T}
	\frac{1}{(4\pi T)^2}
	\int  \mathcal{D}z 
	\int \mathcal{D} \psi 
	e^{-i\int d\tau
		\left[ 
		\frac{1}{4}
		\dot{z}^2
		- \frac{i}{2} \psi_\mu \dot{\psi}^\mu
		\right] }~,
\end{align}
we can write \eqref{eq:funcDetWL} as
\begin{align}
	\text{tr}  \ln
	\left[
	i\slashed{D}_\ell
	- m_\ell 
	\right]		
	=
	\oint d\Gamma_\ell
	e^{
		\frac{i}{2}\int d^4x \Sigma_\ell^{\mu\nu}
		\bm{q}^T_\ell
		\bm{F}_{\bm{A}\mu\nu}
	}+\text{const}~,
\end{align}
where $\slashed{D}_\ell = \gamma^\mu (\partial_\mu + i \bm{q}^T_\ell \bm{A})$, and $\Sigma_\ell = \Sigma_{\ell L} + \Sigma_{\ell S}$ and $\Sigma_{\ell L}$ is supported on a surface spanning the worldline loop.

The generating functional with external worldline currents $J_{i}$ (products of $W_{\gamma_i}$) can then be written as 
\begin{align}
	Z\left[\{J_{i}\}\right]
	&=\frac{1}{Z_0} \int \mathcal{D}\bm{A}
	e^{iS_{\text{gauge}}[\bm{A}]}
	\prod_\ell
	\mathbf{D} \overline{\psi}_\ell
	\mathcal{D} \psi_\ell 
	e^{iS_{\text{Dirac}}[\psi_\ell]}
	\notag\\
	&\times 
	\exp\left\{
	-i \int d^4x \overline{\psi}_\ell
	\gamma^\mu \psi_{\ell} \bm{q}_\ell^T \bm{A}_\mu(x) 
	- i \sum_i \int d^4x J^\mu_{i}
	\bm{q}_i^T \bm{A}(x) 
	\right\}~.
\end{align} 
Performing the fermion integrals, and expressing the external current couplings as Dirac surface couplings, we have 
\begin{align}
	Z =
	\frac{1}{Z_0}
	\int \mathcal{D}\bm A
	e^{i S_{\text{gauge}}[\bm A]} 
	\exp \left\{ \frac{i}{2}
	\sum_i \int d^4x \Sigma_{iL}^{\mu\nu}
	\bm{q}^T_i
	\bm{F}_{\bm{A}\mu\nu}
	\right\}
	\exp \left\{
	\sum_\ell
	\text{tr} \ln 
	\left[
	\frac{
		i\slashed{D}_\ell
		- m_\ell 	
	}{
		i\slashed{\partial} 
		- m_\ell 
	}	
	\right]
	\right\}~.
\end{align}
Expanding the second exponential gives the dyon loop expansion of the theory, which can be expressed in terms of Feynman world-line integrals using \eqref{eq:funcDetWL}.
For each term in this series, performing the gauge field path integrals couples together the $\Sigma^{\mu\nu}$'s by a photon propagator $\langle \bm{F}_{\bm{A} \mu\nu} \bm{F}_{\bm{A} \sigma\rho} \rangle$ within the exponential.

The one-loop corrections can be neatly summarized by the one-loop connected generating functional
\begin{align}
	&W_{1-\text{loop}}[\{J_{i}^\mu\}]
	=
	i\ln Z[\{J_{i}^\mu\}] \big|_{\text{1-loop}}
	=
	\sum_\ell
	\text{tr}
	\ln
	\left[
	\frac{
		i\gamma^\mu \partial_\mu 
		+
		\gamma^\mu
		\bm{q}_\ell^T
		\bm{A}_\mu^{\text{ext}}
		- m_\ell 
	}{
		i\gamma^\mu \partial_\mu 
		- m_\ell 
	}
	\right]
	\notag\\
	&= 
	\sum_\ell \oint d\Gamma_\ell 
	\left(
	\exp \left\{
	-\frac{1}{4}
	\int d^4x d^4y
	\sum_i 
	\Sigma^{\mu\nu}_\ell(x)
	\langle \bm{F}_{\bm{A} \mu\nu}(x) \bm{F}_{\bm{A} \sigma\rho}(y) \rangle
	\Sigma^{\sigma\rho}_{iL}(y)
	\right\}
	- 1
	\right)
	~,
\end{align}
where $\bm{A}^{\text{ext}}[J_i,J_j]$ is the tree-level potential created by the external currents of the scattered particles given by 
\begin{align}
	\bm{A}^{\text{ext}}_\mu(x)
	= 
	-i
	\sum_i 
	\int d^4y
	\langle \bm A_\mu(x) \bm A_\nu(y) \rangle \bm q_i J^\nu_{i}(y)~. 
\end{align}
$W$ can then be computed in the worldline formalism using Eq. \eqref{eq:funcDetWL}.
Because this is the one-loop result the $\ell-\ell$ self contractions are not included, these would contribute to the two-loop result summarized in Sec. \ref{sec:twoLoopDualityPrediction}.

Correlation functions of fermion operators can be treated similarly \cite{Fradkin:1991ci}.
Gauge invariance requires correlation functions be expressed  as $\psi^{\text{dressed}}_i (x) = e^{-i\int_{\gamma_x} \bm{q}_i^T \bm{A}} \psi_i(x) $ where $\gamma_x$ is a path extending from the point $x$ to infinity.
The paths $\gamma_x$ close the open worldlines and allow us to express the bosonic part of the coupling as $\int d^4x \Sigma^{\mu\nu}_L \bm{F}_{\bm{A} \mu\nu}$ for a loop Dirac surface $\Sigma_L$ which extends to infinity. 
Additional technical details are required to define the Grassmann integral for open lines which are beyond the scope of this appendix ( see \cite{Ahmadiniaz:2020wlm,Karanikas:1999hz,Fradkin:1991ci}).

\subsection{Topological Numbers}

We see that all correlation functions can be computed from worldline path integrals of various products of 
\begin{align}
	\exp \left\{
	- \frac{1}{4} 
	\int d^4x d^4y
	\Sigma_{i ,\alpha\beta}(x)
	\bm{q}_i^T
	\langle \mathbf{F}^{\alpha\beta}_{
		\boldsymbol{A}
	}(x)
	\mathbf{F}^{\gamma\delta}_{
		\boldsymbol{A}
	}(y) 
	\rangle
	\bm{q}_j
	\Sigma_{j,\gamma\delta}(y)
	\right\}~,
\end{align}
(for $i\neq j$; for $i = j$ the coefficient is $-\frac{1}{8}$).
In momentum space the argument of this exponential is the tree level amplitude computed in Sec. \ref{sec:treeDyonScattering}. The $n$-independence of this theory follows from the discussion given in section \ref{sec:treeDyonScattering} once we justify the claim that
\begin{align}
	I_1 = 
	-\frac{\Delta}{2}
	\bm{q}_i^T\epsmat \bm{q}_j
	\int \frac{d^4k}{(2\pi)^4}
	\Sigma^{\mu\nu}_i(k)
	{}^\star\Sigma_{j,\mu\nu}^*(k)~,	
\end{align}
and
\begin{align}
	I_2 = 
	-\frac{\Delta}{2}
	\bm{q}_i^T\epsmat \bm{q}_j
	\int \frac{d^4k}{(2\pi)^4}
	\Sigma_{i\alpha\beta}(k)
	\,
	{}_{(k)}^{\alpha\beta}
	\mathcal{P}_{(n)}^{\gamma\delta \star}
	\,
	\Sigma_{j\gamma\delta}^*(k)~,
	\label{eq:nTopTerm}
\end{align}
(when expressed in terms of particle paths) are equal to $2\pi $ times an integer. This contribute a trivial phase, $e^{2\pi i \mathbb{Z}} = 1$, to the worldline path integral. We now proceed to show this.

Moving to position space, from the definition of $\Sigma_{L}^{\mu\nu}$ we have
\begin{align}
	I_1 &= -\frac{\Delta}{2}
	\bm{q}_i^T \epsmat \bm{q}_j
	\int d^4x  
	\Sigma^{\mu\nu}_{iL}(x)
	{}^\star\Sigma_{jL,\mu\nu}(x)
	\notag\\
	&= \Delta
	\bm{q}_i^T \epsmat \bm{q}_j
	\sum_{p \in S_i \cap S_j}
	\text{sign}
	\det 
	\left[
	\partial_u s_i, \partial_v s_i, 
	\partial_{u'}s_j, \partial_{v'} s_j
	\right]
	= \Delta 
	\bm{q}_i^T \epsmat \bm{q}_j
	I(S_i,S_j)~,
\end{align}
where $S_i$ and $S_j$ are the surfaces that support $\Sigma_{iL}^{\mu\nu}$ and $\Sigma_{jL}^{\mu\nu}$, parametrized by $s_i(u,v)$ and $s_j(u',v')$ respectively,
where $I(S_i,S_j)$ is their oriented intersection number. This term counts the number of signed intersections of the two surfaces and so is referred to as an intersection number.

Adding in the spin terms we have 
\begin{align}
	\Sigma^{\mu\nu}_{iL}(x)
	{}^\star\Sigma_{jL,\mu\nu}(x)
	+\Sigma^{\mu\nu}_{iL}(x)
	{}^\star\Sigma_{jS,\mu\nu}(x)
	+
	\Sigma^{\mu\nu}_{iS}(x)
	{}^\star\Sigma_{jL,\mu\nu}(x)
	+
	\Sigma^{\mu\nu}_{iS}(x)
	{}^\star\Sigma_{jS,\mu\nu}(x)~.
\end{align}
Notice that because the spin terms are localized to the worldlines they only contribute when the 1D worldline for one particle intersects the Dirac surface of the other. 
In four spacetime dimensions, this only happens for a measure zero subset of paths in the path integral (as emphasized in \cite{Brandt:1978wc}).\footnote{Strictly speaking, the measure-zero argument applies to smooth paths, so we assume a UV regulator that smooths the worldlines, e.g.\ a cutoff on the Fourier modes of $z(\tau)$. 
	The typical paths of the unregulated worldline integral are Brownian, with Hausdorff dimension two. 
	This indicates that the surfaces for a Brownian loop have Hausdorff dimension three. 
	In four dimensions a Brownian path hits a set of dimension greater than two, such as a surface spanned by another Brownian loop, with nonzero probability~\cite{Morters:2010}. 
	Since the phases described here are equal to one at any finite cutoff, they cannot contribute in the limit where the regulator is removed.} Therefore, for paths that contribute nonzero measure to the worldline path integral we have
\begin{align}
	- \frac{\Delta}{2} 
	\bm{q}_i^T \epsmat \bm{q}_j
	\int d^4x  
	\Sigma^{\mu\nu}_i(x)
	{}^\star\Sigma_{j,\mu\nu}(x)
	\in 2\pi \mathbb{Z}~.
\end{align}
The contribution to the worldline path integral is then $e^{2\pi i \mathbb{Z}} = 1$.

We turn now to $I_2$, which is the $n$-dependent term in the DSZ part of the tree level amplitude. Using $\partial_\mu \Sigma^{\mu\nu} = J^\nu$ we have 
\begin{align}
	{}_{(n)}^{\mu\nu}
	\mathcal{P}_{(k)}^{\gamma\delta}
	\Sigma_{i,\gamma\delta}(k)
	&=
	\Sigma_{i,\text{cyl}}^{\mu\nu}
	\equiv
	-\frac{i}{n\cdot k}
	\left(
	n^\mu J^\nu_i(k)
	- n^\nu J^\mu_i(k)\,\right)~,
	\notag\\
	\Sigma^{\mu\nu}_{i,\text{cyl}}
	&= \frac{1}{n\cdot\partial}
	(n^\mu J_i^\nu - n^\nu J_i^\mu )(x)~.
\end{align}
We can write \eqref{eq:nTopTerm} as 
\begin{align}
	- \frac{\Delta}{2} 
	\bm{q}_i^T \epsmat \bm{q}_j
	\int \frac{d^4k}{(2\pi)^4}
	\Sigma_{i \text{cyl},\mu\nu}(k)
	{}^\star\Sigma_{j}^{*\mu\nu}
	(k) ~.
\end{align}
It should be noted that, going back to position space, $\Sigma_{\text{cyl}}$ is supported on the cylindrical surface
\begin{align}
	s^{\mu}_i(\tau,u)
	= z^{\mu}_i(\tau) + u\, n^\mu~,
\end{align}
where $u \in [0,\infty)$ for a one sided choice of $(n\cdot \partial)^{-1}$. We refer to this surface as the cylinder $C_i$ of particle $i$.

We now plug in the decomposition $\Sigma^{\mu\nu} = \Sigma^{\mu\nu}_L + \Sigma^{\mu\nu}_S$.
It should be noted that because $\Sigma_{S}^{\mu\nu}$ is localized to the worldline, the term proportional
\begin{align}
	\Sigma_{jS,\mu\nu}	
	{}^\star
	\Sigma^{\mu\nu}_{i,\text{cyl}}~,
\end{align}
only contributes for paths where the $j$ particle worldline intersects the cylinder $C_i$ of particle $i$.
As mentioned before, this only happens for a measure zero subset of paths in the worldline path integral. The other term is 
\begin{align}
	&- \frac{\Delta}{2} 
	\bm{q}_i^T \epsmat \bm{q}_j
	\int \frac{d^4k}{(2\pi)^4}
	\Sigma_{i\text{cyl}, \mu\nu}(k)
	{}^\star
	\Sigma_{jL}^{*\mu\nu}\notag\\
	&= 
	\frac{i\Delta}{2}
	\bm{q}_i^T \epsmat \bm{q}_j
	\int \frac{d^4k}{(2\pi)^4}
	\frac{1}
	{n\cdot k}
	\left( 
	n^\mu J_{i L}^\nu(k) 
	- n^\nu J_{i L}^\mu(k) 
	\right) 
	{}^\star
	\Sigma_{jL,\mu\nu}^*(k)\notag\\
	&+
	\frac{i\Delta}{2}
	\bm{q}_i^T \epsmat \bm{q}_j
	\int \frac{d^4k}{(2\pi)^4}
	\frac{1}
	{n\cdot k}
	\left( 
	n^\mu J_{iS}^\nu(k) 
	- n^\nu J_{iS}^\mu(k) 
	\right) 
	{}^\star
	\Sigma_{jL,\mu\nu}^*(k)~.
\end{align}
Notice that the first term contributes 
\begin{align}
	\Delta
	\bm{q}_i^T \epsmat \bm{q}_j
	I(C_i,S_j) \in 2\pi \mathbb{Z}~.
\end{align}

The remaining term is
\begin{align}
	\frac{i\Delta}{2}
	\bm{q}_i^T\epsmat
	\bm{q}_j
	\int 
	\frac{d^4k}{(2\pi)^4}
	\frac{1}
	{n\cdot k}
	\left( 
	n^\mu J_{iS}^\nu(k) 
	- n^\nu J_{iS}^\mu(k) 
	\right) 
	{}^\star
	\Sigma_{jL,\mu\nu}^*(k)~.
\end{align}
Using the following expression for $J_{iS}^\mu$ in momentum space  
\begin{align}
	J^\mu_{iS}(k)
	= 
	-2
	\int d\tau
	\psi^\mu_i(\tau) k_\nu 
	\psi^\nu_i(\tau) 
	e^{-ik\cdot z(\tau)} ~,
\end{align}
we see that the term in question is equal to
\begin{align}
	-i\Delta
	\bm{q}^T_i \epsmat
	\bm{q}_j 
	\int \frac{d^4k}{(2\pi)^4}
	\int d\tau 
	e^{-ik\cdot z(\tau)}	
	\frac{1}{n\cdot k}
	\epsilon^{\alpha\beta\gamma\delta}
	n_\alpha \psi_{i\beta}(\tau)
	k^\mu \psi_{i\mu}(\tau) 
	\Sigma_{jL,\gamma\delta}^*(k)~.
\end{align}
Using the four dimensional Schouten identity 
\begin{align}
	k^\mu\varepsilon^{\alpha\beta\gamma\delta}
	&=
	k^\alpha\varepsilon^{\mu\beta\gamma\delta}
	+k^\beta\varepsilon^{\alpha\mu\gamma\delta}
	+k^\gamma\varepsilon^{\alpha\beta\mu\delta}
	+k^\delta\varepsilon^{\alpha\beta\gamma\mu}~,
\end{align}
we have
\begin{align}
	k^\mu \varepsilon^{\alpha\beta\gamma\delta} n_\mu
	\psi_{i\alpha} \psi_{i\beta} \Sigma_{jL,\gamma\delta}^*
	&=
	k^\alpha\varepsilon^{\mu\beta\gamma\delta} n_\mu
	\psi_{i\alpha} \psi_{i\beta} \Sigma_{jL,\gamma\delta}^*
	+
	k^\beta\varepsilon^{\alpha\mu\gamma\delta} n_\mu
	\psi_{i\alpha} \psi_{i\beta} \Sigma_{jL,\gamma\delta}^*\notag\\
	&+
	k^\gamma\varepsilon^{\alpha\beta\mu\delta} n_\mu
	\psi_{i\alpha} \psi_{i\beta} \Sigma_{jL,\gamma\delta}^*
	+k^\delta\varepsilon^{\alpha\beta\gamma\mu} n_\mu
	\psi_{i\alpha} \psi_{i\beta} \Sigma_{jL,\gamma\delta}^*~.
\end{align}
Simplifying this gives 
\begin{align}
	(n\cdot k)
	\varepsilon^{\alpha\beta\gamma\delta}
	\psi_{i\alpha} \psi_{i\beta} \Sigma_{jL,\gamma\delta}^*(k)
	&=
	-2\varepsilon^{\mu\beta\gamma\delta} n_\mu \psi_{i\beta}
	(k\cdot \psi_i)
	\Sigma_{jL, \gamma\delta}^*(k) \notag\\
	&+2\psi_{i\alpha} \psi_{i\beta} 
	\varepsilon^{\alpha\beta\mu\delta} n_\mu k^\gamma
	\Sigma_{jL,\gamma\delta}^*(k)~.
\end{align}
Solving for the variables in the integrand we have 
\begin{align}
	&\varepsilon^{\alpha\beta\gamma\delta}
	n_\alpha\psi_{i\beta}
	(k\cdot\psi_i)
	\Sigma_{jL,\gamma\delta}^*\nonumber\\
	&\qquad
	=
	-\frac{1}{2}(n\cdot k)\,\varepsilon^{\alpha\beta\gamma\delta}
	\psi_{i\alpha}\psi_{i\beta}
	\Sigma_{jL,\gamma\delta}^*
	+n_\alpha k^\gamma\varepsilon^{\alpha\beta\mu\delta}
	\psi_{i\beta}\psi_{i\mu}
	\Sigma_{jL,\gamma\delta}^*\nonumber\\
	&\qquad
	=
	-\frac{1}{2}(n\cdot k)\,\varepsilon^{\alpha\beta\gamma\delta}
	\psi_{i\alpha}\psi_{i\beta}
	\Sigma_{jL,\gamma\delta}^*
	-in_\alpha\varepsilon^{\alpha\beta\mu\delta}
	\psi_{i\beta}\psi_{i\mu}
	J_{jL,\delta}^* ~,
\end{align}
where we used
\begin{align}
	-ik^\gamma \Sigma_{jL,\gamma\delta}^*(k)
	= J_{jL\delta}^*(k)~.
\end{align}
This means that 
\begin{align}
	&
	-i\Delta \bm{q}_i^T\epsmat\bm{q}_j
	\int 
	\frac{d^4k}{(2\pi)^4}
	\int d\tau\,
	e^{-ik\cdot z}
	\frac{1}{n\cdot k}
	\varepsilon^{\alpha\beta\gamma\delta} n_\alpha\psi_{i\beta} k^\mu\psi_{i\mu}
	\Sigma^{*}_{jL,\gamma\delta}(k) \nonumber\\
	&=
	+\frac{i\Delta}{2}
	\bm{q}_i^T\epsmat\bm{q}_j
	\int 
	\frac{d^4k}{(2\pi)^4}
	\int d\tau\,
	e^{-ik\cdot z}\varepsilon^{\alpha\beta\gamma\delta}
	\psi_{i\alpha}\psi_{i\beta}
	\Sigma^{*}_{jL,\gamma\delta}(k)\nonumber\\
	&\quad
	+\Delta\bm{q}_i^T\epsmat\bm{q}_j
	\int \frac{d^4k}{(2\pi)^4}
	\int d\tau\,
	e^{-ik\cdot z}\frac{n_\alpha}{n\cdot k}\varepsilon^{\alpha\beta\mu\delta}
	\psi_{i\beta}\psi_{i\mu}
	J_{jL\delta}^*(k)\notag\\
	&=
	+\frac{i\Delta}{2}
	\bm{q}_i^T\epsmat\bm{q}_j
	\int 
	\frac{d^4k}{(2\pi)^4}
	\int d\tau\,
	e^{-ik\cdot z}\varepsilon^{\alpha\beta\gamma\delta}
	\psi_{i\alpha}\psi_{i\beta}
	\Sigma^{*}_{jL,\gamma\delta}(k)\nonumber\\
	&\quad
	+\Delta
	\bm{q}_i^T\epsmat\bm{q}_j
	\int \frac{d^4k}{(2\pi)^4}
	\int d\tau\,
	e^{-ik\cdot z}\varepsilon^{\alpha\beta\mu\delta}
	\psi_{i\beta}\psi_{i\mu}
	\Sigma_{j,\text{cyl} L,\alpha\delta}^*~.
\end{align}
Both of these terms only contribute when the worldline of dyon $i$ intersects with a 2-dimensional surface: 
the first  only contributes when the $i$-th particle worldline intersects the $\Sigma_{jL}$ surface, and the second term only contributes when the $i$ particle worldline intersects the cylinder $C_j$.  As explained earlier these terms are only supported on a measure zero set in the path integral.

Therefore, for all paths which contribute a nonzero measure to the worldline path integral we have
\begin{align}
	-\frac{\Delta}
	{2}
	\bm{q}_i^T \epsmat \bm{q}_j
	\int 
	\frac{d^4k}
	{(2\pi)^4}
	\Sigma_{i,\alpha\beta}^* (k)\, 
	{}_{(k)}^{\alpha\beta}
	\mathcal{P}_{(n)}^{\gamma\delta\star }\,
	\Sigma_{j,\gamma\delta}^*(k)
	= 
	\Delta\bm{q}_i^T \epsmat \bm{q}_j
	I(C_i,S_j)
	\in 2\pi \mathbb{Z}~.
\end{align}
The contribution to the worldline path integral is then $e^{2\pi i \mathbb{Z}} = 1$.

\subsection{Vacuum Polarization Surface Correlators} \label{app:SurfCorr}

The vacuum polarization can be computed from the quadratic term in $W_{1-\text{loop}}$.
Specifically, we find that (working with Dirac surfaces) the loop corrections from a single dyon are
\begin{align}
	\frac{\delta^2}
	{\delta \Sigma_{i}^{\mu\nu}(x)
		\delta \Sigma_{j}^{\sigma\rho}(y)}
	W_{1-\text{loop}}
	&=  
	- \frac{i}{16}
	\int d^4u\, d^4v\,
	\bm{q}^T_i
	\langle \bm{F}_{\bm A\mu\nu}(x)
	\bm{F}_{\bm A \alpha\beta}(u)
	\rangle
	\bm{q}_\ell 
	\langle
	\Sigma^{\alpha\beta}_\ell(u)
	\Sigma^{\gamma\delta}_\ell(v)
	\rangle_{wl}
	\notag\\
	&\times\bm{q}_\ell^T 
	\langle \bm{F}_{\bm A \gamma\delta}(v)
	\bm{F}_{\bm A \sigma\rho}(y)
	\rangle \bm{q}_j~, \quad i \neq j
\end{align}
where the worldline surface-surface correlator is defined by 
\begin{align}
	\langle
	\Sigma^{\alpha\beta}_\ell(x)
	\Sigma^{\gamma\delta}_\ell(y)
	\rangle_{wl}
	= \oint d\Gamma_\ell 
	\,
	\Sigma^{\alpha\beta}_\ell(x)
	\Sigma^{\gamma\delta}_\ell(y)~.
	\label{eq:WLsurfCorr}
\end{align} 

We now proceed to identify the ultraviolet-divergent part of this surface-surface correlator using a Lorentz invariant prescription.   
We work in Euclidean signature in the intermediate steps, where  
\begin{align}
	\oint d\Gamma_\ell 
	= - \frac{1}{2}
	\int_0^\infty
	\frac{dT}{T}
	e^{-m_\ell^2 T}
	\frac{1}{(4\pi T)^2}
	\int \mathcal{D}z 
	\mathcal{D}\psi 
	e^{-\int_0^T d\tau \left(
		\frac{1}{4}\dot{z}^2
		+ \frac{1}{2} \psi \cdot \dot{\psi}
		\right)}~.
\end{align}
Let a closed worldline be parametrized by $z^\mu(\tau)$, with $0\leq \tau\leq T$ and $z(T)=z(0)$.  The calculation is also be easiest in momentum space i.e.~with the Fourier transform of the correlator in \eqref{eq:WLsurfCorr}.

As before, we can decompose $\Sigma^{\mu\nu}_\ell = \Sigma_{\ell L}^{\mu\nu}+\Sigma^{\mu\nu}_{\ell S}$, and we remember that $\Sigma^{\mu\nu}_\ell$ is constructed to satisfy
\begin{align}
	\partial_\mu \Sigma^{\mu\nu}_\ell
	= J^\nu_\ell~.
	\label{eq:SigmaWardId}
\end{align} 
For definiteness, an $n$-independent spanning surface can be obtained by contracting the loop to its average position,
\begin{align}
	z_0^\mu&=\frac{1}{T}\int_0^T d\tau\,z^\mu(\tau)~,
	&
	q^\mu(\tau)&=z^\mu(\tau)-z_0^\mu~,
	\notag\\
	s^\mu(\tau,\eta)&=z_0^\mu+\eta q^\mu(\tau)~,
	&0\leq\eta&\leq1~.
	\label{eq:coneSurfaceParam}
\end{align}
With the orientation fixed by $\partial_\mu \Sigma^{\mu\nu} = J^\nu$, its
Fourier-space loop surface, using \eqref{eq:loopSurf}, is
\begin{align}
	\Sigma_{\ell L}^{\mu\nu}(p)
	=
	-\int_0^T d\tau\int_0^1 d\eta\,\eta\,
	\left(
	q^\mu \dot{q}^\nu 
	- q^\nu \dot{q}^\mu
	\right)
	e^{-ip\cdot[z_0+\eta q(\tau)]}~.\label{eq:coneSurfaceFourier}
\end{align}
No fixed vector $n^\mu$ enters this construction, and because of our choice of $z_0$ as the zero-mode this construction is translation covariant.

Define the momentum space surface correlator as
\begin{align}
	G_{\Sigma}^{\alpha\beta,\gamma\delta}(p)
	\equiv
	\int d^4x 
	e^{-ip\cdot x}
	\left\langle
	\Sigma^{\alpha\beta}(x)
	\Sigma^{\gamma\delta}(0)
	\right\rangle_{wl}~.\label{eq:spinorSurfaceCorrelatorDef}
\end{align}
The identity \eqref{eq:SigmaWardId} implies
\begin{align}
	p_\alpha p_\gamma
	G_{\Sigma}^{\alpha\beta,\gamma\delta}(p)
	=\left(p^2\delta^{\beta\delta}-p^\beta p^\delta\right)
	\Pi_{\rm D}(p^2)~,
	\label{eq:spinorSurfaceWardIdentity}
\end{align}
where
\begin{align}
	\Pi_{\rm D}(p^2)
	&=\frac{8}{(4\pi)^2}
	\int_0^\infty\frac{dT}{T}\,e^{-m^2_\ell T}
	\int_0^1du\,u(1-u)e^{-Tu(1-u)p^2}~.\label{eq:spinorWorldlineVacPol}
\end{align}
The right hand side of Eq.~\eqref{eq:spinorSurfaceWardIdentity} is the usual Dirac vacuum polarization and is manifestly transverse.

We now show that \eqref{eq:spinorSurfaceWardIdentity} together with Euclidean invariance determine the complete logarithmically divergent part of the Dirac-surface correlator.  
Separate the translational zero mode and rescale the nonzero modes according to
\begin{align}
	\tau&=Tt~,
	&q^\mu(\tau)&=\sqrt T\,\xi^\mu(t)~,
	&0\leq t\leq1~.
	\label{eq:worldlineSmallTRescaling}
\end{align}
The free bosonic and fermionic worldline actions are then independent of $T$. For the cone surface in Eq.~\eqref{eq:coneSurfaceParam}, the loop and spin parts of the Dirac surface become
\begin{align}
	\Sigma_L^{\mu\nu}(p)
	&=-T e^{-ip\cdot z_0}
	\int_0^1dt\int_0^1d\eta\,\eta\,
	\left(
	\xi^\mu \xi'{}^\nu 
	- \xi^\nu \xi'{}^\mu
	\right)
	e^{-i\sqrt T\,\eta p\cdot\xi(t)}~,\notag\\
	\Sigma_{S}^{\mu\nu}(p)
	&=2T e^{-ip\cdot z_0}
	\int_0^1dt\,
	\psi^\mu(t)\psi^\nu(t)
	e^{-i\sqrt T\,p\cdot\xi(t)}~.\label{eq:worldlineTwoCurrentScaling}
\end{align}
Consequently the insertion of $\Sigma^{\mu\nu}$ has a regular small-$T$ expansion
\begin{align}
	\Sigma_\ell^{\mu\nu}(p)
	=T e^{-ip\cdot z_0}\left[
	\Xi_0^{\mu\nu}
	+\sqrt T\,p_\rho\Xi_1^{\mu\nu;\rho}
	+T p_\rho p_\sigma\Xi_2^{\mu\nu;\rho\sigma}
	+\mathcal O\!\left(T^{3/2}p^3\right)
	\right],
	\label{eq:worldlineTwoCurrentExpansion}
\end{align}
where the $\Xi_r$ are dimensionless functionals of the unit worldline.  
The two-point function is even under $p\to-p$, so the terms odd in $p$ vanish after the worldline average.  
The two insertions therefore scale as
\begin{align}
	\Sigma_{\ell}(p)
	\Sigma_{\ell}(-p)
	=T^2\left[A_0+Tp^2A_1+\mathcal O(T^2p^4)\right]~, \label{eq:worldlineCorrelatorScaling}
\end{align}
where the spacetime indices are suppressed to emphasize the scaling behavior.

In the worldline loop path integral measure, the $ \oint d \Gamma$ term contributes $ \frac{dT}{T} (4\pi T)^{-2} e^{-m_\ell^2 T}$. Thus, the small-$T$ behavior of the surface correlator is
\begin{align}
	-\frac{1}{2}\frac{dT}{T}(4\pi T)^{-2}
	e^{-m_\ell^2 T}
	\Sigma_{\ell}(p)
	\Sigma_{\ell}(-p)
	=
	-\frac{1}{2}
	\frac{dT}{T}
	e^{-m_\ell^2 T}
	\left[
	\frac{A_0}{(4\pi)^2}
	+T p^2\frac{A_1}{(4\pi)^2}
	+\mathcal O(T^2p^4)
	\right]~.\label{eq:worldlineUVPowerCounting}
\end{align}
Only the momentum-independent coefficient $A_0$ can therefore produce a logarithmic ultraviolet divergence.  Every term with explicit external momentum carries at least one additional power of $T$ and is ultraviolet finite.

The coefficient $A_0$ is also independent of the choice of any smooth local spanning surface that shrinks with the loop. 
At $p=0$, two surfaces $S$ and $S'$ with the same boundary differ by a closed surface, and
\begin{align}
	\Sigma_L^{\mu\nu}[S](0)-\Sigma_{L'}^{\mu\nu}[S'](0)
	=-\int_{S-S'}ds^{\mu\nu}=0~.\label{eq:surfaceIndependenceAtZeroMomentum}
\end{align}
The last equality follows because the coordinate two-form $dx^\mu\wedge dx^\nu=d(x^\mu dx^\nu)$ is exact in $\mathbb R^4$, so its integral over a closed surface vanishes.  
Dependence on the detailed spanning surface begins with the expansion of $e^{-ip\cdot s}$ and hence is accompanied by positive powers of $\sqrt T$; by Eq.~\eqref{eq:worldlineUVPowerCounting}, it cannot affect the logarithmic counterterm.  
This statement applies to local spanning surfaces such as the explicit parametrization described above. 
The infinitely extended fixed-string cylinders used in the fixed-string Zwanziger construction do not shrink with the loop and thus include an unphysical $n$-dependent pole \cite{Terning:2020dzg}.

It remains to classify the momentum-independent local tensor.  
There are only two Euclidean-invariant momentum independent tensors with the required antisymmetries, and symmetry under interchange of two index pairs.
They are
\begin{align}
	{}^{\alpha\beta}\mathbb I^{\gamma\delta}
	&=\frac12\left(
	\delta^{\alpha\gamma}\delta^{\beta\delta}
	-\delta^{\alpha\delta}\delta^{\beta\gamma}
	\right),
	\text{  and  }
	\varepsilon^{\alpha\beta\gamma\delta}~.\label{eq:twoFormIdentityAppendix}
\end{align}
Imposing a proper time cutoff $T \geq \Lambda^{-2}$ most general logarithmically divergent local term is therefore
\begin{align}
	G_{\Sigma}^{\alpha\beta,\gamma\delta}(p)
	\Big|_{\log} = 
	\ln \frac{\Lambda^2}{m_\ell^2}
	\left[
	a{}^{\alpha\beta}\mathbb I^{\gamma\delta}
	+
	b\varepsilon^{\alpha\beta\gamma\delta}
	\right]~.\label{eq:surfaceGeneralLocalLog}
\end{align}
Since
\begin{align}
	p_\alpha p_\gamma
	\mathbb I^{\alpha\beta,\gamma\delta}
	&=
	\frac{1}{2}
	\left(p^2\delta^{\beta\delta}-p^\beta p^\delta\right)~,
	\quad
	p_\alpha p_\gamma
	\varepsilon^{\alpha\beta\gamma\delta}=0~,\label{eq:surfaceWardTensorContractions}
\end{align}
Eq.~\eqref{eq:spinorSurfaceWardIdentity} fixes
\begin{align}
	a \ln \frac{\Lambda^2}{m_\ell^2}
	= 2 \left.\Pi_{\rm D}(p^2)\right|_{\log}~.\label{eq:surfaceWardFixesIdentityCoefficient}
\end{align}
The $\varepsilon^{\alpha\beta\gamma\delta}$ is parity odd and would multiply the local density $F_{\mu\nu}\dual F^{\mu\nu}$.  The determinant evaluated here is that of a
vectorlike Dirac fermion with real mass and contains no pseudoscalar insertion, so parity invariance gives $b=0$. This statement concerns the Lorentz tensor of the fermion determinant.  It does not forbid the off-diagonal running of the coupling-space parameter $\theta$ in a dyonic theory, which arises when this parity-even kernel is contracted with electric and magnetic coupling vectors. A complex mass phase or a chiral coupling would require a separate anomalous analysis and is not included here.

Combining ultraviolet power counting, Euclidean invariance, parity, and the identity \eqref{eq:SigmaWardId} therefore gives the precise result
\begin{align}
	\left.
	G_{\Sigma}^{\alpha\beta,\gamma\delta}(p)
	\right|_{\rm UV\,div}
	=
	2\left.\Pi_{\rm D}(p^2)\right|_{\rm UV\,div}
	{}^{\alpha\beta}\mathbb I^{\gamma\delta}~.\label{eq:surfacePhysicalDirac}
\end{align}
The finite correlator may also contain momentum- or surface-dependent terms not determined by these constraints.  
Equation~\eqref{eq:surfacePhysicalDirac} determines the complete ultraviolet-divergent local part, which is all that enters the $\beta$-functions.

For a massive Dirac fermion, we may set the external momentum to zero without introducing an infrared singularity. With a proper-time cutoff $T\geq\Lambda^{-2}$, Eq.~\eqref{eq:spinorWorldlineVacPol} then gives
\begin{align}
	\Pi_{\rm D}(0)\big|_{\log}
	&=\frac{8}{(4\pi)^2}
	\left[\int_0^1du\,u(1-u)\right]
	\ln\frac{\Lambda^2}{m^2_\ell}
	=\frac{4}{3(4\pi)^2}
	\ln\frac{\Lambda^2}{m^2_\ell}~.\label{eq:diracVacPolLog}
\end{align}
Consequently,
\begin{align}
	G_{\Sigma}^{\alpha\beta,\gamma\delta}
	\big|_{\log}
	=\frac{8}{3(4\pi)^2}
	\ln\frac{\Lambda^2}{m^2_\ell}\,
	{}^{\alpha\beta}\mathbb I^{\gamma\delta}
	=
	\frac{1}{6\pi^2}
	\ln\frac{\Lambda^2}{m^2_\ell}\,
	{}^{\alpha\beta}\mathbb I^{\gamma\delta}~.\label{eq:surfaceDiracLogFinal}
\end{align}
After subtraction, the renormalized zero-momentum function can be written as
\begin{align}
	\Pi_{\rm D}(0;\mu)
	&=\frac{1}{12\pi^2}\ln\frac{\mu^2}{m^2_\ell}
	+\text{finite}~.\label{eq:diracVacPolRenormalized}
\end{align}
Therefore,
\begin{align}
	\mu\frac{\partial}{\partial\mu}\Pi_{\rm D}(0;\mu)
	&=\frac{2}{12\pi^2}
	=\frac{1}{6\pi^2}~.\label{eq:diracVacPolDerivativeAppendix}
\end{align}

The zero-momentum evaluation is only a convenient infrared-safe way to read the ultraviolet coefficient when $m\neq0$. For a massless fermion, $\Pi_{\rm D}(0)$ is infrared singular; 
keeping nonzero Euclidean $p^2$ or extracting the coefficient directly from the ultraviolet logarithm gives the same derivative.  This is exactly the normalization of $\Pi_\ell^{\rm ren}$ used in Eq.~\eqref{eq:diracVacPolDerivativeMain}. Thus the explicit spinor worldline calculation gives both the Dirac coefficient used in the $\beta$-functions and the ultraviolet tensor prescription
\begin{align}
	\left.
	{}^{\alpha\beta}\Pi_{\mathrm{loc}}^{\gamma\delta}(p)
	\right|_{\rm UV\,div}
	= 2
	\left.\Pi_{\rm D}(p^2)\right|_{\rm UV\,div}
	{}^{\alpha\beta}\mathbb I^{\gamma\delta}~.\label{eq:worldlinePhysicalPrescription}
\end{align}
Because the worldline path integral measure is Gaussian this calculation can also be verified by a direct calculation of the Feynman integral in \eqref{eq:WLsurfCorr}.
The direct calculation is not included here to save space.

\section{Projector-Linear Two-Potential Theories}\label{app:Zwan}
In this appendix we describe a class of two-potential theories which describe a single photon. 
We assume the $n$-dependence enters only through projection operators ${}_{(n)}^{\alpha\beta}\mathcal{P}{}_{(n)}^{\gamma\delta}$ defined below. As these operators satisfy
\beq
{}_{(n)}^{\alpha\beta}\mathcal{P}{}_{(n)}^{\mu\nu}\,{}^{(n)}_{\mu\nu}\mathcal{P}{}_{(n)}^{\gamma\delta}  = {}_{(n)}^{\alpha\beta}\mathcal{P}{}_{(n)}^{\gamma\delta}~,
\eeq
we need only consider linear terms.
Given this, we refer to these theories as projector-linear. Zwanziger's formalism is one such theory. The generalization of Zwanziger that gives rise to the CP violating Maxwell equations is also a theory of this type. 
We first define this class of theories precisely.
Then, we demonstrate that they are single-photon theories so long as they satisfy a constraint. 
Finally, we determine the form of the gauge field propagators. 

\subsection{Two-Potential Kinetic Terms}
Consider a theory with two vector potentials $\bm{A}_\mu = (A^\mu,B^\mu)^T$. 
We can then define the derivatives
\begin{align}
\bm{F}_{\bm{A}}
&=
\partial_\mu \bm{A}_\nu-\partial_\nu \bm{A}_\mu
~.
\end{align}
The Lagrangian for such a theory is expected to contain a kinetic terms for the potentials that is quadratic in the field strengths. That is, we expect a term with the schematic form
\begin{align}
	\mathcal{L}_2=-\frac14 \bm{F}_{\bm{A}\mu\nu}^T
	\,{}^{\mu\nu}\mathbf{K}^{\rho\sigma}\, 
	\bm{F}_{ \bm{A}\rho\sigma}~.
\end{align}
The kinetic matrix, which we write in components as ${}^{\mu\nu}\!K_{ab}^{\rho\sigma}$, is a constant matrix constructed from the metric, the Levi-Civita tensor, and $n^\mu$. 
Note that $n^\mu$ is a constant space-like vector whose purpose is to ensure (by enabling a constraint to be defined) that the two vector potentials describe a single photon. 

In order to do so we employ a family of operators that map two-forms to two-forms. They are defined by
\begin{align}
 {}^{\mu\nu}_{(k)}\!\mathcal{P}_{(n)}^{\rho\sigma}=\frac{1}{2n\cdot k}\left(\eta^{\mu\rho}k^\nu n^\sigma-\eta^{\nu\rho}k^\mu n^\sigma-\eta^{\mu\sigma}k^\nu n^\rho+\eta^{\nu\sigma}k^\mu n^\rho \right)
\end{align}
in terms of two vectors, such as $k_\mu$ and $n_\mu$. These operators satisfy projector-like relations
\begin{align}
 {}^{\mu\nu}_{(k)}\!\mathcal{P}_{(n)}^{\alpha\beta}\; {}_{\alpha\beta}^{(k)}\mathcal{P}_{(n)}^{\rho\sigma}={}^{\mu\nu}_{(k)}\!\mathcal{P}_{(n)}^{\rho\sigma}~,
\label{eq:appendixProjectorDefinition}
\end{align}
even though they are only true projectors when the same four vector is associated with both pairs of indices.

A star on the left indices indicates a Hodge dual acting on those two indices, and similar for the right two indices. These lead to the identities
\begin{align}
 {}^{\star\mu\nu}_{(k)}\! \mathcal{P}_{(n)}^{\rho\sigma\star}&= {}^{\mu\nu}_{(n)}\!\mathcal{P}_{(k)}^{\rho\sigma}-{}^{\mu\nu}\mathbb{I}^{\rho\sigma}~,&
 {}^{\star\mu\nu}_{(k)}\! \mathcal{P}_{(n)}^{\rho\sigma}+ {}^{\mu\nu}_{(n)}\!\mathcal{P}_{(k)}^{\rho\sigma\star}&={}^{\star\mu\nu}\mathbb{I}^{\rho\sigma}={}^{\mu\nu}\mathbb{I}^{\rho\sigma\star}=\frac12\varepsilon^{\mu\nu\rho\sigma}~.
\label{eq:singleDualProjectorReduction}
\end{align}
The second relation follows by applying a right dual to the first and using ${}^\star{}^\star=-\mathbb{I}$ on Lorentzian two-forms. 

With this in mind we can expand ${}^{\mu\nu}\!K_{ab}^{\rho\sigma}$ in the redundant basis
\begin{align}
{}^{\mu\nu}\mathbb{I}^{\rho\sigma}~,
\qquad
 {}^{\mu\nu}_{(n)}\!\mathcal{P}_{(n)}^{\rho\sigma}~,
\qquad
 {}^{\star\mu\nu}_{(n)}\! \mathcal{P}_{(n)}^{\rho\sigma}~,
\qquad
 {}^{\mu\nu}_{(n)}\!\mathcal{P}_{(n)}^{\rho\sigma\star}~,
\qquad
 {}^{\star\mu\nu}_{(n)}\! \mathcal{P}_{(n)}^{\rho\sigma\star}~.
\label{eq:redundantZwanzigerKernelBasis}
\end{align}
Transposition of the left and right pairs of indices exchanges $ {}^{\star\mu\nu}_{(n)}\!\mathcal{P}_{(n)}^{\rho\sigma}$ and $ {}^{\mu\nu}_{(n)}\!\mathcal{P}_{(n)}^{\rho\sigma\star}$, so the identities in Eq.~\eqref{eq:singleDualProjectorReduction} can be used to determine that
\begin{align}
	{}^{\mu\nu}\!\mathbf{K}_{ab}^{\rho\sigma}=&\,
	\mathbf{X}\,\mathbb {}^{\mu\nu}\mathbb{I}^{\rho\sigma}
	+\mathbf{H}\,\mathcal  {}^{\mu\nu}_{(n)}\!\mathcal{P}_{(n)}^{\rho\sigma}
	+\mathbf{Y}\, {}^{\star\mu\nu}_{(n)}\! \mathcal{P}_{(n)}^{\rho\sigma}~,
\label{eq:redundantZwanzigerQuadraticAction}
\end{align}
where $\mathbf{X}$ and $\mathbf{H}$ may be taken symmetric.
The last term can be separated into symmetric $Y_{(ab)}$ and antisymmetric $Y_{[ab]}$ parts. Then, by using the right identity in Eq.~\eqref{eq:singleDualProjectorReduction}, we find
\begin{align}
	F^a_{\mu\nu} Y_{ab}\, 
	{}^{\star\mu\nu}_{(n)}\! \mathcal{P}_{(n)}^{\rho\sigma} F^{a}_{\rho\sigma}
	={}&
	F^a_{\mu\nu} 
	Y_{[ab]}\, 
	{}^{\star\mu\nu}_{(n)}\! \mathcal{P}_{(n)}^{\rho\sigma}  F^b_{\rho\sigma}
	+\frac12 
	Y_{(ab)}
	F^a_{\mu\nu}{}\dual 
	F^{b\mu\nu}~.
	\label{eq:dualCoefficientDecomposition}
\end{align}

By construction the field strengths satisfy the Bianchi identity so 
\begin{align}
	F_{\mu\nu}^a\dual 
	F^{b\mu\nu}
	&=
	\partial_\mu\left(
	\varepsilon^{\mu\nu\rho\sigma}
	A^a_\nu F^b_{\rho\sigma}
	\right)
	\label{eq:topologicalDensityTotalDerivative}
\end{align}
is a total derivative that can be dropped from the Lagrangian. Any two-by-two antisymmetric matrix has only one component,
\begin{align}
	Y_{[ab]}=\frac{\kappa}{2}\,\epsilon_{ab}~,
\end{align}
so this part of the kinetic term is 
\begin{align}
	\frac{\kappa}{4}\bm{F}_{\bm{A} \mu\nu}^T\,
	\epsmat \,\,
	{}^{\star\mu\nu}_{(n)}\! \mathcal{P}_{(n)}^{\rho\sigma} 
	\,
	\bm{F}_{\bm{A} \rho\sigma}~,
\end{align}
where we have combined terms by again using Eq.~\eqref{eq:singleDualProjectorReduction} and dropping a total derivative.
As we show in the next two sections, including a nonzero $ \mathbf{X}$ term in Eq.~\eqref{eq:redundantZwanzigerQuadraticAction} leads to a theory that violates Lorentz invariance even in a single charge sector.

Therefore, taking $\mathbf{X} = 0$, we consider kinetic terms of the form
\begin{align}
	\mathcal{L}_2=-\frac14 
	\bm{F}_{\bm{A} \mu\nu}^T 
	\mathbf{H}
	\,\,
	{}_{(n)}^{\mu\nu}\!
	\mathcal{P} {}_{(n)}^{\sigma \rho}
	\bm{F}_{\bm{A} \sigma\rho}
	+
	\frac{\kappa}{4}\bm{F}_{\bm{A} \mu\nu}^T\,
	\epsmat \,\,
	{}^{\star\mu\nu}_{(n)}\! \mathcal{P}_{(n)}^{\rho\sigma} 
	\,
	\bm{F}_{\bm{A} \rho\sigma} 
	~.
	\label{eq:LinProjLag}
\end{align}
 This class contains the standard Zwanziger action~\cite{Zwanziger:1970hk} and is closed under constant GL$(2,\mathbb R)$ redefinitions of the two potentials.

 We then proceed to add a current coupling and solve for the equations of motion.
 We can get the equations of motion by enforcing the Bianchi identity for $\mathbf{F}_{\mathbf{A}}^{\mu\nu}$ with a lagrange multiplier field $\boldsymbol{\Lambda} \in \mathbb{R}^2 \otimes \Omega^1(\mathbb{R}^{1,3})$ and then varying the Lagrangian with respect to $\mathbf{F}_{\mathbf{A}}^{\mu\nu}$.
 In order to do vary with respect to $\mathbf{F}_{\mathbf{A}}^{\mu\nu}$ we need to write things in terms of
 \begin{align}
 	\boldsymbol{\Sigma}^{\mu\nu}
 	= 
 	\begin{pmatrix}
 		\Sigma_J^{\mu\nu} \\
 		\Sigma_{K}^{\mu\nu}
 	\end{pmatrix}
 \end{align}
 which satisfies
 \begin{align}
 	\partial_\mu \boldsymbol{\Sigma}^{\mu\nu}
 	= \bm{J}^\nu.
 \end{align}
 Using this the Lagrange multiplier Lagrangian is 
 \begin{align}
 	\mathcal{L}_{Z}
 	&= 
 	-\frac{1}{4}
 	\mathbf{F}_{\mathbf{A}}^T{}_{\alpha\beta}
 	\,
 	{}^{\alpha\beta}
 	\boldsymbol{K}^{\gamma\delta}
 	\mathbf{F}_{\mathbf{A}}{}_{\gamma\delta}
 	+
 	\frac{1}{2}
 	\mathbf{F}_\mathbf{A}^T{}_{\mu\nu} 
 	\mathbf{S}
 	\boldsymbol{\Sigma}^{\mu\nu}
 	+ 
 	\varepsilon^{\mu\nu\alpha\beta}
 	\boldsymbol{\Lambda}_\mu^T 
 	\partial_\nu
 	\mathbf{F}_{\mathbf{A} \alpha \beta},
 \end{align}
 this gives equations of motion
 \begin{align}
 	- \frac{1}{2}
 	{}^{\mu\nu}
 	\mathbf{K}^{\alpha\beta}
 	\mathbf{F}_{\mathbf{A} \alpha\beta}
 	= - \frac{1}{2}
 	\mathbf{S}
 	\boldsymbol{\Sigma}^{\mu\nu}
 	+ {}^\star
 	\mathbf{F}_{\mathbf{\Lambda}}^{\mu\nu},
 \end{align}
 taking the divergence of this equation removes the Lagrange multiplier and leaves us with the equations of motion in terms of currents given by Eq. \eqref{eq:GenEOMZwan}.

\subsection{Single-Photon Condition}

We now consider a general kinetic operator of the form
\begin{align}
	{}^{\mu\nu}\mathbf{K}^{\rho\sigma}
	=
	\mathbf{X}\,{}^{\mu\nu}\mathbb{I}^{\rho\sigma}
	+\mathbf{H}\,{}^{\mu\nu}_{(n)}\!\mathcal{P}_{(n)}^{\rho\sigma}
	-\kappa\,\epsmat\,{}^{\star\mu\nu}_{(n)}\!\mathcal{P}_{(n)}^{\rho\sigma}~,
	\label{eq:SPKgen}
\end{align}
where $\mathbf{X}$ and $\mathbf{H}$ are real symmetric $2\times2$ matrices.
The Lagrangian depends on the potentials only through their exterior derivatives, i.e.\ through $\bm{F}_{\bm{A}}^{\mu\nu}$.
We therefore need to understand how $\mathbf{K}$ acts on exact two-forms.
In momentum space these are the two-forms
\begin{align}
	(k\wedge a)^{\mu\nu}
	=
	k^\mu a^\nu-k^\nu a^\mu~,
\end{align}
which make up the three-dimensional range of ${}^{\mu\nu}_{(k)}\!\mathcal{P}_{(k)}^{\rho\sigma}$.

Writing the Lagrangian as
\begin{align}
	\mathcal{L}_2
	=
	-\frac{1}{2}\,
	\bm{F}_{\bm{A}\mu\nu}^T
	\left(\frac{1}{2}\,{}^{\mu\nu}\mathbf{K}^{\rho\sigma}\right)
	\bm{F}_{\bm{A}\rho\sigma}~,
\end{align}
the propagator is
\begin{align}
	\langle
	\bm{F}_{\bm{A}}^{\mu\nu}
	\bm{F}_{\bm{A}}^{\rho\sigma}
	\rangle
	=
	-2i\,
	\left(\mathbf{K}_{\text{exact}}^{-1}\right)^{\mu\nu\rho\sigma}~,
	\qquad
	{}^{\mu\nu}\mathbf{K}_{\text{exact}}^{\rho\sigma}
	=
	{}^{\mu\nu}_{(k)}\!\mathcal{P}_{(k)}^{\alpha\beta}\;
	{}_{\alpha\beta}\mathbf{K}_{\gamma\delta}\;
	{}^{\gamma\delta}_{(k)}\!\mathcal{P}_{(k)}^{\rho\sigma}
	\Big|_{{}_{(k)}\mathcal{P}_{(k)}}~,
\end{align}
where ${}^{\mu\nu}\mathbf{K}_{\text{exact}}^{\rho\sigma}$ is the kinetic operator with domain and range restricted to the range of ${}_{(k)}^{\alpha\beta}\mathcal{P}_{(k)}^{\gamma\delta}$. In the next section we perform this inversion directly. In this section we find the propagating degrees of freedom by block-diagonalizing ${}^{\mu\nu}\mathbf{K}_{\text{exact}}^{\rho\sigma}$.

We split the three-dimensional range of ${}_{(k)}^{\alpha\beta}\mathcal{P}_{(k)}^{\gamma\delta}$ into a one-dimensional and a two-dimensional space.
First we define the rank-one projector
\begin{align}
	{}^{\mu\nu}\Pi_1^{\rho\sigma}
	=
	\frac{1}{2\left(n^2k^2-(n\cdot k)^2\right)}
	\left(n^\mu k^\nu-n^\nu k^\mu\right)
	\left(n^\rho k^\sigma-n^\sigma k^\rho\right),
\end{align}
which projects onto the one dimensional space spanned by $(k\wedge n)$.
We also define ${}^{\mu\nu}\Pi_2^{\rho\sigma}={}^{\mu\nu}_{(k)}\!\mathcal{P}_{(k)}^{\rho\sigma}-{}^{\mu\nu}\Pi_1^{\rho\sigma}$ which is the projector orthogonal to $\Pi_1$ on the range of ${}_{(k)}^{\alpha\beta}\mathcal{P}_{(k)}^{\gamma\delta}$. The projectors in Eq.~\eqref{eq:SPKgen} satisfy
\begin{align}
	{}^{\mu\nu}_{(k)}\!\mathcal{P}_{(k)}^{\alpha\beta}\;
	{}_{\alpha\beta}^{(n)}\!\mathcal{P}_{(n)}^{\gamma\delta}\;
	{}_{\gamma\delta}^{(k)}\!\mathcal{P}_{(k)}^{\rho\sigma}
	&=
	{}^{\mu\nu}\Pi_1^{\rho\sigma}
	+\frac{(n\cdot k)^2}{n^2k^2}\,{}^{\mu\nu}\Pi_2^{\rho\sigma}~,
	\nonumber\\
	{}^{\mu\nu}_{(k)}\!\mathcal{P}_{(k)}^{\alpha\beta}\;
	{}_{\star\alpha\beta}^{(n)}\mathcal{P}_{(n)}^{\gamma\delta}\;
	{}_{\gamma\delta}^{(k)}\!\mathcal{P}_{(k)}^{\rho\sigma}
	&=
	\frac{(n\cdot k)^2}{n^2k^2}\,{}^{\mu\nu}K_{\text{mix}}^{\rho\sigma}~,
\end{align}
where
\begin{align}
	{}^{\mu\nu}K_{\text{mix}}^{\rho\sigma}
	=
	{}^{\star\mu\nu}_{(n)}\!\mathcal{P}_{(k)}^{\rho\sigma}
	-{}^{\star\mu\nu}_{(k)}\!\mathcal{P}_{(k)}^{\rho\sigma}~.
\end{align}
We also have the identities 
\begin{align}
	{}^{\mu\nu}\Pi_1^{\alpha\beta}\;{}_{\alpha\beta}K_{\text{mix}}^{\rho\sigma}
	&=
	{}^{\mu\nu}K_{\text{mix}}^{\alpha\beta}\;{}_{\alpha\beta}\Pi_1^{\rho\sigma}
	=0~,
	\notag\\
	{}^{\mu\nu}\Pi_2^{\alpha\beta}\;{}_{\alpha\beta}K_{\text{mix}}^{\rho\sigma}
	&=
	{}^{\mu\nu}K_{\text{mix}}^{\alpha\beta}\;{}_{\alpha\beta}\Pi_2^{\rho\sigma}
	={}^{\mu\nu}K_{\text{mix}}^{\rho\sigma}~,
	\notag\\
	{}^{\mu\nu}K_{\text{mix}}^{\alpha\beta}\;{}_{\alpha\beta}K_{\text{mix}}^{\rho\sigma}
	&=
	\left(\frac{n^2k^2}{(n\cdot k)^2}-1\right){}^{\mu\nu}\Pi_2^{\rho\sigma}~,
	\label{eq:KmixSquare}
\end{align}
so ${}^{\mu\nu}K_{\text{mix}}^{\alpha\beta}$ acts entirely within the ${}^{\mu\nu}\Pi_2^{\rho\sigma}$ subspace.
The kinetic term then decomposes as
\begin{align}
	{}^{\mu\nu}\mathbf{K}_{\text{exact}}^{\rho\sigma}
	&=
	{}^{\mu\nu}\mathbf{K}_1^{\rho\sigma}
	+{}^{\mu\nu}\mathbf{K}_2^{\rho\sigma}~,
	\nonumber\\
	{}^{\mu\nu}\mathbf{K}_1^{\rho\sigma}
	&=
	\left(\mathbf{H}+\mathbf{X}\right){}^{\mu\nu}\Pi_1^{\rho\sigma}~,
	\nonumber\\
	{}^{\mu\nu}\mathbf{K}_2^{\rho\sigma}
	&=
	\left(\mathbf{X}+\frac{(n\cdot k)^2}{n^2k^2}\,\mathbf{H}\right){}^{\mu\nu}\Pi_2^{\rho\sigma}
	-\kappa\,\frac{(n\cdot k)^2}{n^2k^2}\,\epsmat\,
	{}^{\mu\nu}\!K_{\text{mix}}^{\rho\sigma}~,
	\label{eq:K1K2decomp}
\end{align}
where each ${}^{\mu\nu}\mathbf{K}_i^{\rho\sigma}$ acts only within the range of ${}^{\mu\nu}\Pi_i^{\rho\sigma}$.

Since ${}^{\mu\nu}\Pi_2^{\rho\sigma}$ is the identity on its own range, diagonalizing ${}^{\mu\nu}\mathbf{K}_2^{\rho\sigma}$ only requires the eigenstates of ${}^{\mu\nu}K_{\text{mix}}^{\rho\sigma}$. For a vector $e^\mu$ orthogonal to the plane spanned by $n^\mu$ and $k^\mu$, we find
\begin{align}
	{}^{\mu\nu}K_{\text{mix}}^{\rho\sigma}\,(
	k_\rho e_\sigma - k_\sigma e_\rho)
	=
	k_\mu e'_\nu
	- k_\nu  e'_\mu ~,
	\qquad
	e'^{\,\nu}
	=
	\frac{1}{n\cdot k}
	\varepsilon^{\mu\nu\rho\sigma}k_\mu n_\rho e_\sigma~.
\end{align}
The map $e^\mu\to e'^{\,\mu}$ changes the direction by a $90^\circ$ rotation in the plane orthogonal to $n^\mu$ and $k^\mu$, and from \eqref{eq:KmixSquare} we see that the eigenvalues of ${}^{\mu\nu}K_{\text{mix}}^{\rho\sigma}$ square to $\frac{n^2k^2}{(n\cdot k)^2} -1$. 

Its eigenstates are the helicity polarizations
\begin{align}
	e_h^\mu
	=
	\frac{1}{\sqrt{2}}\left(e_1^\mu+ih\,e_2^\mu\right),
	\qquad
	h=\pm1~,
\end{align}
where $e_1^\mu$ and $e_2^\mu$ are orthonormal, $e_{i\mu}e_j^\mu=-\delta_{ij}$, span the plane orthogonal to the $n$-$k$ plane,
They are real when $k$ is timelike or null; When $k$ is spacelike, one of them is imaginary.
They are oriented so that
\begin{align}
	\frac{1}{n\cdot k}
	\varepsilon^{\mu\nu \rho \sigma}
	k_\mu n_\rho e_{h\sigma}
	=ih\,\sqrt{1-\frac{n^2k^2}{(n\cdot k)^2}}\;e_h^\mu~.
\end{align}
Explicitly,
\begin{align}
	{}^{\mu\nu}K_{\text{mix}}^{\rho\sigma}\,
	( k_\rho e_{h \sigma}
	- k_\sigma e_{h\rho}
	)
	=
	ih\sqrt{1-\frac{n^2k^2}{(n\cdot k)^2}}\;
	(
	k^\mu e_{h}^{\nu} 
	- k^\nu e_{h}^{\mu})~.
\end{align}

We now expand a general exact two-form in momentum space as
\begin{align}
	\bm{F}_{\bm{A}}^{\mu\nu}(k)
	=
	(k^\mu n^\nu - k^\nu n^\mu)\,\bm{\phi}(k)
	+\sum_{h=\pm1}
	(k^\mu e_h^\nu -k^\nu e_h^\mu )
	\,\bm{a}_h(k)~,
	\label{eq:SPFexpansion}
\end{align}
where $\bm{\phi}$ and $\bm{a}_h$ are two-component vectors in the charge space of the two potentials.
From Eq.~\eqref{eq:K1K2decomp}, ${}^{\mu\nu}\mathbf{K}_{\text{exact}}^{\rho\sigma}$ acts on each piece as
\begin{align}
	{}^{\mu\nu}\mathbf{K}_{\text{exact}}^{\rho\sigma}\,
	(k_\rho n_\sigma - k_\sigma n_\rho)
	\,\bm{\phi}
	&=
	(k^\mu n^\nu  - k^\nu n^\mu )
	\left(\mathbf{X}+\mathbf{H}\right)\bm{\phi}~,
	\nonumber\\
	{}^{\mu\nu}\mathbf{K}_{\text{exact}}^{\rho\sigma}\,
	(k_\rho e_{h\sigma} 
	- k_\sigma e_{h\rho}) \,\bm{a}_h
	&=
	(k^{\mu}e_{h}^{\nu}
	- k^\nu e_{h}^{\mu})
	\,\frac{1}{n^2k^2}\,\mathbf{K}_h\bm{a}_h~,
\end{align}
which defines the $2\times2$ matrix acting on $\bm{a}_h$,
\begin{align}
	\mathbf{K}_h
	=
	n^2k^2\,\mathbf{X}
	+(n\cdot k)^2\left(\mathbf{H}
	-ih\kappa\sqrt{1-\frac{n^2k^2}{(n\cdot k)^2}}\;\epsmat\right).
	\label{eq:SPKh}
\end{align}

Since $\bm{F}_{\bm{A}}^{\mu\nu}=i\left(k^\mu\bm{A}^\nu-k^\nu\bm{A}^\mu\right)$, the coefficients in Eq.~\eqref{eq:SPFexpansion} determine $\bm{A}^\mu$ up to a gauge transformation $\bm{A}^\mu\to\bm{A}^\mu+k^\mu\bm{\alpha}$.
In axial gauge, $n_\mu\bm{A}^\mu=0$, the potential is
\begin{align}
	\bm{A}^\mu(k)
	=
	-i\left[
	\bm{\phi}(k)\left(n^\mu-\frac{n^2}{n\cdot k}\,k^\mu\right)
	+\sum_{h=\pm1}\bm{a}_h(k)\,e_h^\mu
	\right].
	\label{eq:SPAaxial}
\end{align}
We find the following relations from Eq.~\eqref{eq:SPAaxial},
\begin{align}
	\bm{a}_h
	=
	-i\,e^*_{h\mu}\bm{A}^\mu~,
	\qquad
	\left[(n\cdot k)^2-n^2k^2\right]\bm{\phi}
	=
	i\,(n\cdot k)\,k_\mu\bm{A}^\mu~.
	\label{eq:SPcoeffsFromA}
\end{align}
In the specific case of $n^\mu=(0,0,0,1)$ and $k^\mu=(\omega,0,0,p)$ we have $n^\mu-\frac{n^2}{n\cdot k}k^\mu=(-\omega/p,0,0,0)$.
Then $\bm{\phi}$ is, up to normalization, the $\bm{A}_0$ component, and $e_h^\mu=(0,1,ih,0)/\sqrt{2}$ makes $\bm{a}_h$ the transverse helicity polarization.

To obtain the Lagrangian in terms of $\bm{\phi}$ and $\bm{a}_h$ we sandwich ${}^{\mu\nu}\mathbf{K}_{\text{exact}}^{\rho\sigma}$ between two copies of Eq.~\eqref{eq:SPFexpansion}.
Using $n_\mu e_h^\mu=k_\mu e_h^\mu=0$ and $e^*_{h\mu}e_{h'}^\mu=-\delta_{hh'}$, the inner products of the basis two-forms are
\begin{align}
	(k_\mu n_\nu -k_\nu n_\mu)
	(k^\mu n^\nu - k^\nu n^\mu)
	&=
	-2\left[(n\cdot k)^2-n^2k^2\right],
	\notag\\
	(k_\mu n_\nu - k_\nu n_\mu)
	(k^\mu e_h^\nu - k^\nu e_h^\mu)
	&=0~,
	\nonumber\\
	(k_\mu e_{h\nu}^* 
	-k_\nu e_{h\mu}^*)
	(k^\mu e_{h'}^\nu - k^\nu e_{h'}^\mu)
	&=
	-2k^2\,\delta_{hh'}~.
\end{align}
We find
\begin{align}
	\mathcal{L}_2(k)
	&=
	-\frac{1}{4}\,
	\bm{F}_{\bm{A}\mu\nu}^T \,
	{}^{\mu\nu}\mathbf{K}_{\text{exact}}^{\rho\sigma}\,
	\bm{F}_{\bm{A}\rho\sigma}
	\nonumber\\
	&=
	\frac{1}{2}\left[(n\cdot k)^2-n^2k^2\right]
	\bm{\phi}^\dagger\left(\mathbf{X}+\mathbf{H}\right)\bm{\phi}
	+\frac{1}{2n^2}\sum_{h=\pm1}\bm{a}_h^\dagger\,\mathbf{K}_h\,\bm{a}_h~.
	\label{eq:SPL2coeffs}
\end{align}
Every entry is a polynomial in $k^\mu$. The square root in $\mathbf{K}_h$ comes from the eigenvalue of ${}^{\mu\nu}K_{\text{mix}}^{\rho\sigma}$, and equals $1$ on the light cone.
The $\bm{\phi}$ term does not describe a propagating mode.
The equation of motion for $\bm{\phi}$ is, $(n\cdot k)(\mathbf{X}+\mathbf{H})\,k_\mu\bm{A}^\mu=0$, which is the axial-gauge Gauss law. This is a constraint, and it requires $\det(\mathbf{X}+\mathbf{H})\neq0$.

We now study the propagating photon modes by finding the source-free plane-wave solutions.
The momentum space equation of motion for $\bm{a}_h$ is given by
\begin{align}
	\mathbf{K}_h\bm{a}_h=0
	~.
	\label{eq:KhNullSpace}
\end{align} 
Therefore, the propagating photon solutions are given by the null space of $\mathbf{K}_h$.
We can find the dispersion relation for these photons by solving for the zeroes of the determinant of $\mathbf{K}_h$.

For a $2\times2$ matrix $\mathbf{M}$ we denote the adjugate by
\begin{align}
	\tilde{\mathbf{M}}
	=
	-\epsmat\,\mathbf{M}^T\epsmat~,
	\label{eq:adjugateDef}
\end{align}
which satisfies $\mathbf{M}\tilde{\mathbf{M}}=\tilde{\mathbf{M}}\mathbf{M}=\det\mathbf{M}$.
If $\mathbf{M}$ is invertible then $\tilde{\mathbf{M}} = \det(\mathbf{M})\, \mathbf{M}^{-1}$.
For symmetric $2\times2$ matrices $\mathbf{X}$, $\mathbf{H}$, $\mathbf{M}$ and numbers $\alpha$, $\beta$, $b$ we have
\begin{align}
	\det\left(\alpha\mathbf{X}+\beta\mathbf{H}\right)
	&=
	\alpha^2\det\mathbf{X}
	+\alpha\beta\,\text{tr}\left(\tilde{\mathbf{H}}\mathbf{X}\right)
	+\beta^2\det\mathbf{H}~,
	\nonumber\\
	\text{tr}\left(\tilde{\mathbf{H}}\mathbf{X}\right)
	&=
	H_{11}X_{22}+H_{22}X_{11}-2H_{12}X_{12}~,
	\nonumber\\
	\det\left(\mathbf{M}+b\,\epsmat\right)
	&=
	\det\mathbf{M}+b^2~.
\end{align}
Applying these identities to Eq.~\eqref{eq:SPKh} we find the same result for both helicities
\begin{align}
	\det\mathbf{K}_h
	=
	\left(n^2k^2\right)^2\det\mathbf{X}
	+(n\cdot k)^2\,n^2k^2\left[\kappa^2+\text{tr}\left(\tilde{\mathbf{H}}\mathbf{X}\right)\right]
	+(n\cdot k)^4\left[\det\mathbf{H}-\kappa^2\right].
	\label{eq:SPdetKh}
\end{align}
The factor $(n\cdot k)^2$ is not a photon: it vanishes where the axial gauge~\eqref{eq:SPAaxial} breaks down.
The axial gauge constraint is only defined when $n\cdot k \neq 0$, therefore the zero we should consider the zeros of this determinant away from the set $n\cdot k = 0$.  
At fixed $n\cdot k$, Eq.~\eqref{eq:SPdetKh} is quadratic in $n^2k^2$.

If we define $r = \frac{n^2 k^2 }{(n\cdot k)^2}$, then the zeros of $\det \mathbf{K}_h$ satisfy 
\begin{align}
	r^2 \det \mathbf{X}
	+ r \left[
	\kappa^2 + \text{tr}
	\left(
	\tilde{\mathbf{H}} \mathbf{X}
	\right)
	\right]
	+ [\mathbf{H} - \kappa^2] &= 0.
	\label{eq:SPrEquation}
\end{align}
Without loss of generality we take $n^\mu = (0,\mathbf{n})$, and write $k^\mu =(\omega,\mathbf{k})$.
Then if we write $\mathbf{k}=\mathbf{k}_\perp+ k_n\hat{\mathbf{n}}$, the definition of $r$ gives
\begin{align}
	\omega^2=\mathbf{k}^2-r\,k_n^2~,
	\label{eq:SPdispersion}
\end{align}
Any $r\neq0$ makes the string direction observable.
Each root $r$ is a separate dispersion relation, i.e.\ a separate photon.
A single photon on the light cone therefore requires Eq.~\eqref{eq:SPrEquation} to have exactly one root, equal to zero.

The root $r=0$ is present only if
\begin{align}
	\det\mathbf{H}=\kappa^2~.
\end{align}
In order to avoid a second root we must also have
\begin{align}
	\det\mathbf{X}=0~.
\end{align}
Note that the dispersion relation by itself does not rule out a nonzero $\mathbf{X}$.
By the results of the next section, a nonzero $\mathbf{X}$ leads to a propagator that violates Lorentz invariance, so we set $\mathbf{X}=0$ for the remainder of this section.
The single photon condition is then $\det\mathbf{H}=\kappa^2$, and $\det(\mathbf{H}+\mathbf{X})=\det\mathbf{H}>0$ holds automatically.
The only pole is then the light cone, $k^2=0$.

The propagator for a fixed helicity is given by 
\begin{align}
	i\,\mathbf{K}_h^{-1}
	=
	i\,\frac{\tilde{\mathbf{K}}_h}{\det\mathbf{K}_h}
	=
	\frac{i}{ n^2 k^2}
	\left[
	\mathbf{H}^{-1}
	+\frac{ih}{\kappa}\sqrt{1-\frac{n^2k^2}{(n\cdot k)^2}}\;\epsmat
	\right].
\end{align}
For the photon to have positive norm the residue of the photon pole must be positive.
Since $\det\mathbf{H}=\kappa^2>0$, the real symmetric matrix $\mathbf{H}$ is either positive definite or negative definite
On the light cone the square root equals $1$, so the pole at $k^2=0$ multiplies the matrix
\begin{align}
	\mathbf{H}^{-1}+\frac{ih}{\kappa}\,\epsmat~.
\end{align}
It is Hermitian, and its determinant is $1/\det\mathbf{H}-1/\kappa^2=0$, so it has rank one which tells us there is a single propagating mode for a given helicity.
Thus the two potentials therefore describe a single photon with two helicities.
Its single nonzero eigenvalue is therefore its trace, $\text{tr}\,\mathbf{H}^{-1}$ which must be positive.
We therefore see that $\mathbf{H}^{-1}$, and thus $\mathbf{H}$, must be positive definite.

\subsection{Correlation Functions}\label{app:singlePhotonProp}

By the argument of the previous section, $\det \mathbf{X} = 0$, so $\mathbf{X}$ has rank one.
After a rotation of the two potentials, which leaves $\kappa$ unchanged and keeps $\mathbf{H}$ symmetric, we can write it without loss of generality as
\begin{align}
	\mathbf{X}
	=
	\begin{pmatrix}
		x & 0 \\
		0 & 0
	\end{pmatrix}~.
\end{align}
We then need to find the inverse of ${}^{\mu\nu}\mathbf{K}_{\text{exact}}^{\rho\sigma}$.

In order to do this inversion, we make use of the block diagonal decomposition given in  Eq. \eqref{eq:K1K2decomp}.
We can first invert the operator on the $\Pi_1$ subspace and then on the $\Pi_2$ subspace.
The $\Pi_1$ subspace is one dimensional in two-form space, so we need only invert the $2 \times 2$ matrix in charge space.
The inverse on the $\Pi_1$ subspace is then
\begin{align}
	{}^{\mu\nu}\!\left(\mathbf{K}_{\text{exact}}^{-1}\right)^{\rho\sigma} \Big|_{\Pi_1}
	=
	\left(\mathbf{X} + \mathbf{H}\right)^{-1}
	{}^{\mu\nu}\Pi_1^{\rho\sigma}~.
\end{align}

We now turn to the inverse on the $\Pi_2$ subspace.
The operator we must invert is ${}^{\mu\nu}\mathbf{K}_2^{\rho\sigma}$ of Eq.~\eqref{eq:K1K2decomp},
\begin{align}
	{}^{\mu\nu}\mathbf{K}_{\text{exact}}^{\rho\sigma}\Big|_{\Pi_2}
	=
	\left(
	\mathbf{X}
	+\frac{(n\cdot k)^2}{n^2k^2}\,\mathbf{H}
	\right){}^{\mu\nu}\Pi_2^{\rho\sigma}
	-\kappa\,\frac{(n\cdot k)^2}{n^2k^2}\,
	\epsmat\,{}^{\mu\nu}K_{\text{mix}}^{\rho\sigma}~.
\end{align}

Using the notation of \eqref{eq:adjugateDef} for the specific choice $\mathbf{M} = \mathbf{X} + \frac{(n\cdot k)^2}{n^2k^2} \mathbf{H}$, then we have
\begin{align}
	&\left[
	\mathbf{M}\,{}^{\mu\nu}\Pi_2^{\alpha\beta}
	-\kappa\,\frac{(n\cdot k)^2}{n^2k^2}\,
	\epsmat\,{}^{\mu\nu}K_{\text{mix}}^{\alpha\beta}
	\right]
	\left[
	\tilde{\mathbf{M}}\,{}_{\alpha\beta}\Pi_2^{\rho\sigma}
	+\kappa\,\frac{(n\cdot k)^2}{n^2k^2}\,
	\epsmat\,{}_{\alpha\beta}K_{\text{mix}}^{\rho\sigma}
	\right]
	\notag\\
	&\qquad=
	\det\left(\mathbf{X} + \frac{(n\cdot k)^2}{n^2k^2}\mathbf{H}\right)
	{}^{\mu\nu}\Pi_2^{\rho\sigma}
	+\kappa^2\,\frac{(n\cdot k)^4}{n^4k^4}\,
	{}^{\mu\nu}K_{\text{mix}}^{\alpha\beta}\;{}_{\alpha\beta}K_{\text{mix}}^{\rho\sigma}
	\notag\\
	&\qquad=
	\det\left(\mathbf{X} + \frac{(n\cdot k)^2}{n^2k^2}\mathbf{H}\right)
	{}^{\mu\nu}\Pi_2^{\rho\sigma}
	+\kappa^2\,\frac{(n\cdot k)^4}{n^4k^4}
	\left(\frac{n^2k^2}{(n\cdot k)^2}-1\right)
	{}^{\mu\nu}\Pi_2^{\rho\sigma}~,
\end{align}
where we used Eq.~\eqref{eq:K1K2decomp}, and the cross terms cancel due to the identity $\mathbf{M}\epsmat = \epsmat\tilde{\mathbf{M}}$.
We also have
\begin{align}
	\det\left(\mathbf{X} + \frac{(n\cdot k)^2}{n^2k^2}\mathbf{H}\right)
	=
	\frac{(n\cdot k)^2}{n^2k^2}\, xH_{22}
	+\frac{(n\cdot k)^4}{n^4k^4}\det\mathbf{H}~.
\end{align}
Therefore, we have
\begin{align}
	&\left[
	\mathbf{M}\,{}^{\mu\nu}\Pi_2^{\alpha\beta}
	-\kappa\,\frac{(n\cdot k)^2}{n^2k^2}\,
	\epsmat\,{}^{\mu\nu}K_{\text{mix}}^{\alpha\beta}
	\right]
	\left[
	\tilde{\mathbf{M}}\,{}_{\alpha\beta}\Pi_2^{\rho\sigma}
	+\kappa\,\frac{(n\cdot k)^2}{n^2k^2}\,
	\epsmat\,{}_{\alpha\beta}K_{\text{mix}}^{\rho\sigma}
	\right]
	\notag\\
	&\qquad=
	\left[
	\frac{(n\cdot k)^2}{n^2k^2}\left(xH_{22}+\kappa^2\right)
	+\frac{(n\cdot k)^4}{n^4k^4}\left(\det\mathbf{H}-\kappa^2\right)
	\right]
	{}^{\mu\nu}\Pi_2^{\rho\sigma}
	\notag\\
	&\qquad=
	\frac{(n\cdot k)^2}{n^2k^2}\left(xH_{22}+\kappa^2\right)
	{}^{\mu\nu}\Pi_2^{\rho\sigma}~,
\end{align}
where we have simplified this using the single photon condition $\det \mathbf{H} = \kappa^2$.
Then we note that we can write
\begin{align}
	\det\left(\mathbf{X} + \mathbf{H}\right)
	=
	xH_{22}+\kappa^2~.
\end{align}
Rescaling, we find that the inverse of the kinetic term on the $\Pi_2$ subspace is
\begin{align}
	{}^{\mu\nu}\!\left(\mathbf{K}_{\text{exact}}^{-1}\right)^{\rho\sigma}\Big|_{\Pi_2}
	=
	\frac{n^2k^2}{(n\cdot k)^2}\,
	\frac{1}{xH_{22}+\kappa^2}
	\left(
	\tilde{\mathbf{M}}\,{}^{\mu\nu}\Pi_2^{\rho\sigma}
	+\kappa\,\frac{(n\cdot k)^2}{n^2k^2}\,
	\epsmat\,{}^{\mu\nu}K_{\text{mix}}^{\rho\sigma}
	\right).
\end{align}
Note that we have
\begin{align}
	\frac{n^2k^2}{(n\cdot k)^2}\,
	\frac{1}{\det\left(\mathbf{X}+\mathbf{H}\right)}\,
	\tilde{\mathbf{M}}
	&=
	\frac{1}{\det\left(\mathbf{X}+\mathbf{H}\right)}
	\left(
	\tilde{\mathbf{H}}
	+\frac{n^2k^2}{(n\cdot k)^2}\,\tilde{\mathbf{X}}
	\right)
	\notag\\
	&=
	\left(\mathbf{X}+\mathbf{H}\right)^{-1}
	+\frac{n^2k^2-(n\cdot k)^2}{(n\cdot k)^2\left(xH_{22}+\kappa^2\right)}\,
	\tilde{\mathbf{X}}~,
\end{align}
where in the second line we used $\tilde{\mathbf{X}}+\tilde{\mathbf{H}}=\det(\mathbf{X}+\mathbf{H})\,(\mathbf{X}+\mathbf{H})^{-1}$.
Combining the results on the ${}^{\mu\nu}\Pi_1^{\rho\sigma}$ and ${}^{\mu\nu}\Pi_2^{\rho\sigma}$ subspaces, and using ${}^{\mu\nu}\Pi_1^{\rho\sigma}+{}^{\mu\nu}\Pi_2^{\rho\sigma}={}^{\mu\nu}_{(k)}\!\mathcal{P}_{(k)}^{\rho\sigma}$, we find that the correlation functions are given in terms of the inverse kinetic term on the entire range of ${}^{\mu\nu}_{(k)}\!\mathcal{P}_{(k)}^{\rho\sigma}$,
\begin{align}
	\langle
	\bm{F}_{\bm{A}}^{\mu\nu}
	\bm{F}_{\bm{A}}^{\rho\sigma}
	\rangle
	&=
	-2i\,
	{}^{\mu\nu}\!\left(\mathbf{K}_{\text{exact}}^{-1}\right)^{\rho\sigma}
	\notag\\
	&=
	-2i\left(\mathbf{X}+\mathbf{H}\right)^{-1}
	{}^{\mu\nu}_{(k)}\!\mathcal{P}_{(k)}^{\rho\sigma}
	-\frac{2i\kappa}{xH_{22}+\kappa^2}\,
	\epsmat\,{}^{\mu\nu}K_{\text{mix}}^{\rho\sigma}
	\notag\\
	&\quad
	-\frac{2i}{xH_{22}+\kappa^2}\,
	\frac{n^2k^2-(n\cdot k)^2}{(n\cdot k)^2}\,
	\tilde{\mathbf{X}}\,{}^{\mu\nu}\Pi_2^{\rho\sigma}~.
	\label{eq:CFwithX}
\end{align}

This, finally, allows us to rule out a nonzero $\mathbf{X}$.
Notice that
\begin{align}
	\tilde{\mathbf{X}}
	=
	\begin{pmatrix}
		0 & 0\\
		0 & x
	\end{pmatrix},
\end{align}
so the last term in Eq.~\eqref{eq:CFwithX} appears only in $\langle F_B^{\mu\nu}F_B^{\rho\sigma}\rangle$.
Consider then the tree-level amplitude in a purely magnetic theory with current $\bm{J}^\mu = (0,1)^T K^\mu$ and surface operator $\Sigma_K^{\mu\nu}$.
In momentum space= Eq.~\eqref{eq:SigmaDef} gives us $k_\mu\Sigma_K^{\mu\nu}=K^\nu$.
The amplitude is given in terms of the current as $K^\mu$
\begin{align}
	\frac{1}{2}\Sigma_{K\mu\nu}(k)\,
	\langle
	F_B^{\mu\nu}
	F_B^{\rho\sigma}
	\rangle_k \,
	\frac{1}{2}\Sigma_{K\rho\sigma}^*(k)
	&=
	-\frac{H_{11}}{xH_{22}+\kappa^2}\,
	\frac{i}{k^2}\,
	K^\mu K_\mu^*\\
	&\quad
	-\frac{x}{xH_{22}+\kappa^2}\,
	\frac{i}{(n\cdot k)^2}
	\left[
	n^2K^\mu K_\mu^*- (n_\mu K^\mu) 
	(n_\mu K^{*\mu})
	\right]~.\nonumber
\end{align}
We see that there is an $n$-dependent interaction even in a purely magnetic theory.
This term is symmetric in the charges and is present for a single species, so it is not one of the intersection-number terms that exponentiate to a trivial phase.
This $n$-dependence therefore cannot be removed by a quantization condition, and we conclude that the theory truly is not Lorentz invariant.
Therefore we set $\mathbf{X} = 0$ and find the correlators
\begin{align}
	\langle
	\bm{F}_{\bm{A}}^{\mu\nu}
	\bm{F}_{\bm{A}}^{\rho\sigma}
	\rangle_k
	&=
	-2i\,\mathbf{H}^{-1}\,
	{}^{\mu\nu}_{(k)}\!\mathcal{P}_{(k)}^{\rho\sigma}
	-\frac{2i}{\kappa}\,
	\epsmat\,{}^{\mu\nu}K_{\text{mix}}^{\rho\sigma}~,
	\label{eq:CFfinal}
\end{align}
where
\begin{align}
	{}^{\mu\nu}K_{\text{mix}}^{\rho\sigma}
	=
	{}^{\star\mu\nu}_{(n)}\!\mathcal{P}_{(k)}^{\rho\sigma}
	-{}^{\star\mu\nu}_{(k)}\!\mathcal{P}_{(k)}^{\rho\sigma}
	=
	{}^{\mu\nu}_{(k)}\!\mathcal{P}_{(k)}^{\alpha\beta}\;
	{}_{\star\alpha\beta}^{(n)}\mathcal{P}_{(k)}^{\rho\sigma}~.
\end{align}
The last form of ${}^{\mu\nu}K_{\text{mix}}^{\rho\sigma}$ is the one used in the main text.

\section{Degree-Two Argyres--Douglas Isogeny}\label{app:ADNormalization}

This appendix derives the degree-two period and charge map used in Section~\ref{sec:ADComparison}.  
The map keeps the primitive electric generator fixed and doubles the magnetic generator, so it explains the factor of two in one protected contribution without supplying an additivity argument for mutually nonlocal states.

Argyres and Douglas use a primitive electromagnetic charge lattice
\begin{align}
\Gammaem^{\rm AD}
=\mathbb Z\,e_{\rm AD}\oplus\mathbb Z\,m_{\rm AD}~,
\end{align}
with central charge
\begin{align}
Z=q_{\rm m}^{\rm AD}a_D^{\rm AD}+q_{\rm e}^{\rm AD}a^{\rm AD}~.
\end{align}
Consider the index-two magnetic charge lattice
\begin{align}
\Gammaem^{(2)}
=\mathbb Z\,e_{\rm AD}\oplus\mathbb Z(2m_{\rm AD})
\subset\Gammaem^{\rm AD}~,
\qquad
[\Gammaem^{\rm AD}:\Gammaem^{(2)}]=2~.
\label{eq:ADindexTwoLattice}
\end{align}
This choice keeps the unit electric state integral. A dynamical state in $\Gammadyn^{\rm AD}$ with odd $q_{\rm m}^{\rm AD}$ instead has half-integral magnetic coordinates in the new convention.

In special-K\"ahler terminology the antisymmetric pairing of $\Gammaem^{(2)}$ has polarization two.  
For one Abelian vector multiplet, a polarization-$\dlat_{\rm pol}$ description has local kinetic term and metric~\cite{Argyres:2022hel}
\begin{align}
\mathcal L_{\rm kin}^{(\dlat_{\rm pol})}
&=-\frac{\operatorname{Im}\tau}{16\pi \dlat_{\rm pol}}F_h^2
+\frac{\operatorname{Re}\tau}{16\pi \dlat_{\rm pol}}F_h{}^\ast F_h~,
\label{eq:polarizedKineticTerm}\\
ds^2&=2\operatorname{Im}\!\left(da_D\,\dlat_{\rm pol}^{-1}d\bar a\right)~.
\end{align}
The AD periods have $\dlat_{{\rm pol},\rm AD}=1$.  The period map is non-unimodular: in the ordered basis $(a_D,a)$ it is represented by $\operatorname{diag}(2,1)$ and has determinant two. It is therefore not an element of the classical duality group SL$(2,\mathbb R)$ or an integral automorphism of $\Gammaem^{\rm AD}$.  
Instead it maps between period and charge lattices with different polarization.

The isogenous polarization-two variables are
\begin{align}
a&=a^{\rm AD},
& a_D&=2a_D^{\rm AD}~,
& F_h&=F_h^{\rm AD}~,
\nonumber\\
\tau&=2\tau_{\rm AD}~,
& Q_{\rm e}&=q_{\rm e}^{\rm AD}~,
& Q_{\rm m}&=\frac12q_{\rm m}^{\rm AD}~,
& \dlat_{\rm pol}&=2~.\label{eq:ADnormalizationMap}
\end{align}
These transformations leave the BPS masses, the gauge-field kinetic terms, and the geometry of the Coulomb branch unchanged:
\begin{align}
Q_{\rm m}a_D+Q_{\rm e}a
&=q_{\rm m}^{\rm AD}a_D^{\rm AD}+q_{\rm e}^{\rm AD}a^{\rm AD}~,
\nonumber\\
\frac{\tau}{\dlat_{\rm pol}}F_h^2
&=\tau_{\rm AD}(F_h^{\rm AD})^2~,
\nonumber\\
da_D\,\dlat_{\rm pol}^{-1}d\bar a
&=da_D^{\rm AD}d\bar a^{\rm AD}~.
\end{align}
The antisymmetric pairing is correspondingly
\begin{align}
\langle\gamma,\gamma'\rangle_{\rm AD}
=2\left(Q_{\rm m}Q_{\rm e}'-Q_{\rm e}Q_{\rm m}'\right)~.
\end{align}
Thus $\Gammaem^{(2)}$ has polarization two even though the physical DSZ phase of any state pair being compared is unchanged.

The $\beta$-function transforms as
\begin{align}
\beta_\tau^{\rm W}=2\beta_{\tau_{\rm AD}}^{\rm W}~,
\qquad
Q_{\rm e}+Q_{\rm m}\tau
=q_{\rm e}^{\rm AD}+q_{\rm m}^{\rm AD}\tau_{\rm AD}~.
\end{align}
The protected single-hypermultiplet contribution therefore becomes
\begin{align}
\beta_\tau^{\rm W}
=-\frac{i}{\pi}\left(Q_{\rm e}+Q_{\rm m}\tau\right)^2~,
\end{align}
which is the conventional $\dlat=2$ hypermultiplet coefficient.  For a unit electric state, $\tau_{\rm AD}\to\tau_{\rm AD}+1$ becomes $\tau\to\tau+2$, so the monodromy gives the same factor-of-two check.

A degree-two isogeny relates elliptic fibrations whose period lattices differ by precisely such an index-two inclusion~\cite{Argyres:2022hel,Closset:2023isogeny}.  The isogenous curves need not describe the same absolute theory, because their electromagnetic charge lattices, quadratic refinements, and associated one-form symmetry data can differ.  
This is why the map provides a clean local comparison of the protected Wilsonian coefficients in the AD and standard conventions without embedding the full $\Gammadyn^{\rm AD}$ into one common integral charge lattice. If the additive Wilsonian ansatz is adopted, the map may be applied to each protected term, but the isogeny itself supplies no dynamical argument that this sum equals the full mutually nonlocal 1PI $\beta$-function.

\end{document}